%Paper: cond-mat/9406003
%From: zaf@amoco.saclay.cea.fr
%Date: Wed, 1 JUN 94 11:55 GMT

% Plain TeX file (tex it twice)
\def\service{T}
% *************************** output macros ********************************
\catcode`\@=11
%%% Universal (European) A4 format:
\def\unredoffs{\voffset=11mm \hoffset=0.5mm}

%
%---------------------------------------------------------------------%
\newbox\leftpage \newdimen\fullhsize \newdimen\hstitle \newdimen\hsbody
\newdimen\hdim
\tolerance=400\pretolerance=800
%\tolerance=1000\hfuzz=2pt
%\def\fontflag{cm}
%
%
\newif\ifsmall \smallfalse
\newif\ifdraft \draftfalse
\newif\iffrench \frenchfalse
\newif\ifeqnumerosimple \eqnumerosimplefalse
\nopagenumbers
\headline={\ifnum\pageno=1\hfill\else\hfil{\headrm\folio}\hfil\fi}
\def\draftstart{
\magnification=1200 \unredoffs\hsize=130mm\vsize=190mm
\hsbody=\hsize \hstitle=\hsize %take default values for unreduced format
\nolabels
\iffrench
% \fhyph
\dicof
\else
\dicoa
\fi
}
% ****************************************************************************
%**************************** MAC.TEX ***********************************
%***************************************************************************
% origine: harvmac + modifications J. Zinn-Justin
%  + modifications J.-M. Drouffe
% fonts, Dirac slash

\font\elevrm=cmr9

\newdimen\chapskip
\font\twbf=cmssbx10 scaled 1200
\font\ssbx=cmssbx10

\font\caprm=cmr9
\font\capit=cmti9
\font\capbf=cmbx9
\font\capsl=cmsl9
\font\capmi=cmmi9
\font\capex=cmex9
\font\capsy=cmsy9
\chapskip=17.5mm
\def\makeheadline{\vbox to 0pt{\vskip-22.5pt
\line{\vbox to8.5pt{}\the\headline}\vss}\nointerlineskip}
%***************************************************
\font\tbfi=cmmib10
\font\tenbi=cmmib7
\font\fivebi=cmmib5
\textfont4=\tbfi
\scriptfont4=\tenbi
\scriptscriptfont4=\fivebi
\font\headrm=cmr10

%****************************
\font\eightrm=cmr6
\font\sixrm=cmr5
\font\eightmi=cmmi6
\font\sixmi=cmmi5
\font\eightsy=cmsy6
\font\sixsy=cmsy5
\font\eightbf=cmbx6
\font\sixbf=cmbx5
\skewchar\capmi='177 \skewchar\eightmi='177 \skewchar\sixmi='177
\skewchar\capsy='60 \skewchar\eightsy='60 \skewchar\sixsy='60

\def\elevenpoint{
\textfont0=\caprm \scriptfont0=\eightrm \scriptscriptfont0=\sixrm
\def\rm{\fam0\caprm}
\textfont1=\capmi \scriptfont1=\eightmi \scriptscriptfont1=\sixmi
\textfont2=\capsy \scriptfont2=\eightsy \scriptscriptfont2=\sixsy
\textfont3=\capex \scriptfont3=\capex \scriptscriptfont3=\capex
\textfont\itfam=\capit \def\it{\fam\itfam\capit} % \it is family 4
\textfont\slfam=\capsl  \def\sl{\fam\slfam\capsl} % \sl is family 5
\textfont\bffam=\capbf \scriptfont\bffam=\eightbf
\scriptscriptfont\bffam=\sixbf
\def\bf{\fam\bffam\capbf} % \bf is family 6
\textfont4=\tbfi \scriptfont4=\tenbi \scriptscriptfont4=\tenbi
\normalbaselineskip=13pt
\setbox\strutbox=\hbox{\vrule height9.5pt depth3.9pt width0pt}
\let\big=\elevenbig \normalbaselines \rm}

\catcode`\@=11

\font\tenmsa=msam10
\font\sevenmsa=msam7
\font\fivemsa=msam5
\font\tenmsb=msbm10
\font\sevenmsb=msbm7
\font\fivemsb=msbm5
\newfam\msafam
\newfam\msbfam
\textfont\msafam=\tenmsa  \scriptfont\msafam=\sevenmsa
  \scriptscriptfont\msafam=\fivemsa
\textfont\msbfam=\tenmsb  \scriptfont\msbfam=\sevenmsb
  \scriptscriptfont\msbfam=\fivemsb

\def\hexnumber@#1{\ifcase#1 0\or1\or2\or3\or4\or5\or6\or7\or8\or9\or
	A\or B\or C\or D\or E\or F\fi }

%  The following 13 lines establish the use of the Euler Fraktur font.
%  To use this font, remove % from beginning of these lines.
\font\teneuf=eufm10
\font\seveneuf=eufm7
\font\fiveeuf=eufm5
\newfam\euffam
\textfont\euffam=\teneuf
\scriptfont\euffam=\seveneuf
\scriptscriptfont\euffam=\fiveeuf
\def\frak{\ifmmode\let\next\frak@\else
 \def\next{\Err@{Use \string\frak\space only in math mode}}\fi\next}
\def\goth{\ifmmode\let\next\frak@\else
 \def\next{\Err@{Use \string\goth\space only in math mode}}\fi\next}
\def\frak@#1{{\frak@@{#1}}}
\def\frak@@#1{\fam\euffam#1}
%  End definition of Euler Fraktur font.

\edef\msa@{\hexnumber@\msafam}
\edef\msb@{\hexnumber@\msbfam}

\def\Bbb{\ifmmode\let\next\Bbb@\else
 \def\next{\errmessage{Use \string\Bbb\space only in math mode}}\fi\next}
\def\Bbb@#1{{\Bbb@@{#1}}}
\def\Bbb@@#1{\fam\msbfam#1}

\catcode`\@=12
\def\sla#1{\mkern-1.5mu\raise0.4pt\hbox{$\not$}\mkern1.2mu #1\mkern 0.7mu}
\def\Dbar{\mkern-1.5mu\raise0.4pt\hbox{$\not$}\mkern-.1mu {\rm D}\mkern.1mu}
\def\Abar{\mkern1.mu\raise0.4pt\hbox{$\not$}\mkern-1.3mu A\mkern.1mu}
% ****************************************************************************
%       Dictionnaires francais et anglais
\def\dicof{
\gdef\Resume{RESUME}
\gdef\Toc{Table des mati\`eres}
\gdef\soumisa{Soumis \`a:}
}
\def\dicoa{
\gdef\Resume{ABSTRACT}
\gdef\Toc{Table of Contents}
\gdef\soumisa{Submitted to}
}
% ****** extrait de definit.tex (obsolete ?)

\def\uniset{\rlap{\elevrm 1}\kern.15em 1}
\def\bkR{{\rm I\kern-.17em R}}
\def\bkC{{\rm \kern.24em
            \vrule width.05em height1.4ex depth-.05ex
            \kern-.26em C}}
% ********* A few math symbols

\def\frac#1#2{{\textstyle{#1\over#2}}}

\def\leaderfill{\leaders\hbox to 1em{\hss.\hss}\hfill}
% ******************** LOGOS **********************************************
\def\saclay{\if S\service \spec \else \spht \fi}
\def\spht{
\centerline{CEA, Service de Physique Th\'eorique, CE-Saclay}
\centerline{F-91191 Gif-sur-Yvette Cedex, FRANCE}}
\def\spec{
\centerline{CEA, Service de Physique de l'Etat Condens\'e, CE-Saclay}
\centerline{F-91191 Gif-sur-Yvette Cedex, FRANCE}}
\def\logo{
\if S\service % Logo SPEC
\font\sstw=cmss10 scaled 1200
\font\ssx=cmss8
\vtop{\hsize 9cm
{\sstw {\twbf P}hysique de l'{\twbf E}tat {\twbf C}ondens\'e \par}
\ssx SPEC -- DRECAM -- DSM\par
\vskip 0.5mm
\sstw CEA -- Saclay \par
}
\else % Logo SPHT
\vtop{\hsize 9cm
}
\fi }
% *************************************************************************
\catcode`\@=11
% ************** double alignment in eqalignno style **********************
\def\deqalignno#1{\displ@y\tabskip\centering \halign to
\displaywidth{\hfil$\displaystyle{##}$\tabskip0pt&$\displaystyle{{}##}$
\hfil\tabskip0pt &\quad
\hfil$\displaystyle{##}$\tabskip0pt&$\displaystyle{{}##}$
\hfil\tabskip\centering& \llap{$##$}\tabskip0pt \crcr #1 \crcr}}
% ************** double eqalign ******************************************
\def\deqalign#1{\null\,\vcenter{\openup\jot\m@th\ialign{
\strut\hfil$\displaystyle{##}$&$\displaystyle{{}##}$\hfil
&&\quad\strut\hfil$\displaystyle{##}$&$\displaystyle{{}##}$
\hfil\crcr#1\crcr}}\,}
%***************************************************************************
%********* titlepage, headline, section, subsection, sub, appendix *********
%***************************************************************************
%********* introduce equation number file: for non-causal quotation
\openin 1=\jobname.sym
\ifeof 1\closein1\message{<< (\jobname.sym DOES NOT EXIST) >>}\else%
\input\jobname.sym\closein 1\fi
\newcount\nosection
\newcount\nosubsection
\newcount\neqno
\newcount\notenumber
\newcount\figno
\newcount\tabno
\def\content{\jobname.toc}
\def\symbols{\jobname.sym}
\newwrite\toc
\newwrite\sym
%\newwrite\Fig
%\newwrite\Tab
% ******************* titlepage **********************************
%\def\authorname#1{\maketitle{\bf #1}\smallskip}
\def\authorname#1{\centerline{\bf #1}\smallskip}
\def\address#1{ #1\medskip}
\newdimen\hulp
\def\maketitle#1{
\edef\oneliner##1{\centerline{##1}}
\edef\twoliner##1{\vbox{\parindent=0pt\leftskip=0pt plus 1fill\rightskip=0pt
plus 1fill
                     \parfillskip=0pt\relax##1}}
\setbox0=\vbox{#1}\hulp=0.5\hsize
                 \ifdim\wd0<\hulp\oneliner{#1}\else
                 \twoliner{#1}\fi}

\def\submitted#1{{\it {\soumisa} #1}\par}
% **************** beginning
\def\title#1{\gdef\titlename{#1}
\maketitle{
%\ssbx\uppercase\expandafter
\twbf
{\titlename}}
\vskip3truemm\vfill
\nosection=0
\neqno=0
\notenumber=0
\figno=1
\tabno=1
\def\prefix{}
\def\eqprefix{}
\mark{\the\nosection}
\message{#1}
\immediate\openout\sym=\symbols
}
\def\preprint#1{\vglue-10mm
\line{ \logo \hfill {#1} }\vglue 20mm\vfill}
\def\abstract{\vfill\centerline{\Resume} \smallskip \begingroup\narrower
\elevenpoint\baselineskip10pt}
\def\endabstract{\par\endgroup \bigskip}
% ***************** input table of contents
\def\mktoc{\centerline{\bf \Toc} \medskip\caprm
\parindent=2em
\openin 1=\jobname.toc
\ifeof 1\closein1\message{<< (\jobname.toc DOES NOT EXIST. TeX again)>>}%
\else\input\jobname.toc\closein 1\fi
 \bigskip}
%******************************* section ***********************************
\def\section#1\par{\vskip0pt plus.1\vsize\penalty-100\vskip0pt plus-.1
\vsize\bigskip\vskip\parskip
\message{ #1}
\ifnum\nosection=0\immediate\openout\toc=\content%
\edef\ecrire{\write\toc{\par\noindent{\ssbx\ \titlename}
\string\leaderfill{\noexpand\number\pageno}}}\ecrire\fi% ajout
\advance\nosection by 1\nosubsection=0
\ifeqnumerosimple
\else \xdef\eqprefix{\prefix\the\nosection.}\neqno=0\fi
\vbox{\noindent\bf\prefix\the\nosection\ #1}
\mark{\the\nosection}\bigskip\noindent
\xdef\ecrire{\write\toc{\string\par\string\item{\prefix\the\nosection}
#1
\string\leaderfill {\noexpand\number\pageno}}}\ecrire}

% appendix
\def\appendix#1#2\par{\bigbreak\nosection=0
\notenumber=0
\neqno=0
\def\prefix{A}
\mark{\the\nosection}
\message{\appendixname}
\leftline{\ssbx APPENDIX}
\leftline{\ssbx\uppercase\expandafter{#1}}
\leftline{\ssbx\uppercase\expandafter{#2}}
\bigskip\noindent\nonfrenchspacing
\edef\ecrire{\write\toc{\par\noindent{{\ssbx A}\
{\ssbx#1\ #2}}\string\leaderfill{\noexpand\number\pageno}}}\ecrire}%

% **************************** \subsection *************************
\def\subsection#1\par {\vskip0pt plus.05\vsize\penalty-100\vskip0pt
plus-.05\vsize\bigskip\vskip\parskip\advance\nosubsection by 1
\vbox{\noindent\it\prefix\the\nosection.\the\nosubsection\
\it #1}\smallskip\noindent
\edef\ecrire{\write\toc{\string\par\string\itemitem
{\prefix\the\nosection.\the\nosubsection} {#1}
\string\leaderfill{\noexpand\number\pageno}}}\ecrire
}
\def\note #1{\advance\notenumber by 1
\footnote{$^{\the\notenumber}$}{\sevenrm #1}}
% ?????

%\parindent=1em
%\newinsert\margin
%\dimen\margin=\maxdimen
%\count\margin=0 \skip\margin=0pt
% ********************* references harvmac style
\def\nolabels{\def\wrlabel##1{}\def\eqlabel##1{}\def\reflabel##1{}}
\def\writelabels{\def\wrlabel##1{\leavevmode\vadjust{\rlap{\smash%
{\line{{\escapechar=` \hfill\rlap{\sevenrm\hskip.03in\string##1}}}}}}}%
\def\eqlabel##1{{\escapechar-1\rlap{\sevenrm\hskip.05in\string##1}}}%
\def\reflabel##1{\noexpand\llap{\noexpand\sevenrm\string\string\string##1}}}
%*********
%\catcode`\@=11
\global\newcount\refno \global\refno=1
\newwrite\rfile
\def\ref{[\the\refno]\nref}
\def\nref#1{\xdef#1{[\the\refno]}\writedef{#1\leftbracket#1}%
\ifnum\refno=1\immediate\openout\rfile=\jobname.ref\fi
\global\advance\refno by1\chardef\wfile=\rfile\immediate
\write\rfile{\noexpand\item{#1\ }\reflabel{#1\hskip.31in}\pctsign}\findarg}
%	horrible hack to sidestep tex \write limitation
\def\findarg#1#{\begingroup\obeylines\newlinechar=`\^^M\pass@rg}
{\obeylines\gdef\pass@rg#1{\writ@line\relax #1^^M\hbox{}^^M}%
\gdef\writ@line#1^^M{\expandafter\toks0\expandafter{\striprel@x #1}%
\edef\next{\the\toks0}\ifx\next\em@rk\let\next=\endgroup\else\ifx\next\empty%
\else\immediate\write\wfile{\the\toks0}\fi\let\next=\writ@line\fi\next\relax}}
\def\striprel@x#1{}
\def\em@rk{\hbox{}}

\def\addref#1{\immediate\write\rfile{\noexpand\item{}#1}} %now unnecessary
\def\listrefs{
\ifnum\refno=1 \else
\immediate\closeout\rfile\writestoppt\baselineskip=14pt%
\vskip0pt plus.1\vsize\penalty-100\vskip0pt plus-.1
\vsize\bigskip\vskip\parskip\centerline{{\bf References}}\bigskip%
{\frenchspacing%
\parindent=20pt\escapechar=` \input \jobname.ref\vfill\eject}%
\nonfrenchspacing
\fi}
\def\startrefs#1{\immediate\openout\rfile=\jobname.ref\refno=#1}
\def\xref{\expandafter\xr@f}\def\xr@f[#1]{#1}
\def\refs#1{[\r@fs #1{\hbox{}}]}
\def\r@fs#1{\ifx\und@fined#1\message{reflabel \string#1 is undefined.}%
\xdef#1{(?.?)}\fi \edef\next{#1}\ifx\next\em@rk\def\next{}%
\else\ifx\next#1\xref#1\else#1\fi\let\next=\r@fs\fi\next}
%************************
%
\newwrite\lfile
{\escapechar-1\xdef\pctsign{\string\%}\xdef\leftbracket{\string\{}
\xdef\rightbracket{\string\}}}

\def\writestop{\def\writestoppt{\immediate\write\lfile{\string\pageno%
\the\pageno\string\startrefs\leftbracket\the\refno\rightbracket%
\string\def\string\secsym\leftbracket\secsym\rightbracket%
\string\secno\the\secno\string\meqno\the\meqno}\immediate\closeout\lfile}}
\def\writestoppt{}\def\writedef#1{}
%*************************************************************************
%Macro de numerotation automatique
%*************************************************************************
% numbering without naming
\def\eqnn{\global\advance\neqno by 1 \ifinner\relax\else%
\eqno\fi(\eqprefix\the\neqno)}
%
% numbering and attaching a name: \eqnd{\ename}
\def\eqnd#1{\global\advance\neqno by 1 \ifinner\relax\else%
\eqno\fi(\eqprefix\the\neqno)\eqlabel#1
{\xdef#1{($\eqprefix\the\neqno$)}}
\edef\ewrite{\write\sym{\string\def\string#1{($\eqprefix%
\the\neqno$)}}%
}\ewrite%
}
%
% for eqalignno, allows (1a) (1b)...
\def\eqna#1{\wrlabel#1\global\advance\neqno by1
{\xdef #1##1{\hbox{$(\eqprefix\the\neqno##1)$}}}
\edef\ewrite{\write\sym{\string\def\string#1{($\eqprefix%
\the\neqno$)}}%
}\ewrite%
}
\def\em@rk{\hbox{}}
\def\xeqn{\expandafter\xe@n}\def\xe@n(#1){#1}
\def\xeqna#1{\expandafter\xe@na#1}\def\xe@na\hbox#1{\xe@nap #1}
\def\xe@nap$(#1)${\hbox{$#1$}}
% \eqns allows to quote several equations, suppressing unnecessary ()
\def\eqns#1{(\e@ns #1{\hbox{}})}
\def\e@ns#1{\ifx\und@fined#1\message{eqnlabel \string#1 is undefined.}%
\xdef#1{(?.?)}\fi \edef\next{#1}\ifx\next\em@rk\def\next{}%
\else\ifx\next#1\xeqn#1\else\def\n@xt{#1}\ifx\n@xt\next#1\else\xeqna#1\fi
\fi\let\next=\e@ns\fi\next}
%*************************** figure macros ****************************
\def\fig{fig.~\the\figno\nfig}
\def\nfig#1{\xdef#1{\the\figno}%
\immediate\write\sym{\string\def\string#1{\the\figno}}%
\global\advance\figno by1}%
\def\xfig{\expandafter\xf@g}\def\xf@g fig.\penalty\@M\ {}%
\def\figs#1{figs.~\f@gs #1{\hbox{}}}%
\def\f@gs#1{\edef\next{#1}\ifx\next\em@rk\def\next{}\else%
\ifx\next#1\xfig #1\else#1\fi\let\next=\f@gs\fi\next}%
\long\def\figure#1#2#3{\midinsert
#2\par
{\elevenpoint
\setbox1=\hbox{#3}
\ifdim\wd1=0pt\centerline{{\bf Figure\ #1}\hskip7.5mm}%
\else\setbox0=\hbox{{\bf Figure #1}\quad#3\hskip7mm}
\ifdim\wd0>\hsize{\narrower\noindent\unhbox0\par}\else\centerline{\box0}\fi
\fi}
\wrlabel#1\par
\endinsert}
%*************************** table macros ****************************
\def\tab{table~\uppercase\expandafter{\romannumeral\the\tabno}\ntab}
\def\ntab#1{\xdef#1{\the\tabno}
\immediate\write\sym{\string\def\string#1{\the\tabno}}
\global\advance\tabno by1}
\long\def\table#1#2#3{\topinsert
#2\par
{\elevenpoint
\setbox1=\hbox{#3}
\ifdim\wd1=0pt\centerline{{\bf Table
\uppercase\expandafter{\romannumeral#1}}\hskip7.5mm}%
\else\setbox0=\hbox{{\bf Table
\uppercase\expandafter{\romannumeral#1}}\quad#3\hskip7mm}
\ifdim\wd0>\hsize{\narrower\noindent\unhbox0\par}\else\centerline{\box0}\fi
\fi}
\wrlabel#1\par
\endinsert}
%***********************************************************************
\catcode`@=12
\def\draftend{\immediate\closeout\sym\immediate\closeout\toc
}
\draftstart
\preprint{T94/032}
\title{Hyperscaling for polymer rings}
\authorname{Bertrand Duplantier}
\address{\saclay}
\abstract
The statistics of a long closed self-avoiding walk (SAW) or
polymer ring on a $ d $-dimensional lattice obeys hyperscaling.
The
combination $ p_N \left\langle R^2 \right\rangle^{ d/2}_N\mu^{ -N}, $ (where $
p_N $ is the number of configurations of an oriented and
rooted $ N $-step ring, $ \left\langle R^2 \right\rangle_ N $ a typical
average size squared, and $ \mu $ the SAW
effective connectivity constant of the lattice) is equal for $ N
\longrightarrow \infty $ to a
lattice-dependent constant times a universal amplitude $ A(d). $ The latter
amplitude is calculated directly from the minimal continuous Edwards model to
second order in $ \varepsilon \equiv 4-d. $ The case of rings at the upper
critical dimension $ d=4 $ is also studied. The results are checked against
field theoretical calculations, and former simulations. As a consequence, we
show that the universal constant $ \lambda $ appearing to second order in $
\varepsilon $ in all
critical phenomena amplitude ratios is equal to $ \lambda  = {1 \over 18}
\left[\psi^{ \prime}( 1/6)+\psi^{ \prime}( 1/3) \right]-{4\pi^ 2 \over 27}. $
\endabstract
\vfill
\submitted{Nuclear Physics B $\lbrack$FS$\rbrack$ }
\eject
\eject
\input mssymb
\input etmacro.tex
\baselineskip=17pt
\def\superone#1{\buildrel{\circ}\over#1}

\noindent {\bf I. INTRODUCTION}

\vskip 10pt

Hyperscaling is the term which has been coined for the relation holding in a
second
order phase transition between the specific heat exponent $ \alpha $ and the
correlation length exponent $ \nu $
$$ \alpha  = 2-\nu d\ , \eqno (1.1) $$
where $ d $ is the space dimension ($\lbrack$1$\rbrack$ and references
therein). More generally,
the term is also used for any relation between critical exponents involving
explicitly the dimension $ d $ $\lbrack$2$\rbrack$. The validity of relation
(1.1) has been
questioned $\lbrack$3$\rbrack$, but (1.1) is now widely believed to be true.
It can be
derived in a standard way from the field theoretic approach to critical
phenomena $\lbrack$2$\rbrack$.

An associated property holds for the singular part of the free energy per
unit volume $ f_S, $ (in $ k_BT $ units), which links the latter to the
corresponding
diverging correlation length $ \xi $ in the bulk: at the critical point $
T=T_c $ the
product $ (f_S\xi^ d) $ should be a universal number $\lbrack$4,5$\rbrack$
depending only on the
dimension $ d, $ and on the universality class of the physical system:
$$ \lim_{T \longrightarrow T^+_c} \left(f_S\xi^ d \right) = A_f(d)\ . \eqno
(1.2) $$
Of course, many other examples exist of universal combinations of amplitudes
(see the review $\lbrack$6$\rbrack$ and references therein).

Since the work by de Gennes $\lbrack$7$\rbrack$ and des Cloizeaux
$\lbrack$8$\rbrack$, it is quite clear that
the statistics of polymers with excluded volume or, equivalently,
self-avoiding walks (SAW) on a lattice are, in the large size limit,
described by the universal properties of the limit $ n \longrightarrow 0 $ of
the $ O(n) $ model
$\lbrack$9$\rbrack$. The consequences, therefore, of hyperscaling relations
such as
Eqs.(1.1) or (1.2) are worth studying in polymer statistical mechanics. As we
shall see, they translate into universal properties of polymer {\it rings}.
Such closed self-avoiding walks have been studied in the past using field
theory $\lbrack$10$\rbrack$ or enumeration methods for self-avoiding polygons
$\lbrack$11,12,13$\rbrack$. As
noted in $\lbrack$6$\rbrack$, however, it is only in 1985 that the analogue of
the so-called
\lq\lq hyperuniversal\rq\rq\ amplitude (1.2) at a critical point has been
introduced in
polymer physics and studied numerically $\lbrack$11,12$\rbrack$. Quite
recently, a burst of
interest has surged anew, concerning the universal amplitudes associated with
polymer rings in {\it two dimensions}, the exact values of which are now
partly accessible within the framework of conformal invariance
$\lbrack$14,15$\rbrack$. We now
turn to the precise definition of hyperuniversality for polymer rings.

Consider a (closed) self-avoiding ring of $ N $ steps drawn on a $ d
$-dimensional
regular lattice. For later convenience, we consider the ring to result from
the closure of an (oriented) polymer chain. Its origin corresponds to a
marked point along the ring, which is thus {\it rooted} and {\it oriented}.

The total number $ p_N $ of self-avoiding, oriented,  and rooted
configurations of the ring
{\it (per lattice site)} scales like
$$ \doublelow{ p_N \cr N \longrightarrow \infty \cr}  = 2B\ \mu^ N\ N^{\alpha
-2}[1+o(1)]\ , \eqno (1.3) $$
where $ \mu $ is the SAW
connective constant on the considered lattice, where $ B $ is a non-universal
amplitude\footnote{$ ^+ $}{The explicit factor of 2 in (1.3) accounts for the
orientation
of the ring, in order to make contact with the notation of
Ref.$\lbrack$15$\rbrack$, where
rings are {\it unoriented}.}, also lattice dependent,
up to scaling corrections which vanish in the limit $ N \longrightarrow \infty
. $
The exponent $ \alpha $ in (1.3) is the same
as that for the specific heat of the $ n \longrightarrow 0 $ $ O(n) $ model,
since the Feynman graphs
of an interacting ring polymer are exactly those of the free-energy of the $
O(n) $
model, with the $ n=0 $ combinatoric weights $\lbrack$9,10$\rbrack$. A first
consequence of
hyperscaling (1.1) is thus immediate:
$$ p_N = 2B\ \mu^ N\ N^{-\nu d}[1+o(1)]\ . \eqno (1.4) $$

To build the polymer analogue of Eq.(1.2) $\lbrack$11$\rbrack$, one considers
a typical size,
as given by the mean-square end-to-end distance $ \left\langle R^2
\right\rangle_ N $ of an $ N $-step {\it open}
self-avoiding walk, or the mean-square radii of gyration $
\left\langle\superone R^2_G \right\rangle_ N $ of an
$ N $-step {\it polygon}, or $ \left\langle R^2_G \right\rangle_ N $ of an $ N
$-step {\it open walk}. These scale
respectively like:
$$ \eqalignno{ \left\langle R^2 \right\rangle_ N & = C\ a^2N^{2\nu}[ 1+o(1)] &
(1.5a) \cr \left\langle\superone R^2_G \right\rangle_ N & = D\ a^2N^{2\nu}[
1+o(1)] & (1.5b) \cr \left\langle R^2_G \right\rangle_ N & = F\ a^2N^{2\nu}[
1+o(1)] & (1.5c) \cr} $$
where $ C, $ $ D, $ $ F $ are lattice-dependent amplitudes. (We use a notation
similar to
that of Ref.$\lbrack$15$\rbrack$, but make explicit the dimensional lattice
spacing $ a). $

Combining Eqs.(1.4), (1.5), one arrives successively at
$$ \eqalignno{ p_N \left\langle R^2 \right\rangle^{ d/2}_N\mu^{ -N}a^{-d} & =
2BC^{d/2}[1+o(1)] & (1.6a) \cr p_N \left\langle\superone R^2_G \right\rangle^{
d/2}_N\mu^{ -N}a^{-d} & = 2BD^{d/2}[1+o(1)] & (1.6b) \cr p_N \left\langle
R^2_G \right\rangle^{ d/2}_N\mu^{ -N}a^{-d} & = 2BF^{d/2}[1+o(1)]\ . & (1.6c)
\cr} $$

As explained in the original work $\lbrack$11$\rbrack$, (see also
$\lbrack$6,15$\rbrack$) this is not quite
enough for obtaining in (1.6) a universal combination. On a given lattice the
configuration number $ p_N $ depends in a subtle way on \lq\lq parity
effects\rq\rq ; for
instance, on a hypercubic lattice, $ N $ is necessarily even for closing a
polygon (Appendix A). This is accounted for by dividing (1.6) by a
lattice-dependent
factor $ \tau $ $\lbrack$6,11,15$\rbrack$ which corrects for the effect of the
background lattice.
It counts the degeneracy of the possible singularities $ z_c $ of the
generating
function $ \Xi( z)\equiv \sum^{ }_ Np_Nz^n $ of the configuration numbers $
p_N, $ on the circle $ \vert z\vert =\mu^{ -1} $
in the complex plane.

For the triangular lattice in $ d=2 $ or face-centered cubic lattice in $ d=3,
$ $ \tau =1, $
while for the square lattice, simple cubic or body-centered cubic lattices $
\tau =2 $
$\lbrack$6,11,15$\rbrack$.

One is thus led to define the universal amplitude
$$ A(d) \equiv  \lim_{N \longrightarrow \infty}  p_N \left\langle R^2
\right\rangle^{ d/2}_N/(\mu^ Na^d\tau ) = {2BC^{d/2} \over \tau} \eqno (1.7a)
$$

For the radii of gyration (1.5b,c), one defines similarly
$$ \eqalignno{\superone A_G(d) & \equiv  \lim_{N \longrightarrow \infty}  p_N
\left\langle\superone R^2 \right\rangle^{ d/2}_N/(\mu^ Na^d\tau ) = {2BD^{d/2}
\over \tau} &  (1.7b) \cr A_G(d) & \equiv  \lim_{N \longrightarrow \infty}
p_N \left\langle R^2_G \right\rangle^{ d/2}_N/(\mu^ Na^d\tau ) = {2BF^{d/2}
\over \tau} &  (1.7c) \cr} $$

Cardy $\lbrack$14$\rbrack$ has recently been able to derive in two-dimensions
from conformal
invariance the exact value of $ {1 \over 2} \superone A_G(2) = BD/\tau  =
5/32\pi . $
A detailed numerical analysis of the other universal amplitudes $ BC/\tau $
and $ BF/\tau $
can be found in the recent Ref.$\lbrack$15$\rbrack$.

Our aim is to check hyperscaling for self-avoiding rings within the
continuous Edwards model $\lbrack$16$\rbrack$ describing polymers in dimension
$ d. $ We
calculate the universal amplitude $ A(d) $ defined in (1.7a), for $ d\leq 4, $
within the
$ \varepsilon $-expansion scheme to order $ {\cal O} \left(\varepsilon^ 2
\right) $ where $ \varepsilon =4-d. $ The Edwards model is minimal
for calculating universal quantities like amplitudes (1.7), and requires only
direct perturbative calculations, followed by direct renormalisation
$\lbrack$17$\rbrack$, or, as here, {\it
dimensional renormalization} $\lbrack$18$\rbrack$ in polymer theory.

The universal logarithmic corrections
at $ d=4 $ are also given, and should provide a good testing ground of
hyperscaling by numerical simulations.

There exists
an other universal amplitude ratio calculated for polymers to $ {\cal O}
\left(\varepsilon^ 2 \right) $
$\lbrack$19$\rbrack$, namely the well-known mean-square gyration/end-to-end
distance ratio of
an open SAW: $ \lim_{N \longrightarrow \infty} \left\langle R^2_G
\right\rangle_ N/ \left\langle R^2 \right\rangle_ N. $ Its value together with
that of $ A(d) $ given
below also yields
that of $ A_G(d) $ (1.7c) to order $ {\cal O} \left(\varepsilon^ 2 \right) $
for ring polymers.

For an analytical evaluation of $ A(d), $ we shall make use, as
intermediate steps, of some recent results $\lbrack$20,21$\rbrack$ on the
integration of
Bessel functions. This yields the $ {\cal O} \left(\varepsilon^ 2 \right) $
expansion in terms of special
functions.

As mentioned above, there exists a relation $\lbrack$11$\rbrack$ between the
free energy
amplitude ratio (1.2) at a critical point, and the polymer ring amplitude
ratios studied here. More precisely, the well-known amplitude ratio
associated with the singular part of the specific heat above $ T_c $
$\lbrack$22-26$\rbrack$
$$ \eqalignno{ C_S(t) & = {\partial^ 2 \over \partial t^2} f_S(t) & (1.8a) \cr
\doublelow{ C_S(t) \cr t \longrightarrow 0^+ \cr} &  = {A_+ \over \alpha}
t^{-\alpha}( 1+o(1))\ \ ,\ \ \ \ \ \ t=T-T_c\ , & (1.8b) \cr} $$
and with the correlation length divergence
$$ \doublelow{ \xi( t) \cr t \longrightarrow 0^+ \cr}  = \xi_ +t^{-\nu}(
1+o(1))\ , \eqno (1.9) $$
is defined as $\lbrack$22-24$\rbrack$
$$ \eqalignno{ \left(R^+_\xi \right)^d & = \lim_{t \longrightarrow 0^+}
\left[\alpha t^2C_S(t)\xi^ d(t) \right] &  \cr  &  = \alpha( 1-\alpha)(
2-\alpha)  \lim_{t \longrightarrow 0^+}f_S(t)\xi^ d(t) &  \cr  &  = A_+\xi^
d_+\ . & (1.10) \cr} $$
As we shall see, the polymer ring amplitude $ A(d) $ for the end-to-end
distance
(1.7a) is actually related to the limit of $ {1 \over n} \left(R^+_\xi
\right)^d $ when $ n \longrightarrow 0, $ where $ n $ is the
number of components of the general $ O(n) $ model in a $ \left( {\bf \Phi}^ 2
\right)^2 $ field theory.

After completing the calculation within the polymer Edwards model, we
realized that the other universal amplitude ratio (1.10) had also been
calculated
to order $ {\cal O} \left(\varepsilon^ 2 \right) $ in the framework of
perturbative field theory $\lbrack$25,26$\rbrack$. Thus, the
present calculations in polymer theory (which amount to field theory
calculations
off of the critical point) allow for a cross-check of the
polymer and field theoretical approaches, and thus of their results.

In all two-loop results for universal amplitude ratios in the field
theoretical approach to critical phenomena $\lbrack$2$\rbrack$, there appears
a ubiquitous numerical constant $ \lambda , $ which is defined as
$$ \lambda  = -{1 \over 2} \int^ 1_0 {\rm d} x { {\rm \ell n}[x(1-x)] \over
1-x(1-x)} = 1.171953... \eqno (1.11) $$
{}From our approach, especially through the use of
Refs.$\lbrack$20,21$\rbrack$,
it now appears
as a closed form in terms of special functions. Among several possible
forms, given below, let us choose the following
$$ \eqalignno{ \lambda &  = - {1 \over 9} \left[\beta^{ \prime}( 1/3)+\beta^{
\prime}( 2/3) \right] & (1.12a) \cr  &  = \sum^{ }_{ n\geq 0}(-1)^n \left[{1
\over( 3n+1)^2} + {1 \over( 3n+2)^2} \right]\ , & (1.12b) \cr} $$
where the special function $ \beta( x) $ is defined by
$$ \beta( x) = {1 \over 2} \left[\psi \left({x+1 \over 2} \right) - \psi
\left({x \over 2} \right) \right]\ , \eqno (1.13) $$
where $ \psi( x) = { {\rm d} \over {\rm d} x} {\rm \ell n} \ \Gamma( x). $
Hence, another form is
$$ \lambda  = {1 \over 9} \left\{{ 1 \over 2} \left[\psi^{ \prime}( 1/6) +
\psi^{ \prime}( 1/3) \right] - {4 \over 3} \pi^ 2 \right\} \ . \eqno (1.14) $$
In terms of $ \lambda , $ the results for the universal amplitude ratios
(1.7a) and
(1.7c) read as follows:
$$ \eqalignno{ d & = 4-\varepsilon \ , & (1.15) \cr A(d) & = \left({d \over
2\pi} \right)^{d/2} \left[1-{\varepsilon \over 8} + \left({\varepsilon \over
8} \right)^2 \left({23 \over 4} + {\pi^ 2 \over 2} + {14 \over 3} \lambda
\right) + {\cal O} \left(\varepsilon^ 3 \right) \right]\ , & (1.16) \cr A_G(d)
& = \left({d \over 12\pi} \right)^{d/2} \left[1-{7 \over 6} {\varepsilon \over
8} + \left({\varepsilon \over 8} \right)^2 \left({29\times 31 \over( 12)^2} +
{\pi^ 2 \over 2} + {4 \over 3} \lambda \right) + {\cal O} \left(\varepsilon^ 3
\right) \right]\ . & (1.17) \cr} $$

This article is organized as follows. In section II the passage from a number
of distinct discrete walks on a lattice to a partition function in the
continuum formalism is established. Section III deals with the second order
calculation of the partition function of a ring in the continuum limit, while
in section IV the corresponding calculation of the end-to-end distance is
performed.
The amplitude ratios $ A(d) $ and $ A_G(d) $ are calculated to $ {\cal O}
\left(\varepsilon^ 2 \right) $ in section V. The
findings are compared to field theory and numerical simulations in section
VI.

A part of the paper is of technical nature and is placed in Appendices.

\vskip 17pt

\noindent {\bf II. CONTINUOUS LIMIT}

\vskip 10pt

The probability weight of the Edwards model is defined as
$$ {\cal P} \left\{\vec r \right\}  = {\rm exp} \left(-\beta{\cal H}
\left\{\vec r \right\} \right) \eqno (2.1) $$
where the dimensionless Hamiltonian $ \beta{\cal H} $ reads
$$ \beta{\cal H} \left\{\vec r \right\}  = {1 \over 2} \int^ S_0 {\rm d} s
\left({ {\rm d}\vec r(s) \over {\rm d} s} \right)^2 + {b \over 2} \int^ S_0
{\rm d} s \int^ S_0 {\rm d} s^{\prime} \ \delta^ d \left(\vec r(s)-\vec r
\left(s^{\prime} \right) \right)\ , \eqno (2.2) $$
for a $ d $-dimensional configuration $ \vec r(s) \in  \Bbb R^d, $ with
abscissa $ 0 \leq  s \leq  S, $ $ S $
fixing the overall linear size of the polymer. The excluded volume effect is
mimicked by the contact $ \delta $-potential, with coefficient $ b>0. $

The partition function of a {\it rooted} and {\it oriented} polymer ring is
then simply given
by the functional integral
$$ \superone{{\cal Z}}(S) = \int^{ }_{ }{\cal D}\vec r(s)\ \delta^ d
\left(\vec r(S)-\vec r(0) \right)\ {\rm exp} \left(-\beta{\cal H} \left\{\vec
r \right\} \right)\ , \eqno (2.3) $$
normalized so that
$$ 1\equiv \int^{ }_{ }{\cal D}\vec r(s)\ {\rm exp} \left(-\beta{\cal H}_0
\left\{\vec r \right\} \right)\ , $$
with $ {\cal H}_0 $ the Brownian Hamiltonian (2.2) for $ b=0. $ The convention
is therefore
that the partition function of an open Brownian path is unity. With these
definitions, the elementary Brownian correlation function reads
$\lbrack$9$\rbrack$
$$ \eqalignno{ \left\langle {\rm e}^{i\vec q\cdot \left(\vec r(s)-\vec r
\left(s^{\prime} \right) \right)} \right\rangle_ 0 & \equiv  \int^{ }_{ }{\cal
D}\vec r(s)\ {\rm e}^{i\vec q\cdot \left(\vec r(s)-\vec r \left(s^{\prime}
\right) \right)} {\rm e}^{-\beta{\cal H}_0 \left\{\vec r \right\}} &  \cr  &
= {\rm e}^{-q^2 \left\vert s-s^{\prime} \right\vert /2}\ . & (2.4) \cr} $$
The end-to-end distance of a Brownian {\it open} path is then simply
$\lbrack$9$\rbrack$
$$ \left\langle \left[\vec r(S)-\vec r(0) \right]^2 \right\rangle_ 0 = dS\ ,
\eqno (2.5) $$
while the partition function of a Brownian {\it oriented ring} is readily
evaluated to
be:
$$ \superone{{\cal Z}}_B(S) = (2\pi S)^{-d/2} \eqno (2.6) $$
We take the hypercubic lattice $ \Bbb Z^d $ for comparison, with lattice mesh
size
$ a. $ The mean-square end-to-end distance is given trivially, for an $ N
$-step open random
walk, by
$$ \left\langle R^2 \right\rangle_ N = Na^2 = dS \eqno (2.7) $$
leading to the identification $ S\equiv Na^2/d. $

On $ \Bbb Z^d, $ the discrete partition function of a $ N $-step oriented and
rooted polygon reads
asymptotically (Appendix A)
$$ \eqalign{ p_N & = 2(2d)^N \left({d \over 2\pi N} \right)^{d/2}[1+o(1)]\ ,\
\ \ \ \ \ \ \ N \in  2 \Bbb N \cr p_N & = 0\ ,\ \ \ \ \ \ \ \ \ \ \ \ \ \ \ \
\ \ \ N \in  2 \Bbb N+1\ . \cr} \eqno (2.8) $$
For a hypercubic lattice the degeneracy factor $ \sigma $ described in the
introduction is thus precisely 2 and owing to (2.6), (2.7) and (2.8), we are
thus led to define the continuum limit as
$$ \lim_{N \longrightarrow \infty}{ p_N \over( 2d)^Na^d2} = \superone{{\cal
Z}}_B(S)\ , \eqno (2.8a) $$
an identification in the random walk case which parallels the combinations
(1.7), leading in turn to
$$ A_B(d) = (d/2\pi)^{ d/2}\ \ ,\ \ \ \ {\rm with} \ \ \tau =2\ . \eqno (2.8b)
$$

When dealing with self-avoiding polygons on the (e.g. hypercubic) lattice on
the one hand, and with polymer rings in the continuum space on the other
hand, the correct identification is now
$$ p_N/(\mu^ Na^d\tau ) \longrightarrow  \superone{{\cal Z}}(S)\vert_{ {\rm
dim.reg.}}\ , \eqno (2.9) $$
where $ \superone{{\cal Z}}(S) $ is the full continuum partition function
(2.3) of the interacting
closed and oriented paths, calculated in {\it dimensional regularization}
$\lbrack$9,27$\rbrack$. Indeed,
the latter mathematical procedure performs the analytical continuation in
dimension $ d, $ from $ d < 2 $ where the continuum theory is free of
ultraviolet
divergences, to the range $ 2 < d < 4. $ It allows for the direct calculation
of
regularized partition functions, for which an exponentially growing factor $
{\rm exp} \left[C \left(z_0,d \right)S/s_0 \right] $
is divided out, where $ s_0 $ is some short-range cut-off along the chain, and
$ C \left(z_0,d \right) $
a local free-energy depending on $ b $ and $ s_0 $
$\lbrack$9,17$\rbrack$\footnote{$ ^{\dagger} $}{Strictly
speaking, this holds true for $ 2<d<4, $ $ d\not= 4-2/p, $ $ p\in \Bbb N^\ast
. $ For these excluded
values of $ d, $ the scaling of the partition functions is modified in a
calculable way $\lbrack$28$\rbrack$.} through $ z_0 \equiv  (2\pi)^{ -d/2}b\
s^{2-d/2}_0. $

Factorizing out the UV divergent part of the free energy by dimensional
regularization therefore exactly amounts to the same result, with $ \mu^ N $
on the left hand side
of (2.9).

Owing to the hyperscaling relation (1.4) and identification (2.7), the
dimensionally regularized partition function (2.9) scales as
$$ \doublelow{ \superone{{\cal Z}}(S) \cr S \longrightarrow \infty \cr}  \sim
S^{-\nu d} \eqno (2.9b) $$

We arrive finally at the (hyper) universal value of the ring amplitude $ A(d)
$
(1.7a) defined in the introduction, and which is now expressed in the
continuum theory as
$$ \eqalignno{ A(d) & = \lim_{N \longrightarrow \infty}  p_N \left\langle R^2
\right\rangle^{ d/2}_N/(\mu^ Na^d\tau ) &  \cr  &  \equiv  \lim_{S
\longrightarrow \infty}  \superone{{\cal Z}}(S)\vert_{ {\rm dim.reg.}}
\left(R^2 \right)^{d/2} & (2.10) \cr} $$
in terms of the mean-square end-to-end distance
$$ R^2 \equiv  \left\langle \left[\vec r(S)-\vec r(0) \right]^2 \right\rangle
\ . \eqno (2.11) $$
for an {\it open} polymer. It is finally convenient to normalize
it to be unity for a Brownian ring, whence our final definition
$$ {\cal A}(d) \equiv  A(d)/A_B(d) = \lim_{S \longrightarrow \infty}
\superone{{\cal Z}}(S)\vert_{ {\rm dim.reg.}} \left(2\pi R^2/d \right)^{d/2}\
, \eqno (2.12) $$
which is such that in the absence of excluded volume $ (b=0), $ $ {\cal
A}(d)=1. $
\vskip 17pt

\noindent {\bf III. PERTURBATIVE EXPANSION}

\vskip 10pt

The dimensional analysis of Hamiltonian (2.2) shows that $ \left[\vec r
\right]=[S]^{1/2}, $ and $ [b]=[S]^{-(2-d/2)}. $
One is therefore led to introduce the dimensionless coupling parameter for a
finite polymer (corresponding to such a coupling parameter away from the
critical point in the equivalent $ n=0 $ field theory, or to a finite size
scaling if $ \vec r(s) $
itself is viewed as a $ d $-component field living on the one-dimensional
space $ [0,S]) $
$$ z = (2\pi)^{ -d/2}b\ S^{2-d/2}\ \ ,\ \ \ \ \ 2-d/2 = \varepsilon /2\ .
\eqno (3.1) $$
This coupling parameter originated forty years ago in the Fixman expansion of
polymer
theory, modeled as discrete chains in continuous space $\lbrack$29$\rbrack$,
and now plays a
useful and simplifying role in the fully continuous Edwards theory
$\lbrack$17,9$\rbrack$.
Any {\it dimensionally regularized} partition function like (2.3) possesses,
from the expansion of weight (2.1) in powers of $ b, $ and up to a global
dimensional factor like (2.6), a formal perturbative expansion in powers of $
z $
alone, whose coefficients are meromorphic functions of $ d, $ with, of
course, pole singularities at $ d=4. $

The diagrams giving $ \superone{{\cal Z}}(S) $ up to second order in $ b $ (or
$ z) $ are shown in
Fig.1.
Their contributions are calculated using standard \lq\lq Feynman\rq\rq\ rules
for
polymer diagrams $\lbrack$9$\rbrack$, namely by expanding in (2.3) $ {\rm
exp}(-\beta{\cal H}) $ in powers of $ b, $
and writing each $ \delta^ d $ distribution, represented by a dotted line on
the
diagram, by a Fourier integral
$$ \delta^ d \left(\vec r(s) - \vec r \left(s^{\prime} \right) \right) =
\int^{ }_{ \Bbb R^d}{ {\rm d}^dq \over( 2\pi)^ d} {\rm e}^{i\vec q\cdot
\left[\vec r(s)-\vec r \left(s^{\prime} \right) \right]}\ . \eqno  $$

The Brownian averages of the path integral are then performed by use of the
generalization of (2.4)
$$ \eqalignno{ \left\langle {\rm exp} \left\{ -i \sum^ L_{\ell =1}\vec
q_{\ell} \cdot\vec r \left(s_{\ell} \right) \right\} \right\rangle_ 0 & = {\rm
exp} \left\{ -{1 \over 2} \sum^{ L-1}_{\ell =1} \left( \sum^{ \ell}_{ m=1}\vec
q_m \right)^2 \left\vert s_{\ell +1}-s_{\ell} \right\vert \right\} \ , & (3.2)
\cr  &  \ \ \ \ \ \ \ \left( \sum^ L_{\ell =1}\vec q_{\ell} =\vec 0 \right) &
\cr} $$
valid for an arbitrary {\it ordered} set $ \{ \ell\} $ of $ L $ points along
the chain with
insertion of momenta $ \vec q_{\ell} $ at points $ s_{\ell} . $ A Gaussian
integration over momenta $ \left\{\vec q \right\} $
follows. As a simple example, the calculation of diagram (b) is
given explicitly in Appendix B. Notice however that we select the rooted
ring configurations by the very presence in the partition function (2.3) of a
closure distribution $ \delta^ d \left[\vec r(S)-\vec r(0) \right], $averaged
with the path integral measure
of an {\it open} Brownian path. This allows the use of (3.2). We can also
describe a (non rooted) polymer ring as a 1-dimensional closed manifold
(e.g., the sphere $ {\cal S}_1 $ $\lbrack$30$\rbrack$), for which the Gaussian
propagator is more
complicated. However, for Brownian paths, the two methods are equivalent
(Appendix B).

The contributions of the diagrams of Fig.1 read successively
$$ \eqalignno{ \superone{{\cal Z}}^{(0)} & \equiv  \superone{{\cal Z}}_B(S) =
(2\pi S)^{-d/2} & (3.3a) \cr \superone{{\cal Z}}^{(1)} & = (-b) (2\pi)^{ -d}
{1 \over 2} S \int^{ \infty}_ 0 \prod^ 2_{i=1} {\rm d} s_i\ \delta \left(S-
\sum^ 2_{i=1}s_i \right) \left( \prod^ 2_{i=1}s_i \right)^{-d/2} & (3.3b) \cr
\superone{{\cal Z}}^{(2)}_{(a)} & = (-b)^2 (2\pi)^{ -3d/2} {S \over 2} \int^{
\infty}_ 0 \prod^ 4_{i=1} {\rm d} s_i\ \delta \left(S- \sum^ 4_{i=1}s_i
\right) \left( \prod^ 4_{i=1}s_i \right)^{-d/2} \left( \sum^ 2_{j=1}{1 \over
s_j} \right)^{-d/2}\ \ \  & (3.3c) \cr \superone{{\cal Z}}^{(2)}_{(b)} & =
(-b)^2 (2\pi)^{ -3d/2} {S \over 4} \int^{ \infty}_ 0 \prod^ 4_{i=1} {\rm d}
s_i\ \delta \left(S- \sum^ 4_{i=1}s_i \right) \left( \prod^ 4_{i=1}s_i
\right)^{-d/2} \left( \sum^ 4_{j=1}{1 \over s_j} \right)^{-d/2}\ \ \  & (3.3d)
\cr} $$
In these formulae, the variables $ s_i $ correspond to the positive lengths of
successive segments along the loop (i.e. the Fixman representation
in polymer physics $\lbrack$29$\rbrack$ or, in field theory, the so-called $
\alpha $-parameters of
the Schwinger representation $\lbrack$31$\rbrack$). Using the effective
parameter $ z $ (3.1) we get
$$ \superone{{\cal Z}}(S) = (2\pi S)^{-d/2} \left[1-z{\cal J}^{(1)}+z^2{\cal
J}^{(2)}_{(a)}+z^2{\cal J}^{(2)}_{(b)}+{\cal O} \left(z^3 \right) \right]
\eqno (3.4a) $$
with
$$ \eqalignno{{\cal J}^{(1)} & = {1 \over 2} \int^{ \infty}_ 0 \prod^ 2_{i=1}
{\rm d} x_i \left( \prod^ 2_{i=1}x_i \right)^{-d/2}\delta \left(1- \sum^
2_{i=1}x_i \right) & (3.4b) \cr{\cal J}^{(2)}_{(a)} & = {1 \over 2} {\cal
J}_{4,2}\ \ \ \ ,\ \ \ \ {\cal J}^{(2)}_{(b)} = {1 \over 4} {\cal J}_{4,4}\ ,
& (3.4c) \cr} $$
where $ {\cal J}_{n,n^{\prime}} $ is defined for $ n^{\prime}  \leq  n $ as
$$ {\cal J}_{n,n^{\prime}}  \equiv  \int^{ \infty}_ 0 \prod^ n_{i=1} {\rm d}
x_i \left( \prod^ n_{i=1}x_i \right)^{-d/2} \delta \left(1- \sum^ n_{i=1}x_i
\right) \left( \sum^{ n^{\prime}}_{ j=1}{1 \over x_j} \right)^{-d/2}\ . \eqno
(3.4d) $$
As meromorphic functions of $ d, $ $ {\cal J}^{(1)}, $ $ {\cal
J}^{(2)}_{(a,b)} $ have Laurent expansions in $ \varepsilon \equiv 4-d, $
$ {\cal J}^{(1)} $ having a simple pole and $ {\cal J}^{(2)}_{(a,b)} $ double
poles. To calculate a universal
amplitude, we shall need each of these Laurent expansions {\it including} $
{\cal O}(\varepsilon) $
for $ {\cal J}^{(1)} $ and $ {\cal O}(1) $ for $ {\cal J}^{(2)}. $ We always
represent a length constraint $ \delta \left(1- \sum^{ }_ ix_i \right) $
as
$$ \delta( x) = \int^{ }_{{\cal C}}{ {\rm d} a \over 2\pi i} {\rm e}^{ax}\ ,
\eqno (3.5a) $$
and more generally,
$$ {x^{\alpha -1} \over \Gamma( \alpha)}  \theta( x) = \int^{ }_{{\cal C}}{
{\rm d} a \over 2\pi i} { {\rm e}^{ax} \over a^\alpha} \ , \eqno (3.5b) $$
where $ {\cal C} $ is the vertical contour in the complex plane $ (\eta
-i\infty , $ $ \eta +i\infty ), $ $ \eta >0, $ on
the right hand side of the origin.

We also use the basic representation
$$ \left( \sum^{ }_ jx^{-1}_j \right)^{-d/2} = {1 \over \Gamma( d/2)} \int^{
\infty}_ 0 {\rm d} p\ p^{d/2-1}\ {\rm e}^{-p \sum^{ }_ jx^{-1}_j} \eqno (3.6)
$$
to disentangle the various lengths.

Integral $ {\cal J}^{(1)} $ (3.4b) is simply a beta function
$$ {\cal J}^{(1)} = {1 \over 2} B(1-d/2,\ 1-d/2) = {1 \over 2} {\Gamma^
2(1-d/2) \over \Gamma( 2-d)}\ . \eqno (3.7) $$
The family of integrals (3.4d) can be written with the help of (3.5a), (3.6)
as
$$ \eqalignno{{\cal J}_{n,n^{\prime}} &  = {1 \over \Gamma( d/2)} \int^{
\infty}_ 0 {\rm d} p\ p^{d/2-1} \int^{ }_{{\cal C}}{ {\rm d} a \over 2\pi i}
{\rm e}^a \left( \int^{ \infty}_ 0 {\rm d} x\ x^{-d/2} {\rm e}^{-p/x} {\rm
e}^{-ax} \right)^{n^{\prime}} &  \cr  &  \ \ \ \ \ \ \ \ \ \ \ \ \ \times
\left( \int^{ \infty}_ 0 {\rm d} x\ x^{-d/2} {\rm e}^{-ax}
\right)^{n-n^{\prime}} \ . &  \cr} $$
We perform the change of variables in the $ (p,x) $-integrals:
$$ x = p^{\prime} x^{\prime} /a\ \ \ \ ,\ \ \ \ p=p^{\prime} /a\ ; \eqno
(3.8a) $$
and get
$$ \eqalignno{{\cal J}_{n,n^{\prime}} &  = {1 \over \Gamma( d/2)} \int^{
\infty}_ 0 {\rm d} p^{\prime} p^{\prime( 1-d/2) \left(n^{\prime} -1 \right)}
\int^{ }_{{\cal C}}{ {\rm d} a \over 2\pi i} {\rm e}^aa^{-d/2-n(1-d/2)} &  \cr
 &   &  \cr  &  \ \ \ \ \ \ \times  \left( \int^{ \infty}_ 0 {\rm d}
x^{\prime} x^{\prime -d/2} {\rm e}^{-p^{\prime} /x^{\prime}} {\rm
e}^{-x^{\prime}} \right)^{n^{\prime}}[ \Gamma( 1-d/2)]^{n-n^{\prime}} \ . &
(3.8b) \cr} $$
We define the auxiliary function
$$ F(p) = \int^{ \infty}_ 0 {\rm d} x\ x^{-d/2} {\rm e}^{-1/x} {\rm e}^{-px}\
, \eqno (3.8c) $$
and by again using (3.5b) to perform the $ a $-integral we arrive at the
simple form
$$ {\cal J}_{n,n^{\prime}}  = {1 \over \Gamma( d/2)} {[\Gamma(
1-d/2)]^{n-n^{\prime}} \over \Gamma[ d/2+n(1-d/2)]} \int^{ \infty}_ 0 {\rm d}
p\ p^{(1-d/2) \left(n^{\prime} -1 \right)}[F(p)]^{n^{\prime}} \ . \eqno (3.9)
$$
The auxiliary function $ F(p) $ is actually expressible as
$\lbrack$32$\rbrack$
$$ F(p) = 2\ p^{\nu /2}\ K_\nu \ \left(2 \sqrt{ p} \right)\ \ ,\ \ \nu  =
d/2-1\ , \eqno (3.10) $$
where $ K_\nu $ is the Bessel function of the second kind. For $ n^{\prime}
=2, $ the $ p $-integral involved in $ {\cal J}_{4,2} $ is
exactly calculable as
$$ \eqalignno{ \int^{ \infty}_ 0 {\rm d} p\ p^{1-d/2}F^2(p) & = 2 \int^{
\infty}_ 0 {\rm d} x\ x\ K^2_\nu( x) &  \cr  &  = {\pi \nu \over {\rm sin} \
\pi \nu}  = \Gamma( 1-\nu) \Gamma( 1+\nu) \ , & (3.11) \cr} $$
whence
$$ {\cal J}^{(2)}_{(a)} = {1 \over 2} {[\Gamma( 1-d/2)]^2 \over \Gamma(
4-3d/2)} \Gamma( 2-d/2)\ . \eqno (3.12) $$

On the other hand, the last integral $ {\cal J}^{(2)}_{(b)} = {1 \over 4}
{\cal J}_{4,4} $ involves the integral of $ n^{\prime} =4 $ Bessel functions
in (3.9),
which is not exactly calculable. However, a simplification enables us to
calculate its $ 1/\varepsilon $-expansion to the
desired order $ {\cal O}(1). $ Indeed, in (3.9) for $ n=4, $ we observe the
presence of a factor $ 1/\Gamma( 4-3d/2), $ which produces
a (simple) {\it zero} for $ \varepsilon =4-d=0. $ Therefore we need to know
the Laurent expansion of the integral
$$ I \equiv  \int^{ \infty}_ 0 {\rm d} p\ p^{3(1-d/2)}F^4(p) \eqno (3.13) $$
only to orders $ 1/\varepsilon^ 2, $ and $ 1/\varepsilon . $ These poles come
from the neighbourhood of the origin $ p=0 $ in (3.3). Indeed,
from its definition (3.8c), one can calculate the expansion of $ F $ (Appendix
C)
$$ F(p) = F(0) + \hat F(p) + \hhat F(p) + \ttilde F(p)\ , \eqno (3.14) $$
where
$$ \eqalign{ F(0) & = \Gamma( d/2-1)\ , \cr\hat F(p) & = p^{d/2-1}\Gamma(
1-d/2)-p\Gamma( d/2-2)\ , \cr\hhat F(p) & = -p^{d/2}\Gamma( -d/2)+ {p^2 \over
2} \Gamma( d/2-3)\ , \cr{\cal O} \left(\ttilde F(p) \right) & = {\cal O}
\left(p^{d/2+1} \right)\ . \cr} \eqno (3.15) $$
One thus has to expand the fourth power of the sum (3.14) itself and look for
logarithmic divergences in
integral (3.13) for $ \varepsilon \longrightarrow 0. $ This is done in
Appendix C, within the realm of analytic continuation in $ d. $ The
terms contributing poles to (3.13) can be shown to be
$$ \doublelow{ I \cr d \longrightarrow 4^- \cr}  = \int^ 1_0 {\rm d} p\
p^{3(1-d/2)} \left[4F^3(0)\hhat F(p) + \left(^4_2 \right)F^2(0)\hat F^2(p)
\right] + {\cal O}(1)\ . \eqno (3.16) $$
(Notice the upper integral limit now being 1).

Owing to (3.15) we have explicitly
$$ \eqalignno{ \doublelow{ I \cr d \longrightarrow 4 \cr} &  \doteqdot \Gamma^
2 \left({d \over 2}-1 \right)6 \left\{{ 1 \over 2-d/2} \Gamma^ 2 \left(1-{d
\over 2} \right)+{1 \over 6-3d/2} \Gamma^ 2 \left({d \over 2}-2 \right)
\right. &  \cr  & \left.- 2\Gamma \left(1-{d \over 2} \right)\Gamma \left({d
\over 2} -2 \right){1 \over 4-d} \right\} &  \cr  &  + \Gamma^ 3 \left({d
\over 2}-1 \right)4 \left\{ -\Gamma( -d/2) {1 \over 4-d} + {1 \over 2}\Gamma
\left({d \over 2}-3 \right){1 \over 6-3d/2} \right\} &  \cr  &  + {\cal O}(1)\
, & (3.17) \cr} $$
where the symbol $ \doteqdot $ means that the analytic expression is
understood as a Laurent series with respect
to $ \varepsilon =4-d. $ The last integral $ {\cal J}^{(2)}_{(b)} $ has
therefore the explicit expression, owing to (3.4c), (3.9) and (3.13)
$$ {\cal J}^{(2)}_{(b)} = {1 \over 4} {1 \over  \Gamma( d/2)} {1 \over \Gamma(
4-3d/2)} I\ , \eqno (3.18) $$
where $ I $ is given above.

The Laurent expansion of the various coefficients of (3.4a) is performed as
$$ \eqalign{{\cal J}^{(1)} & = {4 \over \varepsilon}  - 2 - \left(1+{\pi^ 2
\over 6} \right)\varepsilon  + {\cal O} \left(\varepsilon^ 2 \right)\ ,
\cr{\cal J}^{(2)} & \equiv  {\cal J}^{(2)}_{(a)}+{\cal J}^{(2)}_{(b)} = {24
\over \varepsilon^ 2} - {19 \over \varepsilon}  - {33 \over 2} - 3\pi^ 2 +
{\cal O}(\varepsilon) \ , \cr} \eqno (3.19) $$
so that the Taylor-Laurent expansion of the partition function reads to the
required order
$$ \eqalign{{\cal X}(z,\varepsilon) \equiv( 2\pi S)^{d/2}\superone{{\cal
Z}}(S) & = 1 -z\ {\cal J}^{(1)}+z^2{\cal J}^{(2)}+{\cal O} \left(z^3 \right)
\cr  &  = 1 + z \left[- {4 \over \varepsilon}  + 2 + \left(1 + {\pi^ 2 \over
6} \right)\varepsilon  + {\cal O} \left(\varepsilon^ 2 \right) \right] \cr  &
+ z^2 \left[{24 \over \varepsilon^ 2} - {19 \over \varepsilon}  - {33 \over 2}
- 3\pi^ 2 + {\cal O}(\varepsilon) \right] + {\cal O} \left(z^3;{1 \over
\varepsilon^ 3} \right).\ \ \ \cr} \eqno (3.20) $$
In Ref.$\lbrack$33$\rbrack$, we have shown that a polymer scaling function of
the general type
$$ {\cal X}(z,\varepsilon)  = 1 + z \left[{A \over \varepsilon}  + A^{\prime}
+{\cal O}(\varepsilon) \right]+z^2 \left[{B \over \varepsilon^ 2} +
{B^{\prime} \over \varepsilon}  + {\cal O}(1) \right] + {\cal O} \left(z^3;{1
\over \varepsilon^ 3} \right) \eqno (3.21) $$
is renormalizable to two-loop order iff its coefficients obey the identity $
8A+2B-A^2=0, $ which of course is
satisfied here for (3.20) for which $ A=-4, $ $ A^{\prime} =2, $ $ B=24, $ $
B^{\prime} =-19. $ Furthermore, it was shown that the scaling of
such a quantity $ {\cal X} $ (3.21) is necessarily of the form $
\doublelow{{\cal X}(z,\varepsilon) \cr z \longrightarrow \infty \cr}  \propto
z^{{2 \over \varepsilon} \sigma}  \propto  S^\sigma , $ with the critical
exponent
$$ \sigma  = {\varepsilon \over 8} {A \over 2} + \left({\varepsilon \over 8}
\right)^2 \left({17 \over 8} A + 8A^{\prime} +B^{\prime} -AA^{\prime} \right)
+ {\cal O} \left(\varepsilon^ 3 \right) \eqno (3.22) $$
calculated within a {\it minimal subtraction} scheme in the variable $ z $
itself $\lbrack$18,33,34$\rbrack$.

Notice that the sub-subleading parts $ {\cal O}(\varepsilon) $ in $ {\cal
J}^{(1)} $ and $ {\cal O}(1) $ in $ {\cal J}^{(2)} $ do not play any role in
the value of $ \sigma $ to $ {\cal O} \left(\varepsilon^ 2 \right) $
(compare (3.20) (3.21)), while they will be crucial for calculating the
universal amplitude ratio $ {\cal A} $ (2.12).

Thus we obtain the scaling behavior
$$ (2\pi S)^{d/2} \doublelow{ \superone{{\cal Z}}(S) \cr S \longrightarrow
\infty \cr}  \sim  S^\sigma \ , $$
$$ \sigma  = (-2) {\varepsilon \over 8} + \left(- {7 \over 2} \right)
\left({\varepsilon \over 8} \right)^2 + {\cal O} \left(\varepsilon^ 3 \right)
= \left({1 \over 2} - \nu \right)d\ . \eqno (3.23) $$
According to the hyperscaling relation (2.9b), we must have $ \sigma =\alpha
-(2-d/2)= \left({1 \over 2}-\nu \right)d, $ from which we
recover indeed the known value $ \nu ={1 \over 2}[1+{\varepsilon \over 8}+{15
\over 4} \left({\varepsilon \over 8} \right)^2+ $$ {\cal O} $$ \left.
\left(\varepsilon^ 3 \right) \right] $ $\lbrack$7$\rbrack$.

\vfill\eject

\noindent {\bf IV. CALCULATION OF }$ {\bf R}^ {\bf 2} $

\vskip 10pt

{\bf 4.1 Perturbative expansion}

\vskip 10pt

To compute the expansion of the amplitude $ {\cal A} $ (2.12) we need to know
the double Taylor-Laurent expansion of $ R^2 $
to the same orders in $ (z,\varepsilon) $ as in the expansion of $
\superone{{\cal Z}}(S) $ given in (3.20). The formal expansion of $ R^2 $ to
order $ {\cal O} \left(z^2 \right) $
can be found in standard works, e.g. $\lbrack$9,17$\rbrack$. However, previous
works addressed the calculation of the critical
exponent $ \nu $ to order $ {\cal O} \left(\varepsilon^ 2 \right), $ and, as
seen in Eqs.(3.21) and (3.22), terms of the $ z $ perturbation expansion
of successive orders $ z{\cal O}_\varepsilon \left({1 \over \varepsilon} ,1
\right) $ and $ z^2{\cal O}_\varepsilon \left({1 \over \varepsilon^ 2},{1
\over \varepsilon} \right) $ are sufficient for this task. Here we need to
know the {\it
next} sub-leading contributions, namely $ z\ {\cal O}_\varepsilon(
\varepsilon) $ and $ z^2{\cal O}_\varepsilon( 1). $ To obtain the second one
is purely technical but
difficult.

The auxiliary partition function required to calculate $ R^2 $ is
$$ {\cal Z} \left(S,\vec k \right) \equiv  \int^{ }_{ }{\cal D}\vec r(s)\ {\rm
e}^{i\vec k\cdot \left[\vec r(S)-\vec r(0) \right]} {\rm e}^{-\beta{\cal H}
\left\{\vec r \right\}} \ , \eqno (4.1) $$
where $ \beta{\cal H} \left\{\vec r \right\} $ is given in (2.2). The
partition function of an open polymer is therefore
$$ {\cal Z}(S) \equiv  {\cal Z} \left(S,\vec k=\vec 0 \right)\ , \eqno (4.2)
$$
while that of a closed polymer ring (2.2) reads accordingly
$$ \superone{{\cal Z}}(S) \equiv  \int^{ }_{ \Bbb R^d}{ {\rm d}^dk \over(
2\pi)^ d} {\cal Z} \left(S,\vec k \right)\ . \eqno (4.3) $$
{}From the definition (4.1) follows the mean-square end-to-end distance
$$ R^2 = -2d {\partial \over \partial k^2} {\rm \ell n} \ \left.{\cal Z}
\left(S,\vec k \right) \right\vert_{\vec k=\vec 0}\ . \eqno (4.4) $$
The perturbative expansion of $ {\cal Z} \left(S,\vec k \right) $ is easily
calculated by following the rules recalled in section 2,
applied here to the set of diagrams given in Fig.2. In agreement with the
expression given in the monograph by
des Cloizeaux and Jannink $\lbrack$9$\rbrack$, one has respectively the loop
expansions:
$$ \eqalignno{{\cal Z} \left(S,\vec k \right) & = {\cal Z}^{(0)} \left(S,\vec
k \right) + {\cal Z}^{(1)} \left(S,\vec k \right)+{\cal Z}^{(2)} \left(S,\vec
k \right)+{\cal O} \left(z^3 \right)\ , & (4.5) \cr{\cal Z}^{(0)} \left(S,\vec
k \right) & = {\rm e}^{-k^2S/2}\ , & (4.6) \cr{\cal Z}^{(1)} \left(S,\vec k
\right) & = -z \int^ 1_0 {\rm d} x\ x^{-d/2}(1-x) {\rm exp} \left[-(1-x){k^2S
\over 2} \right]\ , & (4.7) \cr{\cal Z}^{(2)} \left(S,\vec k \right) & = {\cal
Z}^{(2)}_a + {\cal Z}^{(2)}_b+{\cal Z}^{(2)}_c\ , & (4.8) \cr} $$
$$ \eqalignno{{\cal Z}^{(2)}_a \left(S,\vec k \right) & = {1 \over 2} z^2
\int^{ \infty}_ 0 {\rm d} x_1\ {\rm d} x_2 \left(x_1x_2 \right)^{-d/2}\theta
\left(1-x_1-x_2 \right) \left(1-x_1-x_2 \right)^2 &  \cr  &  \times  {\rm exp}
\left(- \left(1-x_1-x_2 \right){k^2S \over 2} \right)\ , & (4.9) \cr} $$
$$ \eqalignno{{\cal Z}^{(2)}_b \left(S,\vec k \right) & = z^2 \int^{ \infty}_
0 \prod^ 3_{i=1} \left( {\rm d} x_i\ x^{-d/2}_i \right)\theta \left(1- \sum^
3_{i=1}x_i \right) \left(1- \sum^ 3_{i=1}x_i \right) &  \cr  &  \times  \left(
\sum^ 3_{j=1}{1 \over x_j} \right)^{-d/2} {\rm exp} \left\{ -{k^2S \over 2}
\left[1- \sum^ 3_{i=1}x_i+ \left( \sum^ 3_{i=1}{1 \over x_i} \right)^{-1}
\right] \right\} \ ,\ \ \  & (4.10) \cr} $$
$$ \eqalignno{{\cal Z}^{(2)}_c \left(S,\vec k \right) & = z^2 \int^{ \infty}_
0 \prod^ 3_{i=1} {\rm d} x_i \left(x_1+x_3 \right)^{-d/2}x^{-d/2}_2\theta
\left(1- \sum^ 3_{i=1}x_i \right) \left(1- \sum^ 3_{i=1}x_i \right) &  \cr  &
\times  {\rm exp} \left[-{k^2S \over 2} \left(1- \sum^ 3_{i=1}x_i \right)
\right]\ . & (4.11) \cr} $$

To compute (4.4) we need only to know the expansion of $ {\cal Z} \left(S,\vec
k \right) $ to order $ {\cal O} \left(k^2 \right), $ which we write as
$$ \eqalignno{{\cal Z} \left(S,\vec k \right) & = 1 - z\ {\cal I}^{(1)}_1 +
z^2 \left({\cal I}^{(2)}_{a,1}+{\cal I}^{(2)}_{b,1}+{\cal I}^{(2)}_{c,1}+{\cal
O} \left(z^3 \right) \right) &  \cr  &  + \left(-{k^2S \over 2} \right)
\left\{ 1-z\ {\cal I}^{(1)}_2+z^2 \left({\cal I}^{(2)}_{a,2}+{\cal
I}^{(2)}_{b,2}+{\cal I}^{(2)}_{c,2} \right)+{\cal O} \left(z^3 \right)
\right\} &  \cr  &  + {\cal O} \left(k^4 \right)\ . & (4.12) \cr} $$
The coefficients $ \gq\gq{\cal I}\dq\dq $ above are read from
Eqs.(4.9)-(4.11):
$$ \eqalignno{{\cal I}^{(1)}_1 & = \int^ 1_0 {\rm d} x\ x^{-d/2}(1-x) = {1
\over( 2-d/2)(1-d/2)} & (4.7a) \cr{\cal I}^{(1)}_2 & = \int^ 1_0 {\rm d} x\
x^{-d/2}(1-x)^2 = {2 \over( 3-d/2)(2-d/2)(1-d/2)} & (4.7b) \cr} $$
$$ \eqalignno{{\cal I}^{(2)}_{a,1} & = {1 \over 2} \int^{ \infty}_ 0 {\rm d}
x_1 {\rm d} x_2 \left(x_1x_2 \right)^{-d/2}\theta \left(1-x_1-x_2 \right)
\left(1-x_1-x_2 \right)^2 & (4.9a) \cr{\cal I}^{(2)}_{a,2} & = {1 \over 2}
\int^{ \infty}_ 0 {\rm d} x_1 {\rm d} x_2 \left(x_1x_2 \right)^{-d/2}\theta
\left(1-x_1-x_2 \right) \left(1-x_1-x_2 \right)^3 & (4.9b) \cr} $$
$$ \eqalignno{{\cal I}^{(2)}_{b,1} & = \int^{ \infty}_ 0 \prod^ 3_{i=1} \left(
{\rm d} x_ix^{-d/2}_i \right)\theta \left(1- \sum^ 3_{i=1}x_i \right) \left(1-
\sum^ 3_{i=1}x_i \right) \left( \sum^ 3_{j=1}{1 \over x_j} \right)^{-d/2} &
\cr  &  \equiv  {\cal I}_1 & (4.10a) \cr{\cal I}^{(2)}_{b,2} & = \int^{
\infty}_ 0 \prod^ 3_{i=1} \left( {\rm d} x_ix^{-d/2}_i \right)\theta \left(1-
\sum^ 3_{i=1}x_i \right) \left[ \left(1- \sum^ 3_{i=1}x_i \right)^2 \left(
\sum^ 3_{j=1}{1 \over x_j} \right)^{-d/2} \right.  &  \cr  & \left.\ \ \ \ \ \
\ \ \ \ \ \ \ \ \ \ + \left(1- \sum^ 3_{i=1}x_i \right) \left( \sum^ 3_{j=1}{1
\over x_j} \right)^{-{d \over 2}-1} \right] &  \cr  &  \equiv  {\cal
I}_2+{\cal I}_3\ , & (4.10b) \cr} $$
$$ \eqalignno{{\cal I}^{(2)}_{c,1} & = \int^{ \infty}_ 0 \prod^ 3_{i=1} {\rm
d} x_i \left(x_1+x_3 \right)^{-d/2}x^{-d/2}_2\theta \left(1- \sum^ 3_{i=1}x_i
\right) \left(1- \sum^ 3_{i=1}x_i \right) & (4.11a) \cr{\cal I}^{(2)}_{c,2} &
= \int^{ \infty}_ 0 \prod^ 3_{i=1} {\rm d} x_i \left(x_1+x_3
\right)^{-d/2}x^{-d/2}_2\theta \left(1- \sum^ 3_{i=1}x_i \right) \left(1-
\sum^ 3_{i=1}x_i \right)^2\ . & (4.11b) \cr} $$

The Feynman amplitudes $ {\cal I}^{(2)}_{a,1}, $ $ {\cal I}^{(2)}_{a,2} $ and
$ {\cal I}^{(2)}_{c,1}, $ $ {\cal I}^{(2)}_{c,2} $ are easily calculated by
quadrature, once one uses the
contour integral representation (3.5b) to separate the $ x $-variables. One
obtains:
$$ \eqalignno{{\cal I}^{(2)}_{a,1} = {1 \over 2} {\Gamma( 3)\Gamma^ 2(1-d/2)
\over \Gamma( 5-d)}\ \ \ , & \ \ \ {\cal I}^{(2)}_{a,2} = {1 \over 2} {\Gamma(
4)\Gamma^ 2(1-d/2) \over \Gamma( 6-d)} & (4.12) \cr{\cal I}^{(2)}_{c,1} =
{\Gamma( 2-d/2)\Gamma( 1-d/2) \over \Gamma( 5-d)}\ \ , & \ \ \ {\cal
I}^{(2)}_{c,2} = 2{\Gamma( 2-d/2)\Gamma( 1-d/2) \over \Gamma( 6-d)}\ . &
(4.13) \cr} $$

The contributions (4.10a,b) of the last diagram (b) of Fig.2 require much more
work. First, one uses the same
set of integral representations as in section 3 (Eqs.(3.5a,b)), and the same
change of variables (3.8a), and
separation of variables, which lead to representation (3.9). We find for
Feynman amplitudes $ {\cal I}_1 $ (4.10a) and $ {\cal I}_2, $
$ {\cal I}_3 $ (4.10b)
$$ \eqalignno{{\cal I}^{(2)}_{b,1}\equiv{\cal I}_1 & = {1 \over \Gamma( d/2)}
{1 \over \Gamma( 5-d)} {\cal I}^{\prime}_ 2\ ; & (4.14) \cr{\cal
I}^{(2)}_{b,2} & \equiv  {\cal I}_2+{\cal I}_3\ , &  \cr{\cal I}_2 & = {2
\over \Gamma( d/2)} {1 \over \Gamma( 6-d)} {\cal I}^{\prime}_ 2\ ,\ \ \ {\cal
I}_3={1 \over \Gamma \left({d \over 2}+1 \right)\Gamma( 6-d)} {\cal
I}^{\prime}_ 3\ ; & (4.15) \cr} $$
where $ {\cal I}^{\prime}_ 2, $ $ {\cal I}^{\prime}_ 3 $ read
$$ \eqalignno{{\cal I}^{\prime}_ 2 & = \int^{ \infty}_ 0 {\rm d} p\
p^{2-d}F^3(p)\ , & (4.16) \cr{\cal I}^{\prime}_ 3 & = \int^{ \infty}_ 0 {\rm
d} p\ p^{3-d}F^3(p)\ . & (4.17) \cr} $$
Notice that, contrary to the case of the amplitude (3.18) where the singular
integral $ I $ (3.13) of four Bessel
functions was multiplied by a coefficient vanishing when $ d \longrightarrow
4^-, $ there is no such simplification in (4.14)
(4.15). So we cannot use the method of Appendix C, and need to calculate
(4.16) (4.17).

\vskip 17pt

{\bf 4.2. Bessel integrals}

\vskip 10pt

Using the form (3.10) in terms of modified Bessel functions yields explicitly
$$ \eqalignno{{\cal I}^{\prime}_ 2 & = 2^{2+\nu} \int^{ \infty}_ 0t^{1-\nu}
K^3_\nu( t) {\rm d} t\ , & (4.18) \cr{\cal I}^{\prime}_ 3 & = 2^\nu \int^{
\infty}_ 0t^{3-\nu} K^3_\nu( t) {\rm d} t\ , & (4.19) \cr \nu &  = d/2-1 =
1-\varepsilon /2\ . & (4.20) \cr} $$
Our main tool is the set of exact integration formulae recently provided by A.
Gervois and H. Navelet $\lbrack$20,21$\rbrack$ for
products of three Bessel functions:
$$ I(a,b,c) \equiv  \int^{ \infty}_ 0t^{1-\nu} K_\nu( at)K_\nu( bt)K_\nu( ct)
{\rm d} t = \alpha_ 1\Delta_ 1+\alpha_ 2\Delta_ 2\ , \eqno (4.21) $$
where
$$ \eqalignno{ \alpha_ 1 & = - \left({\pi \over 2} \right)^{5/2}2^{\nu
-1/2}{\Gamma( 1/2-\nu) \over \pi \ {\rm sin} \ \pi \nu} \ , & (4.22a) \cr
\alpha_ 2 & = {\pi \over {\rm sin} \ \pi \nu}  2^{2\nu -3} \left({\pi \over 2}
\right)^{1/2}\Gamma( 1-2\nu) \ ; & (4.22b) \cr} $$
$$ \eqalignno{ \Delta_ 1\equiv{ \Delta^{ 2\nu -1} \over( abc)^\nu} \ ;\ \
\Delta_ 2 & \equiv{ \Delta^{ 2\nu -1} \over( abc)^\nu} \left\{ \left( {\rm
sin} \ \varphi_ a \right)^{1/2-\nu} P^{\nu -1/2}_{\nu -1/2} \left( {\rm cos} \
\varphi_ a \right) \right. &  \cr  &  + \left( {\rm sin} \ \varphi_ b
\right)^{1/2-\nu} P^{\nu -1/2}_{\nu -1/2} \left( {\rm cos} \ \varphi_ b
\right) & (4.23) \cr  & \left.+ \left( {\rm sin} \ \varphi_ c
\right)^{1/2-\nu} P^{\nu -1/2}_{\nu -1/2} \left( {\rm cos} \ \varphi_ c
\right) \right\} \ , &  \cr} $$
valid for $ \nu \not= 0, $ $ -1<\nu <{1 \over 2}, $ and for a \lq\lq
triangle\rq\rq\ configuration of \lq\lq lengths\rq\rq\ $ a,b,c $ such that
$$ \eqalignno{ a^2 & = b^2+c^2-2bc\ {\rm cos} \ \varphi_ a\ \ ,\ \
b^2=c^2+a^2-2ac\ {\rm cos} \ \varphi_ b\ , &  \cr c^2 & = a^2+b^2-2ab\ {\rm
cos} \ \varphi_ c\ \ ,\ \ \varphi_ a+\varphi_ b+\varphi_ c=\pi \ , & (4.24)
\cr \Delta &  = {1 \over 2} ab\ {\rm sin} \ \varphi_ c = {1 \over 2}bc\ {\rm
sin} \ \varphi_ a = {1 \over 2} ca\ {\rm sin} \ \varphi_ b\ . & (4.25) \cr} $$
$ P^\mu_ \nu $ in (4.23) is the associated Legendre function
$\lbrack$32$\rbrack$.

The second integral (4.19) is obtained by differentiation with respect to one
of the lengths, say $ a, $ as $\lbrack$21$\rbrack$
$$ \int^{ \infty}_ 0t^{3-\nu} K_\nu( at)K_\nu( bt)K_\nu( ct) {\rm d} t =
D[I(a,b,c)]\ , \eqno (4.26) $$
where $ D $ is the operator
$$ D = \left({\partial \over \partial a} - {\nu -1 \over a} \right)
\left({\partial \over \partial a} + {\nu \over a} \right)\ . \eqno (4.27) $$
Of course, we need simply for (4.18) the value at the symmetric point $
I(1,1,1), $ but calculating (4.19) (4.26) will require to
distinguish $ a $ from $ b=c=1. $ Furthermore, we have $ \nu =d/2-1, $ thus we
want to reach $ \nu \longrightarrow 1, $ by {\it analytic
continution} of the above formulae, defined for $ -1 < \nu  < 1/2, $ $ \nu
\not= 0, $ and compute its Laurent expansion near $ \nu =1. $

In Appendix E, it is shown that the general formula (4.21) (4.23) simplifies
considerably, using properties of
hypergeometric functions $\lbrack$32$\rbrack$, to
$$ \eqalignno{ \Delta_ 2 & = {1 \over 2^{\nu -1/2}} {1 \over \Gamma( 3/2-\nu)}
\left\{{( bc)^{\nu -1} \over a^\nu}  F \left(1,\ 1-2\nu \ ;{3 \over 2}-\nu \
;\ z_a \right) \right. & (4.28) \cr  & \left.+ {(ca)^{\nu -1} \over b^\nu}  F
\left(1,\ 1-2\nu \ ;{3 \over 2}-\nu \ ;\ z_b \right)+{(ab)^{\nu -1} \over
c^\nu}  F \left(1,\ 1-2\nu \ ;{3 \over 2}-\nu \ ;\ z_c \right) \right\} \ , &
\cr} $$
where $ z_a = {1- {\rm cos} \varphi_ a \over 2} = {\rm sin}^2(\varphi_ a/2), $
etc...

Specifying now to our case, we get:
$$ \eqalignno{ \Delta_ 1(a,1,1) & = {a^{\nu -1} \over 2^{2\nu -1}} \left(1 -
{a^2 \over 4} \right)^{\nu -1/2}\ , &  \cr \Delta_ 2(a,1,1) & = {1 \over
2^{\nu -1/2}} {1 \over \Gamma( 3/2-\nu)}  {\cal F}(a)\ , &  \cr} $$
$$ \eqalignno{{\cal F}(a) & = a^{-\nu} F \left(1,\ 1-2\nu \ ;{3 \over 2}-\nu \
;\ z_a \right)+2a^{\nu -1}F \left(1,\ 1-2\nu \ ;{3 \over 2}-\nu \ ;\ z_b
\right)\ , &  \cr} $$
$$ z_a = a^2/4\ \ \ \ ,\ \ \ \ \ \ \ \ \ z_b = {1 \over 2} \left(1-{a \over 2}
\right)\ . $$
This allows the evaluation of the second order operator $ D $ acting on $
\Delta_{( 1,2)}. $ One finds (Appendix E) at the
symmetric point $ a=b=c=1 $
$$ {\cal F}(a=1)=3F,\ \ \ \ \ \ \ D{\cal F}(a=1) = 4(1-2\nu) F-\nu F^{\prime}
$$
where
$$ \eqalignno{ F & \equiv  F \left(1,\ 1-2\nu \ ;{3 \over 2}-\nu \ ;\ z=1/4
\right)\ , & (4.29) \cr F^{\prime} &  \equiv  F^{\prime}_ z \left(1,\ 1-2\nu \
;{3 \over 2}-\nu \ ;\ z=1/4 \right) = {1-2\nu \over 3/2-\nu}  F \left(2,\
2-2\nu \ ;{5 \over 2}-\nu \ ;{1 \over 4} \right)\ .\ \ \ \ \ \ \  & (4.30)
\cr} $$

Finally we collect the above formulae to get
$$ \eqalignno{ \int^{ \infty}_ 0t^{1-\nu} K^3_\nu( t) {\rm d} t & = \alpha_ 1
\left({ \sqrt{ 3} \over 4} \right)^{2\nu -1}+\alpha_ 2\  {1 \over 2^{\nu
-1/2}} {1 \over \Gamma( 3/2-\nu)}  3F\ ; & (4.31) \cr \int^{ \infty}_
0t^{3-\nu} K^3_\nu( t) {\rm d} t & = \alpha_ 1 \left({ \sqrt{ 3} \over 4}
\right)^{2\nu -5}{1 \over 2^6}(1-2\nu)( \nu +3)  &  \cr  &  + \alpha_ 2\ {1
\over 2^{\nu -1/2}} {1 \over \Gamma( 3/2-\nu)}  \left[4(1-2\nu) F-\nu
F^{\prime} \right]\ , & (4.32) \cr} $$
where $ F $ and $ F^{\prime} $ are given explicitly in (4.29) (4.30), and
where $ \alpha_ 1, $ $ \alpha_ 2 $ are coefficients (4.22).

The forms (4.31) (4.32) of the integrals are understood as analytic
continuation in the parameter $ \nu . $ To
perform the Laurent-expansion with respect to $ \nu -1 $ requires further
transformations on the hypergeometric
functions appearing in (4.29) (4.30). The details are given in Appendix F. One
arrives at
$$ \eqalignno{ F & = A \left({3 \over 4} \right)^{- \left({1 \over 2}-\nu
\right)}+B{1 \over 2} \left({3 \over 4} \right)^{-(1-\nu)} F_{1/2}\ , &
(4.33a) \cr A & = {\Gamma \left({3 \over 2}-\nu \right)\Gamma \left({1 \over
2} \right) \over \Gamma( 1-\nu)} \ \ \ ,\ \ \ \ B = -2 \left({1 \over 2}-\nu
\right)\ , &  \cr F_{1/2} & \equiv  F \left({1 \over 2},\ 1-\nu \ ;{3 \over 2}
;\ z=-{1 \over 3} \right)\ ; & (4.33b) \cr} $$
$$ \eqalignno{ F^{\prime} &  = {1-2\nu \over 3/2-\nu}  F_2\ ,\ \ \ \ \ F_2 = F
\left(2,\ 2-2\nu ;{5 \over 2}-\nu \ ;{1 \over 4} \right)\ , &  \cr F_2 & =
A^{\prime} \left({3 \over 4} \right)^{-(1-\nu)} F_{-1/2} + {B^{\prime} \over
2} \left({3 \over 4} \right)^{- \left({3 \over 2}-\nu \right)}\ , & (4.34a)
\cr A^{\prime} &  = 2 \left({3 \over 2}-\nu \right)\ \ \ \ \ \ ,\ \ B^{\prime}
 = {\Gamma \left({5 \over 2}-\nu \right)\Gamma \left(-{1 \over 2} \right)
\over \Gamma( 1-\nu)} \ , &  \cr F_{-1/2} & \equiv  F \left(-{1 \over 2},\
1-\nu \ ;{1 \over 2};\ z=-1/3 \right)\ . & (4.34b) \cr} $$
Only now are the last two remaining hypergeometric functions $ F_{1/2} $
(4.33b) and $ F_{-1/2} $ (4.34b) in a form suitable
for
the expansion with respect to $ \nu =1, $ thanks to the fact that $ \nu $
appears only {\it once} in the argument. The
hypergeometric series have to be expanded to third order in $ \nu -1 $ and
then {\it resummed.} This is done in
Appendix F. The final expansions of $ F_{1/2} $ and $ F_{-1/2} $ are, after
some work:
$$ \eqalignno{ F \left({1 \over 2},1-\nu ;{3 \over 2},z=-1/3 \right) & = 1 +
{1 \over 2} {1 \over \Gamma( 1-\nu)}  \left\{ S^{\prime} +(1-\nu)
S^{\prime\prime} +{\cal O} \left[(1-\nu)^ 2 \right] \right\} \ \ ,\ \ \ \ \ \
& (4.35a) \cr S^{\prime} &  = 2 \left[2-{\pi \over \sqrt{ 3}} - {\rm \ell
n}(4/3) \right]\ , & (4.35b) \cr S^{\prime\prime} &  = (2-\gamma) S^{\prime}
+2S_1-4 \sqrt{ 3}\ S_3\ \ , & (4.35c) \cr S_1 & = {1 \over 2} \left( {\rm \ell
n} \ {4 \over 3} \right)^2\ ,\ \ S_3 = -2L(\pi /6)\ , & (4.36) \cr} $$
where $ L(x) $ is the Lobachevsky function $\lbrack$32$\rbrack$
$$ L(x) \equiv  - \int^ x_0 {\rm \ell n}( {\rm cos} \ t) {\rm d} t\ . \eqno
(4.37) $$
Similarly, we get (Appendix F)
$$ \eqalignno{ F \left(-{1 \over 2},1-\nu ;{1 \over 2};z=-{1 \over 3} \right)
& = 1 - {1 \over 2} {1 \over \Gamma( 1-\nu)} \left\{\tilde S^{\prime}
+(1-\nu)\tilde S^{\prime\prime} +{\cal O} \left[(1-\nu)^ 2 \right] \right\} \
,\ \ \ \ \ \ \ \ \ \ \  & (4.38a) \cr\tilde S^{\prime} &  = 2 \left\{ - {2
\over \sqrt{ 3}} {\pi \over 6} + {\rm \ell n}(4/3) \right\} \ , & (4.38b)
\cr\tilde S^{\prime\prime} &  = -\gamma\tilde S^{\prime}  - 2S_1 - {4 \over 3}
\sqrt{ 3}\ S_3\ , & (4.38c) \cr} $$
with $ S_1 $ and $ S_3 $ the same as in (4.36).

We need the Laurent expansion of Bessel integrals (4.18) (4.19) (4.31) (4.32),
where
$$ \nu  = {d \over 2} - 1 = 1 - \varepsilon /2\ , $$
at orders $ {\cal O} \left(1/\varepsilon^ 2 \right), $ $ {\cal
O}(1/\varepsilon) , $ and $ {\cal O}(1). $ This is done with the help of the
complete set of explicit results (4.33) to (4.38) and
by using {\it Mathematica}$ ^{\circledR}. $ We get
$$ \eqalignno{ \int^{ \infty}_ 0t^{1-\nu} K^3_\nu( t) {\rm d} t & = - {3 \over
4} {1 \over \varepsilon^ 2} + {3 \over 8} (-2+\gamma - {\rm \ell n} \ 2){1
\over \varepsilon} &  \cr  &  + \left(-48+24\gamma -6\gamma^ 2-3\pi^ 2-48
\sqrt{ 3}\ L(\pi /6)-24\ {\rm \ell n} \ 2 \right. &  \cr  &  + \left.12\gamma
\ {\rm \ell n} 2+8 \sqrt{ 3}\ \pi \ {\rm \ell n} \ 2 - 6\ {\rm \ell n}^22
\right){1 \over 64} + {\cal O}(\varepsilon) \ ; & (4.39a) \cr \int^{ \infty}_
0t^{3-\nu} K^3_\nu( t) {\rm d} t & = {1 \over 2\varepsilon}  - {\gamma \over
4} + {4 \over \sqrt{ 3}} L(\pi /6) + {1 \over 4} {\rm \ell n} \ 2 &  \cr  &  -
{2 \over 3} {\pi \over \sqrt{ 3}} {\rm \ell n} \ 2+{\cal O}(\varepsilon) \ . &
(4.39b) \cr} $$

We get for partial integrals (4.16)-(4.19)
$$ \eqalignno{{\cal I}^{\prime}_ 2 & = - {6 \over \varepsilon^ 2} + (3\gamma
-6){1 \over \varepsilon}  - 6 \sqrt{ 3}\ L(\pi /6)+ \sqrt{ 3}\ \pi \ {\rm \ell
n} \ 2 &  \cr  &  - {3 \over 8} \pi^ 2 - {3 \over 4} \gamma^ 2 + 3\gamma  - 6
+ {\cal O}(\varepsilon) \ , & (4.40a) \cr{\cal I}^{\prime}_ 3 & = {1 \over
\varepsilon}  + {8 \over \sqrt{ 3}} L(\pi /6) - {4 \over 3} {\pi \over \sqrt{
3}} {\rm \ell n} \ 2 - {\gamma \over 2} + {\cal O}(\varepsilon) \ . & (4.40b)
\cr} $$

Notice that these integrals (4.18) (4.19) can also be analytically continued
using the technique of section 3
(Eqs.(3.14) and seq.) and of Appendix C. One can easily analytically isolate
the pole part, while the finite
part at $ d=4 $ is given by explicit subtracted integrals, which can be
numerically evaluated. This approach is
followed in Appendix G and allows for a numerical check of Eqs.(4.39) (4.40).

\vskip 17pt

{\bf 4.3. Laurent expansion of Feynman amplitudes}

\vskip 10pt

We are ultimately interested in the Laurent expansions of the set of Feynman
amplitudes associated with the
diagrams of Fig.2, and given by analytical expressions (4.7a,b) at one-loop
order, and by (4.14)-(4.19) at
two-loop order. Using again {\it Mathematica}$ ^{\circledR}, $ as well as the
particular results (4.40) of the last section
for the diagram b of Fig.2, we get
$$ {\cal I}^{(1)}_1= - {2 \over \varepsilon}  - 1 - {\varepsilon \over 2} +
{\cal O} \left(\varepsilon^ 2 \right)\ \ \ ;\ \ \ {\cal I}^{(1)}_2= - {4 \over
\varepsilon}  - \varepsilon  + {\cal O} \left(\varepsilon^ 3 \right)\ , \eqno
(4.41) $$
$$ \eqalignno{{\cal I}^{(2)}_{a,1} & = {4 \over \varepsilon^ 2} + {4 \over
\varepsilon}  + 3 - {\pi^ 2 \over 6} + {\cal O}(\varepsilon) \ , & (4.42a)
\cr{\cal I}^{(2)}_{a,2} & = {12 \over \varepsilon^ 2} + 9 - {\pi^ 2 \over 2} +
{\cal O}(\varepsilon) \ , & (4.42b) \cr{\cal I}^{(2)}_{b,1} & = - {6 \over
\varepsilon^ 2} - {9 \over \varepsilon}  - {21 \over 2} + {\pi^ 2 \over 4} -6
\sqrt{ 3} \left[L(\pi /6) - {\pi \over 6} {\rm \ell n} \ 2 \right]+{\cal
O}(\varepsilon) \ , & (4.43a) \cr{\cal I}^{(2)}_{b,2} & = - {12 \over
\varepsilon^ 2} - {11 \over 2\varepsilon}  - {121 \over 8} + {\pi^ 2 \over 2}
- {32 \over \sqrt{ 3}} \left[L(\pi /6) - {\pi \over 6} {\rm \ell n} \ 2
\right] + {\cal O}(\varepsilon) \ ,\ \  & (4.43b) \cr{\cal I}^{(2)}_{c,1} & =
- {4 \over \varepsilon^ 2} - {2 \over \varepsilon}  - 1+{\pi^ 2 \over 6}+{\cal
O}(\varepsilon) \ , & (4.44a) \cr{\cal I}^{(2)}_{c,2} & = - {8 \over
\varepsilon^ 2} + {4 \over \varepsilon}  -6+{\pi^ 2 \over 3} + {\cal
O}(\varepsilon) \ . & (4.44b) \cr} $$

The final perturbative expansion of $ R^2 $ is obtained from the generating
function (4.4), (4.12) together with the
explicit values (4.41)-(4.44) above. We find in the double Taylor-Laurent $
(z,\varepsilon) $ scheme:
$$ \eqalignno{{\cal X}_0(z,\varepsilon) \equiv R^2/dS & = 1 + z \left[{2 \over
\varepsilon}  - 1 + {\varepsilon \over 2} + {\cal O} \left(\varepsilon^ 2
\right) \right] &  \cr  &  + z^2 \left\{ - {6 \over \varepsilon^ 2} + {11
\over 2} {1 \over \varepsilon}  - {37 \over 8} + {\pi^ 2 \over 12} - {14 \over
\sqrt{ 3}} \left[L(\pi /6) - {\pi \over 6} {\rm \ell n} \ 2 \right]+{\cal
O}(\varepsilon) \right\} \ \ \ \ \ \ \  &  \cr  &  + {\cal O} \left(z^3;{1
\over \varepsilon^ 3} \right)\ . & (4.45) \cr} $$

As already mentioned in section 3, Eqs.(3.21) (3.22), at each order in $ z, $
the two leading pole parts suffice
to give the critical exponent $ \nu $ such that $ R^2\propto S^{2\nu} $ to
order $ {\cal O} \left(\varepsilon^ 2 \right), $ and they are of course
identical to former
results in the direct renormalization approach of des Cloizeaux
$\lbrack$17,9$\rbrack$. The sub-leading terms are new, and
will yield the universal amplitude $ {\cal A}. $ They can also serve to
determine, in the mathematical framework of the
two-parameter model, the {\it non-universal} amplitude of $ R^2 $ itself with
respect to its asymptotic power law
$\lbrack$33$\rbrack$
$$ R^2/dS \equiv  {\cal X}_0(z) \cong \left[{z \over h(\varepsilon)}
\right]^{(2\nu -1)2/\varepsilon} \ ,\ \ \ \ z \longrightarrow \infty \ , \eqno
(4.46) $$
where $ h(\varepsilon) $ is a Taylor series in $ \varepsilon , $ given
explicitly including $ {\cal O} \left(\varepsilon^ 3 \right) $ by the
coefficients of (4.45) $\lbrack$see
Eqs.(A.8)-(A.14) in Ref.$\lbrack$33$\rbrack$). Amplitudes $ h(\varepsilon) $
in (4.46) and $ {\cal A}(\varepsilon) $ (2.12) should be clearly
distinguished: the
former is non universal and is a associated with the Edwards model, while the
latter is {\it universal}
and is obtained here by a calculation within the same minimal two-parameter
Edwards model (see Ref.$\lbrack$35$\rbrack$ for a
recent discussion).

\vskip 17pt

\noindent {\bf V. UNIVERSAL AMPLITUDE}

\vskip 10pt

{\bf 5.1. Formal values}

The ring amplitude has been defined in Eq.(2.12) and is given by the
combination of expressions (3.20) and
(4.45); whence
$$ \eqalignno{{\cal A}(z,\varepsilon) &  = {\cal X}(z,\varepsilon)  {\cal
X}^{d/2}_0(z,\varepsilon) &  \cr  &  = 1 + z \left[-1+ \left({5 \over 2}+{\pi^
2 \over 6} \right)\varepsilon  + {\cal O} \left(\varepsilon^ 2 \right) \right]
&  \cr  &  + z^2 \left\{{ 8 \over \varepsilon} -30-13{\pi^ 2 \over 6} - {28
\over 3} \sqrt{ 3} \left[L(\pi /6) - {\pi \over 6} {\rm \ell n} \ 2 \right]+
{\cal O}(\varepsilon) \right\} &  \cr  &  + {\cal O} \left(z^3;{1 \over
\varepsilon^ 2} \right)\ . & (5.1) \cr} $$

As expected, this expression is more regular than the scaling amplitudes $
{\cal X} $ and $ {\cal X}_0 $ themselves, the order of the
poles in $ \varepsilon $ being lower at each order in $ z. $

It is fully regular (to all orders) when expanded with respect to the
renormalized coupling constant $ z_R $
defined in Appendix D. In this minimal subtraction scheme, adapted to the
two-parameter Edwards model $\lbrack$18$\rbrack$,
the renormalized parameter $ z_R $ reads
$$ \eqalignno{ z_R & = z - z^2 {8 \over \varepsilon}  + {\cal O} \left(z^3;{1
\over \varepsilon^ 2} \right) &  \cr {\rm or} \ \ \  &   & (5.2) \cr z & = z_R
+ {8 \over \varepsilon}  z^2_R + {\cal O} \left(z^3_R;{1 \over \varepsilon^ 2}
\right)\ , &  \cr} $$
with a fixed point value calculated from the next order expansion
($\lbrack$18$\rbrack$ and Appendix D)
$$ z^\ast_ R = {\varepsilon \over 8} + {17 \over 4} \left({\varepsilon \over
8} \right)^2 + {\cal O} \left(\varepsilon^ 3 \right)\ . \eqno (5.3) $$
We therefore have
$$ \eqalignno{{\cal A} \left[z_R,\varepsilon \right] & = 1 + z_R \left[-1+
\left({5 \over 2}+{\pi^ 2 \over 6} \right)\varepsilon  + {\cal O}
\left(\varepsilon^ 2 \right) \right] &  \cr  &  + z^2_R \left\{ -10-5{\pi^ 2
\over 6} - {28 \over 3} \sqrt{ 3} \left[L(\pi /6) - {\pi \over 6} {\rm \ell n}
\ 2 \right]+ {\cal O}(\varepsilon) \right\} &  \cr  &  + {\cal O}
\left(z^3_R;\ 1 \right)\ . & (5.4) \cr} $$
For very large chains, $ z \longrightarrow \infty $ while $ z_R
\longrightarrow  z^\ast_ R, $ and the universal amplitude $ {\cal A} $ reaches
its fixed point value
$$ \eqalignno{{\cal A} \left[z^\ast_ R,\varepsilon \right] & \equiv  {\cal
A}(d) = 1 - {\varepsilon \over 8}+ \left({\varepsilon \over 8} \right)^2
\left\{{ 23 \over 4}+{\pi^ 2 \over 2}-{28 \over \sqrt{ 3}} \left[L(\pi
/6)-{\pi \over 6} {\rm \ell n} \ 2 \right] \right\} &  \cr  &  + {\cal O}
\left(\varepsilon^ 3 \right)\ . & (5.5) \cr} $$

As mentioned above, the Lobachevsky function (4.37) can be expressed for
rational angles in terms of
derivatives of the $ \psi $-function, $ \psi( x) = { {\rm d} \over {\rm d} x}
{\rm \ell n} \ \Gamma( x). $ Specifically, we first use the duality relation
$\lbrack$32$\rbrack$
$$ L(\pi /6) = {2 \over 3} L(\pi /3) - {\pi \over 18} {\rm \ell n} \ 2 $$
and then the sine representation of $ L(x) $ $\lbrack$32$\rbrack$
$$ L(x) = x\ {\rm \ell n} \ 2 - {1 \over 2} \sum^{ \infty}_{ k=1}(-1)^{k-1} {
{\rm sin} \ 2kx \over k^2} $$
to arrive at
$$ 2 \sqrt{ 3} \left[{\pi \over 6} {\rm \ell n} \ 2 - L \left({\pi \over 6}
\right) \right] = \sum^{ }_{ n\geq 0}(-1)^n \left[{1 \over( 3n+1)^2} + {1
\over( 3n+2)^2} \right]\ . \eqno (5.6) $$
Using the series representation $\lbrack$32,36$\rbrack$
$$ \sum^{ }_{ n\geq 0}{(-1)^n \over( xn+y)^2} = -x^{-2}\beta^{ \prime}( y/x)
\eqno (5.7) $$
yields
$$ L(\pi /6) = {\pi \over 6} {\rm \ell n} \ 2 + {1 \over 18 \sqrt{ 3}}
\left[\beta^{ \prime}( 1/3)+\beta^{ \prime}( 2/3) \right]\ , \eqno (5.8) $$
where the $ \beta $-function is $\lbrack$32$\rbrack$
$$ \beta( x) = {1 \over 2} \left[\psi  \left({x+1 \over 2} \right)-\psi
\left({x \over 2} \right) \right]\ . \eqno (5.9) $$
Thus
$$ L(\pi /6) = {\pi \over 6} {\rm \ell n} \ 2 + {1 \over 8\times 9 \sqrt{ 3}}
\left[\psi^{ \prime}( 2/3)+\psi^{ \prime}( 5/6)-\psi^{ \prime}( 1/6)-\psi^{
\prime}( 1/3) \right]\ . \eqno (5.10) $$
Using the doubling formula for the $ \psi $-function $\lbrack$32$\rbrack$, we
arrive at
$$ L(\pi /6) = {\pi \over 6} {\rm \ell n} \ 2 + {1 \over 2 \sqrt{ 3}\ 9}
\left\{{ 4 \over 3}\pi^ 2 - {1 \over 2} \left[\psi^{ \prime}( 1/6)+\psi^{
\prime}( 1/3) \right] \right\} \ . \eqno (5.11) $$
Thus another expression for the amplitude is
$$ \eqalignno{{\cal A}(d) & \equiv  \lim_{S \longrightarrow \infty}
\superone{{\cal Z}}(S) \left(2\pi R^2/d \right)^{d/2} &  \cr  &  = 1 -
{\varepsilon \over 8} + \left({\varepsilon \over 8} \right)^2 \left\{{ 23
\over 4} - {31 \over 2\times 3^4} \pi^ 2 + {7 \over 27} \left[\psi^{ \prime}(
1/6)+\psi^{ \prime}( 1/3) \right] \right\} &  \cr  &  + {\cal O}
\left(\varepsilon^ 3 \right)\ . & (5.12) \cr} $$
\vskip 17pt

{\bf 5.2. Four dimensions}

\vskip 10pt

To find the expression for the universal amplitude in four dimensions, it is
sufficient to take the regular
limit of Eq.(5.4) for $ \varepsilon =0: $
$$ {\cal A}_{d=4} \left[z_R \right] = 1 -z_R + z^2_R \left\{ -10-5{\pi^ 2
\over 6} - {28 \over \sqrt{ 3}} \left[L(\pi /6)-{\pi \over 6} {\rm \ell n} \ 2
\right] \right\} +{\cal O} \left(z^3_R \right)\ .\ \ \eqno (5.13) $$

For {\it infinite} chains in four dimensions the renormalized $ z $-parameter
$ z_R $ vanishes $\lbrack$39$\rbrack$, as does any
renormalized coupling constant at the critical temperature in a continuum
field theory, and Eq.(5.13) will
actually describe the approach to the asymptotic Brownian value $ {\cal A}=1.
$ Indeed the asymptotic expression of $ z_R $ in
$ d=4 $ is $\lbrack$37$\rbrack$
$$ z_R = {1 \over 4\ {\rm \ell n} \left(S/s_0 \right)} + {17 \over 4} { {\rm
\ell n} \left(4\ {\rm \ell n} \ S/s_0 \right) \over \left(4\ {\rm \ell n} \
S/s_0 \right)^2} + {\cal O} \left({1 \over {\rm \ell n}^2S/s_0} \right)\ ,
\eqno (5.14) $$
where of course now a microscopic short range cut-off $ s_0 $ is needed along
the chains. Notice that the
sub-subleading correction term $ {\cal O} \left( {\rm \ell n}^{-2}S/s_0
\right) = {\cal O} \left(z^2_R \right) $ in Eq.(5.14) can be calculated
$\lbrack$37$\rbrack$, but is of the
same order of magnitude as the effect of a change of cut-off in the leading
term $ {\cal O} \left( {\rm \ell n}^{-1}S/s_0 \right). $ For this reason
it won't be written explicitly here, and similarly, terms of order $ z^2_R $
in (5.13) will also be left aside. We get
therefore to this order
$$ {\cal A}_{d=4} = 1 - {1 \over 4\ {\rm \ell n} \left(S/s_0 \right)} - {17
\over 4} { {\rm \ell n} \left(4\ {\rm \ell n} \ S/s_0 \right) \over \left(4\
{\rm \ell n} \ S/s_0 \right)^2} + {\cal O} \left( {\rm \ell n}^{-2}
\left(S/s_0 \right) \right)\ . \eqno (5.15) $$
It would be of interest to try and test this asymptotic behavior by
enumeration series on a four-dimensional
lattice.

\vfill\eject

\noindent {\bf VI. LINK TO FIELD THEORY}

\vskip 10pt

{\bf 6.1 Field theory}

As mentioned in the introduction, the amplitude ratio $ A(d) $ (1.7a) (2.10)
for polymer rings and the specific
heat/correlation length amplitude ratio $ {1 \over n} \left(R^+_\xi \right)^d
$ (1.10) of the $ O(n) $ model, $ n \longrightarrow 0, $ are related to one
another.
Let us establish the precise correspondence within the framework of Lagrangian
field theory.

The Lagrangian of the $ O(n) $ model is defined as
$$ {\cal L}( {\bf \Phi})  = \int^{ }_{ \Bbb R^d} {\rm d}^dx \left\{{ 1 \over
2} \left[\vec \nabla {\bf \Phi} \left(\vec x \right) \right]^2 + {1 \over 2}
m^2 \left[ {\bf \Phi} \left(\vec x \right) \right]^2 + {b \over 2} \left[ {\bf
\Phi}^ 2 \left(\vec x \right) \right]^2 \right\} \ , \eqno (6.1) $$
where
$$ {\bf \Phi}( x) = \left(\varphi_ 1 \left(\vec x \right),...,\varphi_ n
\left(\vec x \right) \right) \eqno (6.2) $$
is an $ n $-component real-valued field depending on the position $ \vec x \in
 \Bbb R^d, $ of (bare) mass $ m, $ and where
$$ \eqalignno{ \left[\vec \nabla {\bf \Phi} \left(\vec x \right) \right]^2 &
\equiv  \sum^ d_{i=1} \sum^ n_{\alpha =1} \left[{\partial \over \partial x_i}
\varphi_ \alpha \left(\vec x \right) \right]^2 & (6.3) \cr \left[ {\bf \Phi}
\left(\vec x \right) \right]^2 & \equiv  \sum^ n_{\alpha =1} \left[\varphi_
\alpha \left(\vec x \right) \right]^2\ . & (6.4) \cr} $$

We consider here the {\it bare} field theory, where the presence of a
short-range cut-off is always understood.
We shall need the two-point vertex function $\lbrack$2$\rbrack$
$$ \Gamma^{( 2)} \left(\vec k,m,b,d \right) = 1/G^{(2)} \left(\vec k,m,b,d
\right)\ , \eqno (6.5) $$
which is defined in terms of the correlation function in momentum space:
$$ G^{(2)} \left(\vec k,m,b,d \right) \equiv  \int^{ }_{ \Bbb R^d} {\rm d}^dx\
{\rm e}^{i\vec k\cdot\vec x} \left\langle \varphi_ \alpha \left(\vec x
\right)\varphi_ \alpha \left(\vec 0 \right) \right\rangle \eqno (6.6) $$
for a fixed component index $ \alpha $ arbitrarily chosen in $ \{ 1,n\} . $

The vertex function $ \Gamma^{( 2,0)} $ with insertion of two composite
operators $ {\bf \Phi}^ 2 \left(\vec x \right) $ reads similarly
$\lbrack$2$\rbrack$
$$ \eqalignno{ \Gamma^{( 2,0)} \left(\vec k,m,b,d \right) & = \int^{ }_{ \Bbb
R^d} {\rm d}^dx\ {\rm e}^{i\vec k\cdot\vec x} \left\langle{ 1 \over 2} {\bf
\Phi}^ 2 \left(\vec x \right){1 \over 2} {\bf \Phi}^ 2 \left(\vec 0 \right)
\right\rangle &  (6.7) \cr} $$
In terms of these vertex functions, the correlation length $ \xi $ is defined
by $\lbrack$2$\rbrack$
$$ \xi^ 2 \equiv  {\partial \over \partial k^2} \Gamma^{( 2)} \left(\vec
k=\vec 0 \right)/\Gamma^{( 2)} \left(\vec k=\vec 0 \right)\ , \eqno (6.8) $$
while the specific heat is
$$ C \equiv  \Gamma^{( 2,0)} \left(\vec k=\vec 0 \right)\ . \eqno (6.9) $$

The important property is the singular behavior of the vertex functions near
the critical point $ m^2_c $ where $ \Gamma^{( 2)} \left(\vec k=\vec 0 \right)
$
vanishes, measured in terms of the \lq\lq reduced temperature\rq\rq\ or mass
shift:
$$ t \equiv  m^2-m^2_c \eqno (6.10) $$
One has respectively $\lbrack$2$\rbrack$
$$ \eqalignno{ \matrix{  \cr \Gamma^{( 2)} \cr t \longrightarrow 0^+ \cr}
\left(\vec k=\vec 0,m,b,d \right) & \simeq  \Gamma_ +t^\gamma &  (6.11a) \cr
\matrix{  \cr {\displaystyle{\partial \over \partial k^2}} \Gamma^{( 2)} \cr t
\longrightarrow 0^+ \cr} \left(\vec k=\vec 0,m,b,d \right) & \simeq  \Gamma^{
\prime}_ +t^{\gamma -2\nu} &  (6.11b) \cr \matrix{  \cr \Gamma^{( 2,0)} \cr t
\longrightarrow 0^+ \cr} \left(\vec k=\vec 0,m,b,d \right) & \simeq  {A_+
\over \alpha}  t^{-\alpha} +C_2\ , & (6.11c) \cr} $$
where $ \gamma , $ $ \nu $ and $ \alpha $ are the standard critical exponents
of the $ O(n) $ model, and where $ \Gamma_ +, $ $ \Gamma^{ \prime}_ +, $ $ A_+
$ are coefficients
independent of $ t. $
$ C_2 $ is an additive constant, corresponding to the analytic part of the
specific heat.

We therefore arrive in the critical regime at the leading behaviors
$$ \eqalignno{ \xi^ 2(t) & \simeq  (\Gamma^{ \prime}_ +/\Gamma_ +)t^{-2\nu} \
,\ \ \ \ \ t \longrightarrow  0^+ & (6.12a) \cr C_S(t) & \simeq  {A_+ \over
\alpha}  t^{-\alpha} \ ,\ \ \ \ \ \ \ \ t \longrightarrow  0^+,\ \ \alpha
=2-\nu d & (6.12b) \cr} $$
when we consider the {\it singular} (non-analytic) part $ C_S $ of the
specific heat.

The amplitude ratio $ \left(R^+_\xi \right)^d $ defined in the introduction
(Eq.(1.10)) is therefore, in this notation
$$ \eqalignno{ \left(R^+_\xi \right)^d & = \lim_{t \longrightarrow 0^+}
\left\{ \alpha \ t^2C_S(t)\xi^ d(t) \right\} &  \cr  &  = A_+ \left(\Gamma^{
\prime}_ +/\Gamma_ + \right)^{d/2} & (6.13) \cr} $$

\vskip 17pt

{\bf 6.2 Polymer theory}

The partition function $ {\cal Z} \left(S,\vec k \right) $ defined by (4.1) is
the inverse Laplace transform of the (bare) correlation
function (6.5), (6.6) $\lbrack$7,8,9$\rbrack$
$$ {\cal Z} \left(S,\vec k,b,d \right) = {1 \over 2\pi i} \int^{ }_{{\cal
C}^{\prime}} {\rm d} m^2\ {\rm e}^{ \left(m^2-m^2_c \right)S/2}
\left[\Gamma^{( 2)} \left(\vec k,m,b,d \right) \right]^{-1}_{ \left\vert_{
n=0} \right.} \eqno (6.14) $$
along the vertical contour $ {\cal C}^{\prime} $ of integration in the complex
plane $ (\sigma -i\infty ,\sigma +i\infty) , $ $ \sigma \in \Bbb R^+, $ $
\sigma  > m^2_c. $ The mass
shift (i.e. free energy shift) in the exponential removes the assumed
underlying cut-off of the field
theory, yielding the dimensionally regularized polymer partition function for
$ 2 < d < 4 $ (and $ d\not= 4-2/p, $
$ p\in \Bbb N^\ast $ $\lbrack$9,28$\rbrack$), which corresponds to a trivial
free energy shift in the polymer partition function.

The mean squared end-to-end distance (4.4) reads therefore explicitly from
Eq.(6.14)
$$ R^2 = 2d { \int^{ }_{{\cal C}} {\rm d} t\ {\rm e}^{tS/2}{\partial \over
\partial k^2} \Gamma^{( 2)} \left(\vec 0 \right) \left[\Gamma^{( 2)}
\left(\vec 0 \right) \right]^{-2} \over \int^{ }_{{\cal C}} {\rm d} t\ {\rm
e}^{tS/2} \left[\Gamma^{( 2)} \left(\vec 0 \right) \right]^{-1}}\ , \eqno
(6.15) $$
where we shifted from variable $ m^2 $ to $ t=m^2-m^2_c, $ and where $ {\cal
C} $ is now a vertical contour in $ \Bbb C, $ $ (0^+-i\infty , $ $ 0^++i\infty
). $

Expressing the partition function of a polymer ring in terms of the field
theory vertex functions must be done
with some care since here we look for amplitude ratios. With the help of
Ref.$\lbrack$38$\rbrack$ we get for a rooted and
oriented continuous ring, a partition function $ \superone{{\cal Z}}(S) $ such
that
$$ {1 \over 2} \left({S \over 2} \right)\superone{{\cal Z}}(S,b,d) = {1 \over
2\pi i} \int^{ }_{{\cal C}} {\rm d} m^2\ {\rm e}^{ \left(m^2-m^2_0
\right)S/2}{\partial \over \partial n} \Gamma^{( 2,0)} \left. \left(\vec
0,m,b,d \right) \right\vert_{ n=0} \eqno (6.16) $$

By Laplace transformation, the large $ S $ behavior of partition functions
(6.14) (6.16) is of course given by the
small $ t = m^2-m^2_c $ behavior of the vertex functions (6.11). We find
asymptotically, for the end-to-end squared
distance (6.15)

$$ \doublelow{ R^2 \cr S \longrightarrow \infty \cr}  \simeq  2d {\Gamma^{
\prime}_ + \over \Gamma_ +} { \int^{ }_{{\cal C}}{ {\rm d} t \over 2\pi i}
{\rm e}^{tS/2}t^{-\gamma -2\nu} \over \int^{ }_{{\cal C}}{ {\rm d} t \over
2\pi i} {\rm e}^{tS/2}t^{-\gamma}} \eqno (6.17a) $$
and for the ring partition function (6.16)
$$ {S \over 2^2} \doublelow{ \superone{{\cal Z}}(S) \cr S \longrightarrow
\infty \cr}  \simeq  \lim_{n \longrightarrow 0} \left({1 \over n} {A_+ \over
\alpha} \right)\times  \int^{ }_{{\cal C}}{ {\rm d} t \over 2\pi i} {\rm
e}^{tS/2}t^{-\alpha} \ , \eqno (6.17b) $$
where the critical exponents are from now on those of the $ n=0 $ model.
We now use the complex integral (3.5b) to get for $ S $ large
$$ \eqalignno{ R^2 & = 2d {\Gamma^{ \prime}_ + \over \Gamma_ +} \left({S \over
2} \right)^{2\nu}{ \Gamma( \gamma) \over \Gamma( \gamma +2\nu)} &  (6.18a) \cr
\superone{{\cal Z}}(S) & = 2 \lim_{n \longrightarrow 0} \left({1 \over n} {A_+
\over \alpha} \right) {1 \over \Gamma( \alpha)}  \left({S \over 2}
\right)^{\alpha -2}\ . & (6.18b) \cr} $$

The universal amplitude ratio $ A(d) $ defined in Eq.(2.10) therefore becomes
$$ \eqalignno{ A(d) & = \lim_{S \longrightarrow \infty}  \superone{{\cal
Z}}(S) \left(R^2 \right)^{d/2} &  \cr  &  = 2 \lim_{n \longrightarrow 0}
\left({1 \over n} {A_+ \over \alpha} \right) {1 \over \Gamma( \alpha)}
\left(2d {\Gamma^{ \prime}_ + \over \Gamma_ +} {\Gamma( \gamma) \over \Gamma(
\gamma +2\nu)} \right)^{d/2}\ . & (6.19) \cr} $$
Recalling the precise definition (6.13) of the field theoretical amplitude
ratio $ \left(R^+_\xi \right)^d, $ we finally arrive at
$$ A(d) = \left[{2 \over \Gamma( \alpha +1)} \left({2d\ \Gamma( \gamma) \over
\Gamma( \gamma +2\nu)} \right)^{d/2}{1 \over n} \left(R^+_\xi \right)^d
\right]_{n=0}\ . \eqno (6.20) $$
This agrees with the formula of Ref.$\lbrack$11$\rbrack$, but with the proper
factor $ {1 \over n} $ reinstated. The factor 2 in front
simply corresponds to the orientation of the ring.

\vskip 17pt

{\bf 6.3 Universal values}

The amplitude ratio $ \left(R^+_\xi \right)^d $ has been calculated for the $
O(n) $ model to order $ {\cal O} \left(\varepsilon^ 2 \right) $ in
Refs.$\lbrack$25$\rbrack$ and $\lbrack$26$\rbrack$,
correcting the results of Refs.$\lbrack$22$\rbrack$, $\lbrack$23$\rbrack$, as
$$ \eqalignno{ \left(R^+_\xi \right)^d & = {n \over 4} K_d \left\{
1+\varepsilon  {n-1 \over n+8} + \varepsilon^ 2 \left[{(n+2) \left(3n^2+46n-4
\right) \over 4(n+8)^3} \right. \right. &  \cr  & \left. \left.+ {\zeta( 2)
\over 4} + {7(n+2)^{ } \over 3(n+8)^2} \lambda \right] + {\cal O}
\left(\varepsilon^ 3 \right) \right\} &  (6.21a) \cr} $$
where $ K_d = {2\pi^{ d/2} \over \Gamma( d/2)} {1 \over( 2\pi)^ d}, $ $ \zeta(
2) = {\pi^ 2 \over 6}, $ and where $ \lambda $ is given as a numerical
constant
$$ \lambda  = - {1 \over 2} \int^ 1_0 {\rm d} x\ { {\rm \ell n}[x(1-x)] \over
1-x(1-x)}\ . \eqno (6.21b) $$
For $ n=0, $ we therefore have
$$ \left.{1 \over n} \left(R^+_\xi \right)^d \right\vert_{ n=0} = {1 \over 4}
K_d \left\{ 1-{\varepsilon \over 8} + \left({\varepsilon \over 8} \right)^2
\left(-{1 \over 4}+16{\pi^ 2 \over 6} + {14 \over 3} \lambda \right) + {\cal
O} \left(\varepsilon^ 3 \right) \right\} \ . \eqno (6.21c) $$

Using the formal relation (6.20) gives therefore
$$ \eqalignno{ A(d) & = \left({d \over 2\pi} \right)^{d/2}{1 \over \Gamma(
d/2)} {1 \over \Gamma( 3-\nu d)} \left[{\Gamma( \gamma) \over \Gamma( \gamma
+2\nu)} \right]^{d/2} &  \cr  &  \times  \left\{ 1-{\varepsilon \over 8}+
\left({\varepsilon \over 8} \right)^2 \left(-{1 \over 4}+8{\pi^ 2 \over 3} +
{14 \over 3} \lambda \right)+{\cal O} \left(\varepsilon^ 3 \right) \right\} \
. & (6.21d) \cr} $$

The full $ \varepsilon $-expansion is now performed for the {\it normalized}
amplitude ratio (2.12) $ {\cal A}(d) = \left({2\pi \over d} \right)^{d/2}A(d),
$
using the expansions $ \nu ={1 \over 2} \left(1+{\varepsilon \over 8} + {15
\over 4} \left({\varepsilon \over 8} \right)^2+... \right) $ and $ \gamma
=1+{\varepsilon \over 8}+{13 \over 4} \left({\varepsilon \over 8}
\right)^2+... $ We thus finally get from field theory
$$ {\cal A}(d=4-\varepsilon)  = 1 - {\varepsilon \over 8} + \left({\varepsilon
\over 8} \right)^2 \left({23 \over 4} + {\pi^ 2 \over 2} + {14 \over 3}
\lambda \right) + {\cal O} \left(\varepsilon^ 3 \right)\ , \eqno (6.22) $$
which is to be compared to the polymer result (5.5).

These results are identical {\it provided} that the numerical constant $
\lambda $ (6.21b) is identical to
$$ \lambda  \equiv  2 \sqrt{ 3} \left[{\pi \over 6} {\rm \ell n} \ 2 - L(\pi
/6) \right]\ , \eqno (6.23) $$
where $ L $ stands for the Lobachevsky function (4.37). This equality can be
verified numerically to arbitrary
precision.

Of course the two methods for finding the universal amplitude ratio, (5.5)
from the Edwards model, and
(6.22) from field theory, provide an indirect proof of our equality (6.23).
A direct calculation of $ \lambda $ to get (6.23) is feasible but long. The
main steps are indicated in
Appendix H.

{}From Eqs.(5.6), (5.8)-(5.11), one gets alternative forms for $ \lambda $ in
terms of special functions
$$ \eqalignno{ \lambda &  = {4 \over \sqrt{ 3}} \left[{\pi \over 3} {\rm \ell
n} \ 2 - L(\pi /3) \right] &  \cr  &  = \sum^{ }_{ n\geq 0}(-1)^n \left[{1
\over( 3n+1)^2} + {1 \over( 3n+2)^2} \right] &  \cr  &  = - {1 \over 9}
\left[\beta^{ \prime}( 1/3)+\beta^{ \prime}( 2/3) \right] & (6.24) \cr  &  =
{1 \over 36} \left[\psi^{ \prime}( 1/6)+\psi^{ \prime}( 1/3)-\psi^{ \prime}(
2/3)-\psi^{ \prime}( 5/6) \right] &  \cr  &  = {1 \over 9} \left\{{ 1 \over 2}
\left[\psi^{ \prime}( 1/6)+\psi^{ \prime}( 1/3) \right] - {4 \over 3} \pi^ 2
\right\} \ . &  \cr} $$
Notice that similar combinations appear elsewhere $\lbrack$39$\rbrack$.

\vskip 17pt

{\bf 6.4. Universal ratio with the radius of gyration}

The universal ratio of the squared radius of gyration to the squared
end-to-end distance was calculated in
Ref.$\lbrack$19$\rbrack$ to second order in $ \varepsilon $
$$ \aleph = \lim_{S \longrightarrow \infty}  6R^2_G/R^2 = 1 - {1 \over 12}
{\varepsilon \over 8} - \left({\varepsilon \over 8} \right)^2 {1 \over 6}
\left({1 \over 24} + 5\xi \right) + {\cal O} \left(\varepsilon^ 3 \right)
\eqno (6.25) $$
where
$$ \xi  = {2\pi \over \sqrt{ 3}} {\rm \ell n} \left({4 \over 3} \right) - 6
\int^ 1_0 {\rm d} u { {\rm \ell n} \left(1-u^2 \right) \over 1+3u^2} =
2.343907235... \eqno  $$
Numerically, one finds that
$$ \xi  \equiv  2\lambda \ , \eqno (6.26) $$
where $ \lambda $ is given in (1.11) and (6.23). So again, the same {\it
ubiquitous} constant appears to second
order, which is now identified by Eqs.(6.24). To prove (6.26) is possible
along the same lines as followed in
Appendix H for (6.23), but is probably quite long.

{}From the universal ratio (6.25), (6.26), we arrive at the second universal
amplitude ratio $ A_G(d) $ defined in
(1.7c)
$$ \eqalignno{ A_G(d) & = \lim_{S \longrightarrow \infty}  \superone{{\cal
Z}}(S) \left(R^2_G \right)^{d/2} = \left({1 \over 6} \aleph \right)^{d/2}A(d)
&  \cr  &  = \left({d \over 12\pi} \right)^{d/2} \aleph^{d/2} {\cal A}(d)\ . &
(6.27) \cr} $$
Normalized to that of the Brownian motion, this amplitude ratio reads:
$$ {\cal A}_G(d) \equiv  \aleph^{d/2}{\cal A}(d) = 1 - {7 \over 6}
{\varepsilon \over 8} + \left({\varepsilon \over 8} \right)^2 \left({29\times
31 \over( 12)^2} + {\pi^ 2 \over 2} + {4 \over 3} \lambda \right)+{\cal O}
\left(\varepsilon^ 3 \right)\ . \eqno (6.28) $$
\vskip 17pt

{\bf 6.5. Comparison to numerical results}

\vskip 10pt

The $ \varepsilon $-expansion of $ {\cal A} $ (5.8) reads in numerical figures
$$ {\cal A}(d) = 1 - {\varepsilon \over 8} + 16.1539... \left({\varepsilon
\over 8} \right)^2 + {\cal O} \left(\varepsilon^ 3 \right)\ . \eqno (6.29) $$
Of course, this expansion is only {\it asymptotic.} We note that the
corrections $ {\cal O}(\varepsilon) $ and $ {\cal O} \left(\varepsilon^ 2
\right) $ have
alternate signs. This is the signal of a {\it non monotonous} function $ {\cal
A}(d) $ of $ d. $

The values obtained from numerical simulations confirm this picture of a
non-monotonous concave function $ {\cal A}(d). $
These are usually given for the absolute self-avoiding ring amplitude $ A $
(1.7a) but the two orientations of the
ring are not distinguished, when counting the configurations.
We use the notation of Refs.$\lbrack$11$\rbrack$ and $\lbrack$15$\rbrack$ for
the amplitude of {\it non oriented} rings:
$$ \eqalignno{ W_R & = \lim_{N \longrightarrow \infty}  {1 \over 2} p_N
\left\langle R^2 \right\rangle^{ d/2}_N/ \left(\mu^ Na^d\tau \right) =
{BC^{d/2} \over \tau} &  \cr  &  \equiv  {1 \over 2} A(d) = {1 \over 2}
\left({d \over 2\pi} \right)^{d/2}{\cal A}(d)\ . & (6.30) \cr} $$
{}From the $ {\cal O} \left(\varepsilon^ 2 \right) $ expansion (6.29) of $
{\cal
A}(d) $ we find
$$ \eqalignno{ W_R(\varepsilon =1) & = 0.186... & (6.31) \cr {\rm and} \ \ \ \
\ \ \ \ \ \ \ \ \ \ \ \ \ \ \ \ \ \  &   &  \cr W_R(\varepsilon =2) & =
0.280...\ , & (5.32) \cr} $$

For $ d=4, $ $ W_R $ is formally defined from (6.30) and $ {\cal A}(4)\equiv
1, $ hence
$$ W_R(d=4) = {1 \over 2} \left({2 \over \pi} \right)^2=0.2026... \eqno (6.33)
$$
So we again observe a non-monotonous concave function.

These values have been measured from numerical simulations on various
three-dimensional lattices $\lbrack$11$\rbrack$:
$$ \eqalignno{ W_R( {\rm SC}) = { {\rm BC}^{3/2} \over 2} & = 0.08\pm 0.015\ \
\ \ {\rm (Simple\ Cubic)\ ,} &  \cr W_R( {\rm FCC}) = {\rm BC}^{3/2} & =
0.06\pm 0.02\ \ \ \ \ {\rm (Face\ Centered\ Cubic)\ ,} & (6.34) \cr W_R( {\rm
BCC}) = { {\rm BC}^{3/2} \over 2} & = 0.08\pm 0.02\ \ \ \ \ {\rm (Body\
Centered\ Cubic)\ ;} &  \cr} $$
while in $ d=2 $ $\lbrack$15,11$\rbrack$
$$ \eqalign{ W_R( {\rm Square}) & = { {\rm BC} \over 2} = 0.2168\ , \cr W_R(
{\rm Triangular}) & = {\rm BC}  = 0.2167\ . \cr} \eqno (6.35) $$
These values, together with the boundary value (6.33), indeed indicate a
non-monotonous concave function, but they are
quite far from the results (6.31), (6.32), especially for $ d=3. $

The discrepancy with the numerical values (6.34), (6.35) is puzzling. One
could think that it comes from using
an $ {\cal O} \left(\varepsilon^ 2 \right) $ expansion. However, usually
results to $ {\cal O} \left(\varepsilon^ 2 \right) $ for critical exponents
yield reasonable predictions
$\lbrack$40,2$\rbrack$, as well as for critical amplitude ratios, where they
can be surprisingly good. In particular, the
amplitude ratio $ R^2_G/R^2 $ (6.25) calculated in $\lbrack$19$\rbrack$ fits
reasonably well with the numerical result $\lbrack$41$\rbrack$.
Furthermore, one can compare for a number of components $ n=1,2,3 $ the $
{\cal O} \left(\varepsilon^ 2 \right) $ values of $ \left(R^+_\xi \right)^d $
(6.21a) with those
obtained from the renormalization group at $ d=3 $ (Refs.$\lbrack$42$\rbrack$,
$\lbrack$43$\rbrack$).

{}From (6.21a) we find
$$ R^+_\xi( n=1) \simeq  0.2702\ ,\ \ \ R^+_\xi( n=2)\simeq 0.3519\ ,\ \ \
R^+_\xi( n=3) \simeq  0.4132...\ , $$
while the $ d=3 $ RG values given in Table IV of Ref.$\lbrack$42$\rbrack$ are
quite close
$$ R^+_\xi( n=1) \simeq  0.2700\ ,\ \ \ R^+_\xi( n=2)\simeq 0.3606\ ,\ \ \
R^+_\xi( n=3) \simeq  0.4347. $$
For the $ n=0 $ vector model, we can nevertheless get a check of our
prediction (6.31) directly in $ d=3. $
Indeed, a simple phenomenological expression representing in three dimensions
$ R^+_\xi $ quite reasonably for the $ O(n) $
model has been proposed in Ref.$\lbrack$43$\rbrack$
$$ R^+_\xi  \simeq  \nu \left({n \over 4\pi} \right)^{1/3}\ . \eqno (6.36) $$
Using this directly in (6.20) for the best known 3D-values for $ n=0, $ $ \nu
=0.5880\pm 0.0015, $ and $ \gamma =1.1615\pm 0.002 $ $\lbrack$2$\rbrack$,
we get the phenomenological value
$$ W_R(d=3) = 0.17959\ , \eqno (6.37) $$
in good agreement with (6.31).

Hence the discrepancy with the lattice simulation values (6.34) remains. We
note that the concavity of the
polymer ring amplitude ratio $ W_R $ appears to be linked with the {\it
non-unitarity} of the theory. Indeed, the
$ \varepsilon $-expansion of the amplitude ratio $ \left[R^+_\xi( n) \right]^d
$ (6.21a) becomes {\it concave} when $ n< 1, $ while it is monotonously
increasing for $ n\geq 1. $

We also notice that the gyration amplitude ratio $ \aleph $ (6.25) is
monotonously decreasing with respect to $ \varepsilon $
and does not present the concavity accident, while, as said above, agreeing
with numerical results $\lbrack$19,41$\rbrack$.

It therefore seems that the expected concavity of the hyperscaling function $
{\cal A}(d) $ for polymers does reflect
itself in a concave asymptotic $ \varepsilon $-expansion, but also perhaps
implies highly unstable results
already at $ {\cal O} \left(\varepsilon^ 2 \right). $ Hence more elaborate
methods like Borel resummation and study of large order behavior appear to be
necessary for fully exploiting analytical results such as (5.8). Another
possibility is for the numerical
simulations to have underestimated this universal ratio.

A refined {\it numerical} study of the amplitude ratio on a lattice in three
dimensions is therefore called for,
considering the discrepancy between the FCC value and the other two in (6.34).

\vskip 17pt

\noindent {\bf Acknowledgements}

\vskip 10pt

It is a pleasure to thank D.A. Kosower and J.-M. Luck for quite useful
discussions and suggestions, and for a
generous help in dealing with calculations with the {\it Mathematica}
and {\it Macsyma} softwares.
I also thank C. Bagnuls and C. Bervillier for extensive informations about the
literature of universal
amplitude ratios, and V. Privman for correspondence. I am grateful to T.C.
Halsey for his careful reading of the manuscript.

\vfill\eject

\centerline{{\bf APPENDIX A}}

\vskip 17pt

On the hypercubic lattice $ \Bbb Z^d, $ the walk executes discrete steps $
\vec \ell , $ of
coordinates $ (0,...,\pm 1,0,...,0)\in \Bbb Z^d. $ Following
Ref.$\lbrack$44$\rbrack$, if one defines
the Fourier transform on the lattice by $ \lambda \left(\vec \theta \right) =
\sum^{ }_{\vec \ell} p \left(\vec \ell \right) {\rm e}^{i\vec \ell \cdot\vec
\theta} , $ $ \vec \theta \in  [0,2\pi]^ d, $ where the
sum extends over all elementary steps with probability $ p \left(\vec \ell
\right), $ then the
probability to have travelled a distance $ \vec \ell $ after $ N $ steps is
just
$$ P_N \left(\vec \ell \right) = {1 \over( 2\pi)^ d} \int^{ 2\pi}_ 0 {\rm d}
\theta_ 1... \int^{ 2\pi}_ 0 {\rm d} \theta_ d\ {\rm e}^{-i\vec \ell \cdot\vec
\theta} \left[\lambda \left(\vec \theta \right) \right]^N\ . \eqno (A.1) $$
For the elementary steps described above, the associated probability is $ {1
\over 2d} $
and therefore
$$ \lambda \left(\vec \theta \right) = {1 \over d} \left( {\rm cos} \ \theta_
1 +...+ {\rm cos} \ \theta_ d \right)\ . $$
The closure probability we are looking for is
$$ P_N \left(\vec 0 \right) = {1 \over( 2\pi)^ d} \int^{ 2\pi}_ 0 \prod^
d_{i=1} {\rm d} \theta_ i \left({1 \over d} \sum^ d_{i=1} {\rm cos} \ \theta_
i \right)^N\ , \eqno (A.2) $$
with an associated total number of ring configurations
$$ p_N=(2d)^NP_N \left(\vec 0 \right)\ . $$

By parity
$$ p_N=0\ \ \ \ \ \ \ {\rm if} \ \ \ \ \ \ N \in  2\Bbb N+1\ . \eqno (A.3) $$
For $ N $ even, the large $ N $ behavior is obtained by a saddle-point method,
but,
since for $ N\in 2\Bbb N $ there are {\it two} such saddle-points, namely
$$ \vec \theta =(0,...,0) {\rm mod}(2\pi) \ \ \ \ {\rm and} \ \ \ \ \vec
\theta  = (\pi ,...,\pi) \ . $$
Both contribute the same amount and for $ N $ large we find by expanding the
cosines in their neighbourhood
$$ p_N = 2(2d)^N \left({d \over 2\pi N} \right)^{d/2}(1+o(1))\ \ \ ,\ \ \ \
N\in 2\Bbb N\ . \eqno (A.4) $$

We therefore see the expected doubling in the hypercubic lattice $ (\tau =2),
$ since
the measure is concentrated on even numbers.

\vfill\eject

\centerline{{\bf APPENDIX B}}

\vskip 17pt

Symmetry factors in rooted cyclic diagrams can sometimes be a bit tricky. So
we give a rather trivial example, which illustrates the formulae of section 3.

With the diagram 1) of Fig.1 is associated a contribution
$$ \superone{{\cal Z}}^{(1)} = (-b) \left\langle \delta^ d \left(\vec
r(S)-\vec r(0) \right) \int^ S_0 {\rm d} s \int^ S_0 {\rm d} s^{\prime}
\delta^ d \left(\vec r(s)-\vec r \left(s^{\prime} \right) \right)
\right\rangle_ 0 \eqno (B.1) $$
which strictly speaking, should be represented as in Fig.3 below.

We can write (B.1) trivially as
$$ \eqalignno{ \superone{{\cal Z}}^{(1)} & = (-b) \int^{ }_{ }{ {\rm d}^dq
\over( 2\pi)^ d} \int^{ }_{ }{ {\rm d}^dq^{\prime} \over( 2\pi)^ d} \int^ S_0
{\rm d} s \int^ S_s {\rm d} s^{\prime} \left\langle {\rm e}^{i\vec q\cdot
\left[\vec r(S)-\vec r(0) \right]} {\rm e}^{i\vec q^{\ \prime} \cdot
\left[\vec r(s)-\vec r \left(s^{\prime} \right) \right]} \right\rangle_ 0.\ \
\ \ \ \  & (B.2) \cr} $$
Once ordered, the exponential, once averaged, reads as in (3.2)
$$ {\rm e}^{-{1 \over 2} \left[q^2s+ \left(\vec q+\vec q^{\ \prime} \right)^2
\left(s^{\prime} -s \right)+q^2 \left(S-s^{\prime} \right) \right]}\ . $$
One defines the segment lengths along the chain $ s_1\equiv s+S-s^{\prime} , $
$ s_2=s^{\prime} -s=S-s_1, $ and the
momenta flowing along those segments $ \vec q_1=\vec q, $ $ \vec q_2=\vec
q+\vec q^{\ \prime} . $ The integrand depends only
on $ s_1 $ and the measure reads
$$ \int^ S_0 {\rm d} s \int^ S_s {\rm d} s^{\prime} ... = \int^ S_0 {\rm d}
s_1\ s_1... $$
We can therefore rewrite (B.2) in a more symmetric way
$$ \eqalignno{ \superone{{\cal Z}}^{(1)} & = (-b) \int^{ }_{ }{ {\rm d}^dq_1
\over( 2\pi)^ d} { {\rm d}^dq_2 \over( 2\pi)^ d} \int^{ \infty}_ 0 {\rm d}
s_1\ s_1 \int^{ \infty}_ 0 {\rm d} s_2\ \delta \left(S-s_1-s_2 \right) {\rm
e}^{-{1 \over 2} \left(q^2_1s_1+q^2_2s_2 \right)}  &  \cr  &  = (-b)(2\pi)^{
-d} \int^{ \infty}_ 0 {\rm d} s_1\ s_1\ {\rm d} s_2\ \delta \left(S-s_1-s_2
\right) \left(s_1s_2 \right)^{-d/2}\ . &  \cr} $$
Notice that the factor $ s_1 $ in the integrand comes from the integration of
the
position of the root anywhere in the segment 1. We can restore the $ (1,2) $
symmetry
by adding a similar term in $ s_2 $ while dividing by 2. Taking advantage of
the
presence of a $ \delta $ distribution, we thus obtain the symmetric formula
(3.3b)
$$ \superone{{\cal Z}}^{(1)} = (-b)(2\pi)^{ -d}{S \over 2} \int^{ \infty}_ 0
{\rm d} s_1 {\rm d} s_2\ \delta \left(S-s_1-s_2 \right) \left(s_1s_2
\right)^{-d/2}\ . $$
Of course, the final factor $ {S \over 2} $ can also be understood from the
rooted diagram 1) of
Fig.1, the factor $ S $ coming from the gliding of the root along the cycle,
and the
factor $ 1/2 $ being the symmetry weight of the diagram.

Similar considerations apply to diagrams (2,a) and (2,b) of Fig.1. In $
\superone{{\cal Z}}^{(2)}_{(a)} $ (3.3c), the
factor $ S $ corresponds to the {\it sum} of the two diagrams (2,a), both
having a symmetry
factor 1/2. In $ \superone{{\cal Z}}^{(2)}_{(b)} $ the symmetry factor 1/4 is
that of the cyclic permutations.

In all of this, we have used the Brownian correlation functions (3.2) of an
open
path, since we have chosen to represent the closure of the ring by enforcing
the
distribution $ \delta^ d \left(\vec r(S)-\vec r(0) \right) $ {\it in the
measure of the partition function} (2.3). The symmetry
factors are thus recovered only at the end. Alternately we could have chosen
to
use uniquely the Brownian correlators along a {\it ring}, by defining from the
start the abscissa as belonging to the 1-sphere $ {\cal S}_1, $ and using the
different path integral
on this {\it closed manifold} $\lbrack$30$\rbrack$. The correlator on $ {\cal
S}_1 $ is then
$$ \left\langle {\rm e}^{i\vec q\cdot \left[\vec r(s)-\vec r \left(s^{\prime}
\right) \right]} \right\rangle_{{\cal S}_1} = {\rm exp} \left\{ -{1 \over
2}q^2 \left\vert s-s^{\prime} \right\vert \left(S- \left\vert s-s^{\prime}
\right\vert \right) \right\} \ . $$
One can check that the final formulae for $ \superone{{\cal
Z}}^{(1,2)}_{(a,b)} $ in (3.3) coincide
when working on the 1-sphere.

\vfill\eject

\centerline{{\bf APPENDIX C}}

\vskip 17pt

\centerline{{\bf Laurent expansion of integral I (3.13)}}

\vskip 10pt

We have to evaluate
$$ I = \int^{ \infty}_ 0 {\rm d} p\ p^{3(1-d/2)}F^4(p)\ , \eqno (C.1) $$
where
$$ F(p) \equiv  \int^{ \infty}_ 0 {\rm e}^{-1/x} {\rm e}^{-px}x^{-d/2} {\rm d}
x\ . \eqno (C.2) $$

The expansion of $ F(p) $ near $ p=0 $ reads explicitly from (C.2):
$$ \eqalignno{ F(p) & = F(0) + \hat F(p) + \tilde F(p)\ , & (C.3) \cr  &  \ \
\ \ \ {\rm or} &  \cr  &  = F(0) + \hat F(p) + \hhat F(p) + \ttilde F(p)\ , &
(C.4) \cr} $$
where
$$ \eqalignno{ F(0) & = \Gamma \left({d \over 2}-1 \right)\ , & (C.5) \cr\hat
F(p) & = p^{d/2-1}\Gamma \left(1-{d \over 2} \right)-p\Gamma \left({d \over
2}-2 \right)\ , & (C.6) \cr\hhat F(p) & = -p^{d/2}\Gamma( -d/2)+{p^2 \over
2}\Gamma \left({d \over 2}-3 \right)\ , & (C.7) \cr\tilde F(p) & \equiv
\int^{ \infty}_ 0 \left( {\rm e}^{-1/x}-1 \right) \left( {\rm e}^{-px}-1+px
\right)x^{-d/2} {\rm d} x\ , & (C.8) \cr\ttilde F(p) & \equiv  \int^{ \infty}_
0 \left( {\rm e}^{-1/x}-1+{1 \over x} \right) \left( {\rm
e}^{-px}-1+px-{p^2x^2 \over 2} \right)x^{-d/2} {\rm d} x\ . & (C.9) \cr} $$
When compared to (C.3) (C.6), expansion (C.4) (C.7) has been pushed one step
further. Notice also that when expanded for $ d \longrightarrow 4, $ the
apparent $ 1/\varepsilon $ poles in $ \hat F, $ $ \hhat F $
disappear and those functions are {\it regular} at $ d=4 $
$$ \eqalign{ \doublelow{\hat F(p) \cr d \longrightarrow 4 \cr} &  = p\ {\rm
\ell n} \ p + (2\gamma -1)p\ , \cr \doublelow{\hhat F(p) \cr d \longrightarrow
4 \cr} &  = {1 \over 2} p^2 {\rm \ell n} \ p + p^2 \left(\gamma -{5 \over 4}
\right)\ . \cr} \eqno (C.10) $$
In the same way the functions $ F, $ $ \tilde F $ and $ \ttilde F $ are
regular at $ d=4. $ We are interested in
the vicinity of $ d=4, $ actually $ d \longrightarrow 4^-. $

Thus (C.1) has to be defined by {\it analytic continuation} as it diverges at
the
{\it origin} $ p=0. $ This means that one has to subtract from $ F^4(p) $ the
first
terms of its expansion for $ p \longrightarrow 0, $ in such a way that the
integral converges near $ p=0. $
Owing to the expansion (C.3), the dimensionally regularized explicit
expression of
(C.1) is therefore for $ d<4 $
$$ I \equiv  \int^{ \infty}_ 0 {\rm d} p\ p^{3(1-d/2)}\delta F^4(p) \eqno
(C.11) $$
with
$$ \delta F^4(p) \equiv  F^4(p)-F^4(0)-4F^3(0)\hat F(p)\ . \eqno (C.12) $$
The integral now converges for $ {10 \over 3} < d < 4, $ the lower limit being
obtained from the
behavior at infinity $ \hat F(p) \sim  p $ for $ p \longrightarrow \infty . $

We can further write
$$ \doublelow{ I \cr d \longrightarrow 4^- \cr}  = \int^ 1_0 {\rm d} p\
p^{3(1-d/2)}\delta F^4(p)+{\cal O}(1)\ , \eqno (C.13) $$
since the integral over $ [1,\infty) $ is regular when $ d \longrightarrow 4 $
(see (C.10)) and contributes
only an $ {\cal O}(1) $ term.

By using (C.3) and (C.4), the subtracted integrand (C.12) can be formally
written
as
$$ \delta F^4 = 4F^3(0) \left[\hhat F+\ttilde F \right]+ \left(^4_2 \right)
F^2(0) \left(\hat F+\tilde F \right)^2 +4F(0)[F-F(0)]^3+[F-F(0)]^4\ . \eqno
(C.14) $$

For $ p \longrightarrow 0, $ we have in a symbolic way
$$ \eqalign{{\cal O}(F-F(0)) & = {\cal O} \left(\hat F \right) = {\cal O}
\left(p^{{d \over 2}-1},p \right) \cr{\cal O} \left(\tilde F
\right)\equiv{\cal O} \left(\hhat F \right) & = {\cal O} \left(p^{d/2},p^2
\right) = p{\cal O} \left(\hat F \right) \cr{\cal O} \left(\ttilde F \right) &
= {\cal O} \left(p^{{d \over 2}+1},p^3 \right) = p^2{\cal O} \left(\hat F
\right)\ . \cr} $$
Therefore, by systematic inspection, and recalling the regularity of $ \hat F,
$ $ \hhat F $ (C.10)
and $ F, $ $ \tilde F, $ $ \ttilde F, $ for $ d=4, $ one can check that the
only terms contributing poles to
integral (C.13) near the origin, are those remaining in the integrand of:

$$ \doublelow{ I \cr d \longrightarrow 4 \cr}  \doteqdot \int^ 1_0 {\rm d} p\
p^{3(1-d/2)} \left[4F^3(0)\hhat F+ \left(^4_2 \right)F^2(0)\hat F^2 \right] +
{\cal O}(1)\ , \eqno (C.15) $$
which is calculated to yield (3.17).

\vfill\eject
\centerline{{\bf APPENDIX D}}

\vskip 17pt

The scaling behaviour of the quantity $ {\cal X}(z,\varepsilon) $ can be
obtained by calculating in a
standard way $\lbrack$17$\rbrack$ the scaling function
$$ \sigma( z,\varepsilon)  \equiv  \left.S {\partial \over \partial S} {\rm
\ell n} \ {\cal X} \right\vert_{ b,\varepsilon}  = \left.{\varepsilon \over 2}
z {\partial \over \partial z} {\rm \ell n} \ {\cal X} \right\vert_ \varepsilon
\ \ . $$
The expansion (3.20) gives
$$ \sigma( z,\varepsilon)  = \left(-2+\varepsilon +{\cal O} \left(\varepsilon^
2 \right) \right) z + z^2 \left({16 \over \varepsilon}  - 11 + {\cal
O}(\varepsilon) \right) + {\cal O} \left(z^3 \right)\ . \eqno (D.1) $$
In Ref.$\lbrack$18,33$\rbrack$, we have shown that for the Edwards model there
exists a {\it minimal
subtraction scheme in the $ z $ variable} itself, namely a
substitution of the minimal form
$$ z = z_R + \sum^{ }_{ n\geq 2}z^n_R \sum^{ n-1}_{p=1}{\beta_{ n,p} \over
\varepsilon^ p}\ , \eqno (D.2) $$
where the $ \beta_{ n,p} $ are pure numbers, and $ z_R $ is the {\it
renormalized} $ z $-parameter.
Substituting $ z_R $ for $ z $ renormalizes to all orders any polymer scaling
function such
as $ \sigma( z,\varepsilon) , $ i.e. when expressed in terms of $ z_R, $ $
\sigma \left[z_R,\varepsilon \right] $ becomes a double Taylor
series in $ z_R $ {\it and} $ \varepsilon , $ and is therefore pole-free when
$ \varepsilon \longrightarrow 0 $ $\lbrack$17,18$\rbrack$.

For polymers, the minimal series (D.2) reads $\lbrack$18,34$\rbrack$
$$ z = z_R + {8 \over \varepsilon}  z^2_R + \left({64 \over \varepsilon^ 2} -
{17 \over \varepsilon} \right) z^3_R + {\cal O} \left(z^4_R \right)\ , $$
with a fixed point value for $ z \longrightarrow  \infty $
$\lbrack$18,34$\rbrack$
$$ z^\ast_ R = {\varepsilon \over 8} + {17 \over 4} \left({\varepsilon \over
8} \right)^2 + {\cal O} \left(\varepsilon^ 3 \right)\ . \eqno (D.3) $$
This gives the regular scaling function
$$ \sigma \left[z_R,\varepsilon \right] = (-2+\varepsilon)  z_R - 3z^2_R +
{\cal O} \left(z^3_R \right) $$
with a fixed point value
$$ \sigma =(-2) {\varepsilon \over 8} + \left(- {7 \over 2} \right)
\left({\varepsilon \over 8} \right)^2 + {\cal O} \left(\varepsilon^ 3 \right)\
, \eqno (D.4) $$
such that $ (2\pi S)^{d/2} \superone{{\cal Z}}(S) \sim  S^\sigma $ for $ S
\longrightarrow \infty . $

\vfill\eject

\centerline{{\bf APPENDIX E}}

\vskip 17pt

Our aim is first to simplify $ \Delta_ 2 $ (4.23). One has
$\lbrack$32,45$\rbrack$
$$ P^{\nu -1/2}_{\nu -1/2}(x) = {1 \over \Gamma \left({3 \over 2}-\nu \right)}
\left({1+x \over 1-x} \right)^{ \left(\nu -{1 \over 2} \right){1 \over 2}}F
\left({1 \over 2}-\nu ,{1 \over 2}+\nu \ ;{3 \over 2}-\nu ,{1-x \over 2}
\right) \eqno (E.1) $$
where $ F $ is the usual hypergeometric series $\lbrack$45$\rbrack$
$$ \eqalign{ F(\alpha ,\beta ;\gamma ;z) & = \sum^{ \infty}_{ k=0}{(\alpha)_
k(\beta)_ k \over( \gamma)_ k} {z^k \over k!}\ , \cr( \alpha)_ k & \equiv
{\Gamma( \alpha +k) \over \Gamma( \alpha)} \ . \cr} \eqno (E.2) $$
It is convenient to use the transformation formula
$$ F(\alpha ,\beta ;\gamma ;z)=(1-z)^{\gamma -\alpha -\beta} F(\gamma -\alpha
,\gamma -\beta ;\gamma ;z) \eqno (E.3) $$
in (E.1), so as to get
$$ P^{\nu -1/2}_{\nu -1/2}(x) = {1 \over \Gamma \left({3 \over 2}-\nu \right)}
\left(1-x^2 \right)^{ \left({1 \over 2}-\nu \right)/2}2^{\nu -{1 \over 2}}F
\left(1,1-2\nu ;{3 \over 2}-\nu ;{1-x \over 2} \right)\ . \eqno (E.4) $$
Using the expression for $ x = {\rm cos} \ \varphi , $ we get
$$ P^{\nu -1/2}_{\nu -1/2}( {\rm cos} \ \varphi)  = {1 \over \Gamma \left({3
\over 3}-\nu \right)} ( {\rm sin} \ \varphi)^{{ 1 \over 2}-\nu} 2^{\nu -{1
\over 2}}F \left(1,1-2\nu ;{3 \over 2}-\nu ;{1- {\rm cos} \ \varphi \over 2}
\right)\ . $$
The expression $ \Delta_ 2 $ (4.23) we seek is then simplified with the use of
the latter
equation and of (4.25) to be
$$ \eqalignno{ \Delta_ 2(a,b,c) & = {1 \over 2^{\nu -1/2}} {1 \over \Gamma
\left({3 \over 2}-\nu \right)} \left\{{( bc)^{\nu -1} \over a^\nu}  F
\left(1,1-2\nu ;{3 \over 2}-\nu ;z_a \right) \right. &  \cr  &  + {(ca)^{\nu
-1} \over b^\nu}  F \left(1,1-2\nu ;{3 \over 2}-\nu ;z_b \right) & (E.5) \cr
&  + \left.{(ab)^{\nu -1} \over c^\nu}  F \left(1,1-2\nu ;{3 \over 2}-\nu ;z_c
\right) \right\}  , &  \cr} $$
where
$$ z_a = {1 \over 2} (1- {\rm cos} \ \varphi_ a)\ , \eqno (E.6) $$
with corresponding definitions for $ b,c. $ The equations (4.24) are the
parametric
representation of the elements of a Euclidean triangle. For calculating the
integrals (4.18) (4.19), we shall need only the isoceles case $ (a;b=c=1), $
before
reaching at the end the symmetric point $ (a=b=c=1). $ In the isoceles case we
have
from (4.24)
$$ z_a = {1 \over 2} \left(1- {\rm cos} \ \varphi_ a \right) = {1 \over 4}
a^2\ \ ,\ \ \ \ z_b = z_c = {1 \over 2} -{a \over 4}\ , \eqno (E.7) $$
and the quantities $ \Delta_{( 1,2)} $ of (4.21) (4.23) read
$$ \eqalignno{ \Delta_ 1(a,1,1) & = {a^{\nu -1} \over 2^{2\nu -1}} \left(1 -
{a^2 \over 4} \right)^{\nu -1/2}\ , & (E.8) \cr \Delta_ 2(a,1,1) & = {1 \over
2^{\nu -1/2}} {1 \over \Gamma( 3/2-\nu)}  {\cal F}(a)\ , & (E.9) \cr{\cal
F}(a) & \equiv  a^{-\nu} F \left(1,1-2\nu ;{3 \over 2}-\nu ;z_a \right)+2\
a^{\nu -1}F \left(1,1-2\nu ;{3 \over 2}-\nu ;z_b \right)\ .\ \ \ \ \ \ \ \ \
& (E.10) \cr} $$
The integrals (4.18) (4.19) therefore correspond to evaluating (4.21) and
(4.26)
$$ \eqalignno{ \int^{ \infty}_ 0t^{1-\nu} K^3_\nu( t) {\rm d} t & = \alpha_
1\Delta_ 1(1,1,1)+\alpha_ 2\Delta_ 2(1,1,1) & (E.11) \cr \int^{ \infty}_
0t^{3-\nu} K^3_\nu( t) {\rm d} t & = \alpha_ 1\ \left.D\Delta_ 1(a,1,1)
\right\vert_{ a=1} + \alpha_ 2\ \left.D\Delta_ 2(a,1,1) \right\vert_{ a=1}\ .
& (E.12) \cr} $$

We have
$$ \eqalignno{ \Delta_ 1(1,1,1) & = \left({ \sqrt{ 3} \over 4} \right)^{2\nu
-1}\ , & (E.13) \cr \Delta_ 2(1,1,1) & = {1 \over 2^{\nu -1/2}} {1 \over
\Gamma( 3/2-\nu)}  3F \left(1,1-2\nu ;{3 \over 2}-\nu ;z={1 \over 4} \right)\
. & (E.14) \cr} $$
The action of the second order operator $ D $ (4.27)
$$ D = {\partial^ 2 \over \partial a^2} + {1 \over a} {\partial \over \partial
a} - {\nu^ 2 \over a^2} \eqno (E.15) $$
in (E.12) requires evaluating $ D\Delta_{( 1,2)}, $ where expressions (E.8)
(E.9) are to be
used. The first one is
$$ D\Delta_ 1(1,1,1) = {(-1) \over 2^{2\nu +1}} \left({3 \over 4} \right)^{\nu
-5/2}(2\nu -1)(\nu +3)\ , \eqno (E.16) $$
whereas the second one
$$ D\Delta_ 2(1,1,1) = {1 \over 2^{\nu -1/2}\Gamma \left({3 \over 2}-\nu
\right)} D{\cal F}(a) \left\vert_{ a=1} \right. \eqno (E.17) $$
requires evaluating $ D{\cal F}(a), $ with $ {\cal F}(a) $ given by (E.10) and
(E.7). The calculation
involve successive derivatives and we give here only the result at the
symmetric
point $ a=1 $
$$ D{\cal F}(a) \left\vert_{ a=1} \right. = (2-4\nu) F+ \left({3 \over 2} -
2\nu \right) F^{\prime} +{3 \over 8}F^{\prime\prime} \eqno (E.18) $$
where
$$ \left(F,F^{\prime} ,F^{\prime\prime} \right) = \left(1,{\partial \over
\partial z},{\partial^ 2 \over \partial z^2} \right) \left.F \left(1,1-2\nu
;{3 \over 2}-\nu ;z \right) \right\vert_{ z=1/4} \eqno (E.19) $$
in an obvious short-hand notation. One finally uses the constitutive equation
of
the hypergeometric series $ F(\alpha ,\beta ;\gamma ;z) $ $\lbrack$45$\rbrack$
$$ z(1-z){\partial^ 2 \over \partial z^2} F + [\gamma -(\alpha +\beta
+1)]{\partial \over \partial z} F - \alpha \beta F = 0 $$
to simplify (E.18) (E.19) into
$$ D{\cal F}(a) \left\vert_{ a=1} \right. = 4(1-2\nu) F-\nu F^{\prime} \ ,
\eqno (E.20) $$
which is the result given in the text above Eq.(4.29). The latter equation
(4.30)
comes from the standard recursion formula
$$ {\partial F \over \partial z} (\alpha ,\beta ;\gamma ;z) = {\alpha \beta
\over \gamma}  F(\alpha +1,\beta +1;\gamma +1;z)\ . $$
Formulae (E.11,12) together with (E.13,14), (E.16,17) and (E.20), yield the
announced analytical expressions (4.31) (4.32).

\vfill\eject

\centerline{{\bf APPENDIX F}}

\vskip 17pt

\centerline{{\bf Laurent expansion of Bessel integrals}}

\vskip 10pt

We have to expand the hypergeometric functions (4.29) (4.30) involved in the
analytic expressions of integrals (4.31) (4.32)
$$ \eqalignno{ F_1 & \equiv  F \left(1,1-2\nu ;{3 \over 2}-\nu ;z=1/4 \right)\
, & (F.1) \cr F^{\prime} &  = {1-2\nu \over 3/2-\nu}  F \left(2,2-2\nu ;{5
\over 2}-\nu ;z=1/4 \right)\ , & (F.2) \cr} $$
in powers of
$$ 1-\nu  \equiv  2-d/2\ . \eqno (F.3) $$
For this, one has first to reduce the number of places where $ \nu $ appears
in the
argument of the hypergeometric series. We take advantage of the particular
identity for hypergeometric functions $\lbrack$32,45$\rbrack$
$$ F \left(2\alpha ,2\beta ;\alpha +\beta +{1 \over 2};{1- \sqrt{ z^{\prime}}
\over 2} \right) = AF \left(\alpha ,\beta ;{1 \over 2};z^{\prime} \right) + B\
\sqrt{ z^{\prime}}  F \left(\alpha +{1 \over 2},\beta +{1 \over 2};{3 \over
2};z^{\prime} \right)\ , \eqno (F.4) $$
where
$$ A = {\Gamma \left(\alpha +\beta +{1 \over 2} \right)\Gamma \left({1 \over
2} \right) \over \Gamma \left(\alpha +{1 \over 2} \right)\Gamma \left(\beta
+{1 \over 2} \right)}\ \ \ ,\ \ \ \ B = {\Gamma \left(\alpha +\beta +{1 \over
2} \right)\Gamma \left(-{1 \over 2} \right) \over \Gamma( \alpha) \Gamma(
\beta)} \ . \eqno (F.5) $$
Luckily enough, the hypergeometric functions (F.1), (F.2) are precisely of the
form (F.4), with $ \alpha ={1 \over 2}, $ $ \beta ={1 \over 2}-\nu $ in the
former case, and $ \alpha^{ \prime} =1, $ $ \beta^{ \prime} =1-\nu , $ in the
second
case. We treat in detail the case of (F.1).

We have
$$ F_1=AF \left({1 \over 2},{1 \over 2}-\nu ;{1 \over 2};z^{\prime} =1/4
\right) + B {1 \over 2} F \left(1,1-\nu ;{3 \over 2};z^{\prime} =1/4 \right)
\eqno (F.6) $$
with
$$ A = {\Gamma \left({3 \over 2}-\nu \right)\Gamma \left({1 \over 2} \right)
\over \Gamma( 1-\nu)} \ \ \ ,\ \ \ B = 2\nu -1\ . \eqno (F.7) $$

The first hypergeometric series above is exactly calculable since
$\lbrack$32,45$\rbrack$:
$$ F(b,a;b;z) \equiv  (1-z)^{-a}\ ; \eqno (F.8) $$
while the second one has to be transformed with $\lbrack$32,45$\rbrack$
$$ F \left(\alpha ,\beta ;\gamma ;z^{\prime} \right) = \left(1-z^{\prime}
\right)^{-\beta}  F \left(\gamma -\alpha ,\beta ;\gamma ;{z^{\prime} \over
z^{\prime} -1} \right)\ . \eqno (F.9) $$
We so arrive at
$$ F_1 = A(3/4)^{- \left({1 \over 2}-\nu \right)}+ B {1 \over
2}(3/4)^{-(1-\nu)} F \left({1 \over 2},1-\nu ;{3 \over 2};z=-1/3 \right)\ .
\eqno (F.10) $$
The interest of the last form is that two arguments, $ \alpha $ and $ \gamma ,
$ are exactly
separated by one unit. We have now explicitly a much simplified form (see
(E.2))
$$ F_{1/2}\equiv  F \left({1 \over 2},1-\nu ;{3 \over 2};z=-1/3 \right) = 1 +
{1 \over 2\Gamma( 1-\nu)}  S(z=-1/3) \eqno (F.11) $$
with
$$ S(z) \equiv  \sum^{ }_{ n\geq 1}{\Gamma( n+1-\nu) \over n+1/2} {z^n \over
n!}\ . \eqno (F.12) $$
The presence in the denominator of $ \Gamma( 1-\nu) $ in (F.11) already
produces in the
numerator a factor $ (1-\nu) $ when $ \nu \longrightarrow 1. $ We have to
expand $ S(z) $ only one step further,
using
$$ \Gamma( n+1-\nu)  = (n-1)! \left\{ 1+(1-\nu) \psi( n)+{\cal O}
\left[(1-\nu)^ 2 \right] \right\} \ . $$
so that
$$ S(z) = S^{\prime}( z) + (1-\nu) S^{\prime\prime}( z) + {\cal O}
\left[(1-\nu)^ 2 \right] \eqno (F.13) $$
with the new series
$$ \eqalignno{ S^{\prime}( z) & \equiv  \sum^{ }_{ n\geq 1}{1 \over n+1/2} {1
\over n} z^n\ , & (F.14) \cr S^{\prime\prime}( z) & \equiv  \sum^{ }_{ n\geq
1}\psi( n) {1 \over n+1/2} {1 \over n} z^n\ . & (F.15) \cr} $$
The second series can be written as a double series, owing to $ \psi( n) =
-\gamma + \sum^{ n-1}_{k=1}{1 \over k}. $
After decomposing the fractions into simple elements and after some
reassembling
of the terms, we arrive at
$$ \eqalignno{ S^{\prime\prime}( z) & = -\gamma \ S^{\prime}( z) + 2S_1(z) -
2S_2(z)\ ; & (F.16) \cr S_1(z) & = \sum^{ }_{ n\geq 2}z^n {1 \over n} \sum^{
n-1}_{k=1}{1 \over k}\ ; & (F.17) \cr S_2(z) & = -S^{\prime}( z) + {2 \over x}
S_3(x)\ \ \ ,\ \ \ x = \sqrt{ -z}\ ; & (F.18) \cr S_3(x) & \equiv  \sum^{ }_{
n\geq 1}{(-1)^nx^{2n+1} \over 2n+1} \sum^ n_{k=1}{1 \over k}\ . & (F.19) \cr}
$$
Thus
$$ S^{\prime\prime}( z) = (2-\gamma) S^{\prime}( z) + 2S_1(z) - {4 \over x}
S_3(x) \eqno (F.20) $$
which is taken for $ z=-1/3 $ and $ x= \sqrt{ -z} = 1/ \sqrt{ 3}\ . $

The simple series $ S^{\prime}( z) $ (F.14), or double series $ S_1(z) $
(F.17) and $ S_3(x) $ (F.19) can
be found in the \lq\lq Table of Series and Products\rq\rq\ by E.R. Hansen
$\lbrack$36$\rbrack$. We have
respectively
$$ \doublelow{ S^{\prime}( z) \cr z<0 \cr}  = 2 \left[2- {2 \over \sqrt{ -z}}
{\rm Arctg} \ \sqrt{ -z} - {\rm \ell n}(1-z) \right]\ , $$
$$ \eqalignno{ S_1(z) & = {1 \over 2} [ {\rm \ell n}(1-z)]^2\ \ \ ,\ \ \ \ -1
< z < 1\ , &  \cr S_3(x) & = - {\rm \ell n} \left(1+x^2 \right) {\rm Arctg} \
x + x( {\rm Arctg} \ x)^2 &  \cr  &  - \int^ x_0( {\rm Arctg} \ t)^2 {\rm d}
t\ \ \ \ \ ,\ \ \ \vert x\vert  < 1\ . & (F.21) \cr} $$
The formula (F.21) for the double series (F.19) corresponds to the formula
(5.4.25) given in Ref.$\lbrack$36$\rbrack$, p.24, but with a {\it global sign
error} corrected.

Actually, the integral which remains in that formula can be transformed by a
double integration by parts. The integrated terms simplify considerably and we
get
$$ S_3(x) = -2\ L( {\rm Arctg} \ x)\ , \eqno (F.22a) $$
where $ L(\theta) $ is the {\it Lobachevsky} function
$$ L(\theta)  = - \int^ \theta_ 0 {\rm \ell n}( {\rm cos} \ t) {\rm d} t\ .
\eqno (F.22b) $$
For our numerical case $ z=-1/3, $ $ x=1/ \sqrt{ 3}, $ $ \theta = {\rm Arctg}
\ x = {\pi \over 6}, $ and we finally have
$$ \eqalignno{ S^{\prime} &  \equiv  S^{\prime}( z=-1/3) = 2 \left[2-{\pi
\over \sqrt{ 3}} - {\rm \ell n} \left({4 \over 3} \right) \right] &  \cr S_1 &
\equiv  S_1(z=-1/3) = {1 \over 2} \left( {\rm \ell n}{4 \over 3} \right)^2 &
(F.23) \cr S_3 & \equiv  S_3 \left(x=1/ \sqrt{ 3} \right) = -2L(\pi /6)\ , &
\cr} $$
so that we can write the expansion (F.11) (F.13) (F.20)
$$ \eqalign{ F_{1/2} & = 1 + {1 \over 2\Gamma( 1-\nu)}  \left\{ S^{\prime}
+(1-\nu) S^{\prime\prime} +{\cal O} \left[(1-\nu)^ 2 \right] \right\} \cr
S^{\prime\prime} &  = (2-\gamma)  S^{\prime} +2S_1-4 \sqrt{ 3}\ S_3\ , \cr}
\eqno (F.24) $$
as given in section 4.

The calculation of (F.2) $ F_2\equiv F \left(2,2-2\nu ;{5 \over 2}-\nu ;z=1/4
\right) $ proceeds along the same lines.
Using the transformation formula (F.4) for $ \alpha =1, $ $ \beta =1-\nu , $
we get
$$ F_2 = A^{\prime} F \left(1,1-\nu ;{1 \over 2};z^{\prime} =1/4 \right) +
B^{\prime}{ 1 \over 2}F \left({3 \over 2},{3 \over 2}-\nu ;{3 \over
2};z^{\prime} =1/4 \right) \eqno (F.25) $$
with
$$ A^{\prime}  = 2 \left({3 \over 2}-\nu \right)\ \ \ \ \ ,\ \ \ \ \ \
B^{\prime}  = {\Gamma \left({5 \over 2}-\nu \right)\Gamma( -1/2) \over \Gamma(
1-\nu)} \ . \eqno (F.26) $$

The first hypergeometric series is transformed with the help of Eq.(F.9),
whereas
the second is of the exact form (F.8), giving altogether
$$ F_2=A^{\prime} \left({3 \over 4} \right)^{-(1-\nu)} F \left(-{1 \over
2},1-\nu ;{1 \over 2};z=-1/3 \right) + B^{\prime}{ 1 \over 2} \left({3 \over
4} \right)^{- \left({3 \over 2}-\nu \right)}\ . \eqno (F.27) $$
It remains to calculate the expansion of $ F_{-1/2}\equiv F \left(-{1 \over
2},1-\nu ;{1 \over 2};z=-1/3 \right) $ with respect to $ 1-\nu . $

Following the same lines as in the beginning of this appendix (Eq.(F.11)), we
have
the simple expression
$$ F_{-1/2} = 1 - {1 \over 2\Gamma( 1-\nu)}  \tilde S(z)\ \ \ \ \ \ \ \  (z =
- 1/3)\ , \eqno (F.28) $$
with
$$ \eqalignno{\tilde S(z) & = \sum^{ }_{ n\geq 1}{1 \over n-1/2} {\Gamma(
n+1-\nu) \over n!} z^n &  \cr  &  = \tilde S^{\prime}( z) + (1-\nu)  \tilde
S^{\prime\prime}( z) + {\cal O} \left[(1-\nu)^ 2 \right]\ , & (F.29) \cr\tilde
S^{\prime}( z) & \equiv  \sum^{ }_{ n\geq 1}{z^n \over( n-1/2)n}\ \ \ \ ,\ \ \
\ \tilde S^{\prime\prime}( z) \equiv  \sum^{ }_{ n\geq 1}{1 \over n-1/2} {1
\over n} \psi( n) z^n\ . & (F.30) \cr} $$
Both $ \tilde S^{\prime} $ and $ \tilde S^{\prime\prime} $ are calculable.
Using formula (5.12.36) of Ref.$\lbrack$36$\rbrack$, we have
$$ \tilde S^{\prime}( z) = 2 \left\{ -2 \sqrt{ -z}\ {\rm Arctg} \sqrt{ -z} +
{\rm \ell n}(1-z) \right\} \ \ \ \ (z<0)\ . \eqno (F.31) $$
The second one is split into various double series using the series for $
\psi( n): $
$$ \tilde S^{\prime\prime}( z) = -\gamma \ \tilde S^{\prime}( z) + 2\tilde
S_2(z) - 2S_1(z)\ , \eqno (F.32) $$
where $ S_1 $ is the same as in Eq.(F.17), while $ \tilde S_2 $ reads in terms
of the double series
(F.19)
$$ \tilde S_2(z) = -2x\ S_3(x)\ \ \ ,\ \ \ \ \ \ \ x = \sqrt{ -z}\ \ , $$
which has been resummed explicitly in (F.22).

Numerically, for $ z=-1/3, $ $ x={1 \over \sqrt{ 3}}, $ we finally get from
the above the expansion
$$ F_{-1/2} = 1 - {1 \over 2\Gamma( 1-\nu)}  \left\{\tilde S^{\prime}(
-1/3)+(1-\nu)\tilde S^{\prime\prime}( -1/3)+{\cal O} \left[(1-\nu)^ 2 \right]
\right\} \eqno (F.33) $$
with
$$ \eqalign{\tilde S^{\prime}( -1/3) & = 2 \left[-{2 \over \sqrt{ 3}} {\pi
\over 6} + {\rm \ell n} \left({4 \over 3} \right) \right]\ , \cr\tilde
S^{\prime\prime}( -1/3) & = -\gamma \ \tilde S^{\prime}( -1/3)-{4 \over \sqrt{
3}} S_3 \left({1 \over \sqrt{ 3}} \right) -2S_1(-1/3)\ , \cr} \eqno (F.34) $$
together with the values (F.23) $ S_3 = -2L(\pi /6), $ $ S_1={1 \over 2}
\left( {\rm \ell n}{4 \over 3} \right)^2; $ which leads to the
results (4.38).

\vfill\eject

\centerline{{\bf APPENDIX G}}

\vskip 17pt

\centerline{{\bf Numerical check of the Laurent expansion}}

\vskip 10pt

The integrals (4.16) (4.17), whose Bessel form (4.18) (4.19) has been
calculated
in Appendices E and F, can also be expanded formally near $ d=4, $ using
subtracted
integrals in analytical continuation. This allows for a numerical check of the
results (4.40) obtained from analytic continuation of Bessel integrals.
This check helped to discover a global sign error in a formula of
Ref.$\lbrack$36$\rbrack$,
(Appendix E), which otherwise would probably have stayed unnoticed.

We start from
$$ \eqalignno{{\cal I}^{\prime}_ 2 & = \int^{ \infty}_ 0 {\rm d} p\
p^{2-d}F^3(p) & (G.1) \cr{\cal I}^{\prime}_ 3 & = \int^{ \infty}_ 0 {\rm d} p\
p^{3-d}F^3(p)\ , & (G.2) \cr} $$
where
$$ \eqalignno{ F(p) & = \int^{ \infty}_ 0 {\rm e}^{-1/x} {\rm e}^{-px}x^{-d/2}
{\rm d} x & (G.3) \cr  &  = 2p^{\nu /2}K_\nu \left(2 \sqrt{ p} \right)\ \ \ ,\
\ \ \nu  = {d \over 2} - 1\ . & (G.4) \cr} $$
We follow the same method as in Appendix C. Near $ p=0, $ $ F $ reaches a
constant value $ F(0) = \Gamma \left({d \over 2}-1 \right), $
whereas it decays exponentially at infinity. The proper analytical
continuation for $ 3 < d < 4 $ of integral (G.1) is therefore, in this domain
$$ {\cal I}^{\prime}_ 2 = \int^{ \infty}_ 0 {\rm d} p\ p^{2-d}
\left[F^3(p)-F^3(0) \right]\ , \eqno (G.5) $$
whereas $ {\cal I}^{\prime}_ 3 $ stays unchanged. We represent $ F $ as $ F(p)
= F(0)+\bar F(p), $ such that
trivially
$$ {\cal I}^{\prime}_ 2 = \int^{ \infty}_ 0 {\rm d} p\ p^{2-d}
\left[3F^2(0)\bar F(p)+3\bar F^2(p)F(0)+\bar F^3(p) \right]\ . \eqno (G.6) $$
Near $ p=0, $ owing to the expansion (C.3) of Appendix C, we have $ \bar
F={\cal O} \left(p^{d/2-1} \right), $
therefore the two last terms in the integrand of (G.6) give {\it convergent}
integrals at $ d=4. $

This is not true of the first term in the integrand and we split the
corresponding
integral in (G.6) into,
$$ \int^{ \infty}_ 0 {\rm d} p\ p^{2-d}\bar F(p) = \int^ 1_0 {\rm d} p\
p^{2-d} \left[\hat F(p)+\tilde F(p) \right] + \int^{ \infty}_ 1 {\rm d} p\
p^{2-d}\bar F(p)\ , \eqno (G.7) $$
where we have made use of the expansion (C.3) (C.6) (C.8) of $ \bar F, $ such
that $ \hat F(p) = p^{d/2-1}\Gamma( 1-d/2)-p\Gamma( d/2-2). $
Its integral in (G.7) can be exactly calculated, whereas the other two terms
yield
finite integrals for $ d=4. $ At this dimension, we can substitute from (G.4)
and
(C.10)
$$ \eqalign{\bar F_{d=4}(p) & = 2 \sqrt{ p}\ K_1 \left(2 \sqrt{ p} \right) - 1
\cr\tilde F_{d=4}(p) & = 2 \sqrt{ p}\ K_1 \left(2 \sqrt{ p} \right) - 1 - [p\
{\rm \ell n} \ p + (2\gamma -1)p] \cr} $$
in those integrals in (G.6) (G.7), that are convergent at $ d=4. $ The final
result
is therefore
$$ \eqalignno{{\cal I}^{\prime}_ 2 & \doteqdot 3\Gamma^ 2 \left({d \over 2}-1
\right) \left[{\Gamma( 1-d/2) \over 2-d/2} - {\Gamma( d/2-2) \over 4-d}
\right]+{\cal S}_2+{\cal O}(\varepsilon) \ , & (G.8) \cr{\cal S}_2 & = 3 \int^
1_0 {\rm d} p\ p^{-2} \left\{ 2 \sqrt{ p}\ K_1 \left(2 \sqrt{ p} \right)-1-[p\
{\rm \ell n} \ p+(2\gamma -1)p] \right\} &  \cr  &  + 3 \int^{ \infty}_ 1 {\rm
d} p\ p^{-2} \left\{ 2 \sqrt{ p}\ K_1 \left(2 \sqrt{ p} \right)-1 \right\} &
\cr  &  + \int^{ \infty}_ 0 {\rm d} p\ p^{-2} \left\{ 3 \left[2 \sqrt{ p}\ K_1
\left(2 \sqrt{ p} \right)-1 \right]^2+ \left[2 \sqrt{ p}\ K_1 \left(2 \sqrt{
p} \right)-1 \right]^3 \right\} \ , & (G.9) \cr} $$
where of course the first term (which have double poles at $ d=4) $ has also
to be
expanded in Laurent series of $ \varepsilon =4-d. $

A similar but simpler procedure is followed for the less divergent integral $
{\cal I}^{\prime}_ 3 $
(G.3) which leads to
$$ \eqalignno{{\cal I}^{\prime}_ 3 & = \Gamma^ 3(d/2-1) {1 \over 4-d} + {\cal
S}_3 + {\cal O}(\varepsilon) \ , & (G.10) \cr{\cal S}_3 & = \int^ 1_0 {\rm d}
p\ p^{-1} \left\{ \left[2 \sqrt{ p}\ K_1 \left(2 \sqrt{ p} \right) \right]^3-1
\right\} &  \cr  &  + \int^{ \infty}_ 1 {\rm d} p\ p^{-1} \left[2 \sqrt{ p}\
K_1 \left(2 \sqrt{ p} \right) \right]^3\ . & (G.11) \cr} $$
The only utility of those results is the numerical check they offer. The
meromorphic functions in (G.8) (G.10) are easily expanded:
$$ \eqalignno{{\cal I}^{\prime}_ 2 & = - {6 \over \varepsilon^ 2} + {3\gamma
-6 \over \varepsilon}  - {3\pi^ 2 \over 8} + {21 \over 4} \gamma^ 2 - 3\gamma
- 3 + {\cal S}_2+{\cal O}(\varepsilon) \ , & (G.12) \cr{\cal I}^{\prime}_ 3 &
= {1 \over \varepsilon}  + {3\gamma \over 2} + {\cal S}_3+{\cal
O}(\varepsilon) \ . & (G.13) \cr} $$
The identification of those results to the analytical ones given in Eq.(4.40)
yields
$$ \eqalignno{{\cal S}_2 & = -6\gamma^ 2+6\gamma -3-6 \sqrt{ 3} \left[L(\pi
/6)-{\pi \over 6} {\rm \ell n} \ 2 \right] & (G.14) \cr{\cal S}_3 & = -2\gamma
 + {8 \over \sqrt{ 3}} \left[L(\pi /6)-{\pi \over 6} {\rm \ell n} \ 2 \right]\
. & (G.15) \cr} $$
Using the numerical value of the Lobachevsky function (4.37) $ L(\pi /6)\simeq
0.02461..., $
we get from (G.14) (G.15) $ {\cal S}_2 = 1.98008..., $ $ {\cal
S}_3=-2.71703...\ . $ Those values are to be
compared to the independent ones obtained from the integral expressions (G.9)
and
(G.11) involving the special Bessel function $ K_1, $ which can be integrated
numerically:
$$ {\cal S}_2 \simeq  1.98009\ \ \ \ \ \ \ \ \ ,\ \ \ \ \ \ \ \ \  {\cal S}_3
\simeq  -2.717036\ ; $$
thereby achieving the check of the validity of analytical results (4.40) and
(4.43), QED.

\vfill\eject
\centerline{{\bf APPENDIX H}}

\vskip 15pt

\centerline{{\bf Calculation of }$ \lambdab $}

\vskip 10pt

We present here the set of transformations leading from the numerical
expression
(6.21b) for the constant $ \lambda , $
$$ \lambda  = - {1 \over 2} \int^ 1_0 {\rm d} x { {\rm \ell n}[x(1-x)] \over
1-x(1-x)}\ , \eqno (H.1a) $$
obtained and used in field theory, to the one (6.23) obtained in the polymer
approach
$$ \lambda  = 2 \sqrt{ 3} \left[{\pi \over 6} {\rm \ell n} \ 2 - L(\pi /6)
\right]\ . \eqno (H.1b) $$
For the segment $ x \in  [0,1/2], $ one goes over to the variable\footnote{$
^{\dagger} $}{The first
steps were suggested by J.M. Luck.}
$$ \eqalign{ u & = x(1-x)\ \ \ ,\ \ \ \ \ \ \ x \in  [0,1/2]\ ,\ \ \ u \in
[0,1/4] \cr du & = (1-2x) {\rm d} x = (1-4u)^{1/2} {\rm d} x \cr} $$
so that
$$ \lambda  = - \int^{ 1/2}_0 {\rm d} x { {\rm \ell n}[x(1-x)] \over 1-x(1-x)}
= - \int^{ 1/4}_0{du \over( 1-4u)^{1/2}} { {\rm \ell n} \ u \over 1-u}\ .
\eqno (H.2) $$
One defines the auxiliary function
$$ h(s) = \int^{ 1/4}_0{u^s \over 1-u} {1 \over( 1-4u)^{1/2}} du\ , \eqno
(H.3) $$
such that
$$ \lambda  = -h^{\prime}( 0)\ . \eqno (H.4) $$
This function can be written as a Laplace transform
$$ h(s) = {1 \over 2} \int^{ \infty}_ 0 {\rm e}^{-p} {\rm d} p\ ({\cal
L}f)(-p)\ , \eqno (H.5) $$
where
$$ \eqalignno{ f(u) & = u^s \left({1 \over 4}-u \right)^{-1/2}\ , & (H.6)
\cr({\cal L}f)(p) & = \int^{ 1/4}_0 {\rm e}^{-pu}f(u)du\ . & (H.7) \cr} $$
For the real function $ f(u) = (u-a)^{2\mu -1}(b-u)^{2\nu -1}, $ defined on
the segment $ [a,b], $
and equal to zero elsewhere, the Laplace transform reads $\lbrack$46$\rbrack$
$$ \eqalign{({\cal L}f)(p) & = \int^ b_a {\rm e}^{-pu}f(u)du = B(2\mu ,2\nu)(
b-a)^{\mu +\nu -1}p^{-\mu -\nu} {\rm e}^{-{1 \over 2}p(a+b)}M_{\mu -\nu ,\mu
+\nu -1/2}(bp-ap), \cr} $$
where $ M_{\kappa ,\mu^{ \prime}} $ is Whittaker's function
$$ M_{\kappa ,\mu^{ \prime}}( z) = z^{{1 \over 2}+\mu^{ \prime}} {\rm
e}^{-z/2}\ _1F_1 \left({1 \over 2}+\mu^{ \prime} -\kappa ;2\mu^{ \prime} +1;z
\right)\ . $$
We therefore arrive at
$$ ({\cal L}f)(p) = B(2\mu ,2\nu)( b-a)^{2\mu +2\nu -1} {\rm e}^{-bp}\
_1F_1(2\nu ;2\mu +2\nu ;(b-a)p)\ , \eqno (H.8a) $$
where, to recover (H.7), we have to set
$$ a=0\ ,\ \ \ b = 1/4\ ,\ \ \ 2\mu -1=s\ ,\ \ \ \nu =1/4\ \ . \eqno (H.8b) $$
The second Laplace transform (H.5) now reads
$$ h(s) = {1 \over 2}B(2\mu ,2\nu) b^{2\mu +2\nu -1} \int^{ \infty}_ 0 {\rm d}
p\ {\rm e}^{-(1-b)p}\ _1F_1(2\nu ;2\mu +2\nu ;-bp)\ . \eqno (H.9) $$
It can be performed explicitly with the help of $\lbrack$Ref.32, p.860,
formula (7.621.4)$\rbrack$
$$ \int^{ \infty}_ 0 {\rm e}^{-s^{\prime} t}t^{b^{\prime} -1}\ _1F_1(a;c;kt)
{\rm d} t = \Gamma \left(b^{\prime} \right)s^{-b^{\prime}} F
\left(a,b^{\prime} ;c;k/s \right)\ ,\ (\vert s\vert >\vert k\vert) \ . \eqno
(H.10) $$
Here: $ k = -b=-1/4, $ $ s^{\prime} =3/4, $ $ b^{\prime} =1, $ $ a=2\nu =1/2,
$ $ c = 2\mu +2\nu =s+3/2. $ Hence, we find
$$ h(s) = {1 \over 2}B \left(1+s,{1 \over 2} \right) \left({1 \over 4}
\right)^{s+{1 \over 2}}{4 \over 3}F \left({1 \over 2},1;{3 \over 2}+s;-{1
\over 3} \right)\ . \eqno (H.11) $$
We now have to calculate the small-$ s $ expansion of the latter
hypergeometric
series, in a way analogous to, but not identical to that used in Appendix F.

Using the definition (E.2) of the hypergeometric series, we get
$$ h(s) = {1 \over 2} \Gamma( 1+s) \left({1 \over 4} \right)^{s+{1 \over 2}}{4
\over 3} \sum^{ }_{ n\geq 0}{\Gamma \left(n+{1 \over 2} \right) \over \Gamma
\left({3 \over 2}+n+s \right)} z^n\ , \eqno (H.12a) $$
where
$$ z = -1/3\ . \eqno (H.12b) $$
Taking advantage of the doubling formula for the $ \Gamma $-function, we have
$$ 4^s\Gamma \left({3 \over 2}+n+s \right) = {\Gamma( 2+2n+2s) 2 \sqrt{ \pi}
\over \Gamma( 1+n+s) 4^{1+n}}\ ,  $$
which can now be expanded for $ s \longrightarrow  0 $ as
$$ = \Gamma \left({3 \over 2}+n \right) \left[1+2s\ \psi( 2+2n)-s\ \psi(
n+1)+{\cal O} \left(s^2 \right) \right]\ . \eqno (H.13) $$
Overall, this leads to an expression for the derivative of the series (H.12a)
$$ -h^{\prime}( 0) = {1 \over 3} \sum^{ }_{ n\geq 0}z^n {1 \over n+1/2}
[2\psi( 2n+2)-\psi( n+1)-\psi( 1)]\ , \eqno (H.14) $$
where
$$ \doublelow{ \psi( n+1) \cr n\geq 1 \cr}  = -\gamma  + \sum^ n_{k=1}{1 \over
k} \ ,\ \ \ \ \ \ \psi( 1) = -\gamma \ . $$

Isolating the $ n=0 $ term in (H.14) yields
$$ -h^{\prime}( 0) = {1 \over 3}(A+4) \eqno (H.15a) $$
in terms of the double series
$$ A = \sum^{ }_{ n\geq 1}{z^n \over n+{1 \over 2}} \left(2 \sum^{
2n+1}_{k=1}{1 \over k} - \sum^ n_{k^{\prime} =1}{1 \over k^{\prime}} \right)\
. \eqno (H.15b) $$
As in Appendix F, we define the auxiliary series (F.19)
$$ S_3(x) = \sum^{ }_{ n\geq 1}(-1)^n {x^{2n+1} \over 2n+1} \sum^ n_{k=1}{1
\over k}\ , \eqno (H.16) $$
and
$$ S_4(x) = \sum^{ }_{ n\geq 1}(-1)^{n-1} {x^{2n+1} \over 2n+1} \sum^{
2n}_{k=1}{1 \over k}\ , \eqno (H.17) $$
and finally the $ {\rm Ti}_2 $ function $\lbrack$47$\rbrack$
$$ {\rm Ti}_2(x) = \sum^{ }_{ n\geq 0}(-1)^n {x^{2n+1} \over( 2n+1)^2}\ .
\eqno (H.18) $$
With this notation, we have for $ z = -x^2, $ $ x=1/ \sqrt{ 3} $
$$ A = {2 \over x} \left[-2S_4(x)+2 \left[ {\rm Ti}_2(x)-x \right]-S_3(x)
\right]\ . \eqno (H.19) $$
The series (F.19) (H.16) was calculated explicitly in (F.22a), in terms of
the Lobachevsky function
$$ S_3(x) = -2L( {\rm Arctg} \ x) \eqno (H.20) $$
while $ S_4 $ is $\lbrack$36$\rbrack$
$$ S_4(x) = {1 \over 2} {\rm \ell n} \left(1+x^2 \right) {\rm Arctg} \ x\ .
\eqno (H.21) $$
The special function $ {\rm Ti}_2 $ possesses the integral representation
$\lbrack$47$\rbrack$
$$ {\rm Ti}_2(x) = \int^ x_0 {\rm Arctg} \ y { {\rm d} y \over y}\ . \eqno
(H.22) $$
Setting $ x= {\rm tg} \ \theta , $ we get
$$ {\rm Ti}_2(x) = \int^ \theta_ 0\theta^{ \prime} {\rm d} \left[ {\rm \ell n}
\left( {\rm tg} \ \theta^{ \prime} \right) \right]\ . $$
Integrating by parts and using the definition of the Lobachevsky function
$$ L(\theta)  = - \int^ \theta_ 0 {\rm \ell n} \left( {\rm cos} \ \theta^{
\prime} \right) {\rm d} \theta^{ \prime} $$
yields
$$ {\rm Ti}_2(x= {\rm tg} \ \theta)  = \theta \ {\rm \ell n}( {\rm tg} \
\theta)  - L(\theta)  + L \left({\pi \over 2} \right) - L \left({\pi \over 2}
- \theta \right)\ . \eqno (H.23) $$
The particular value of $ x $ is $ x = 1/ \sqrt{ 3}, $ for which Eqs.(H.20),
(H.21), and (H.23) give
$$ S_3 \left(1/ \sqrt{ 3} \right) = -2L(\pi /6)\ ,\ \ S_4 \left(1/ \sqrt{ 3}
\right) = {1 \over 2} {\rm \ell n}(4/3){\pi \over 6}\ , \eqno (H.24) $$
and
$$ {\rm Ti}_2 \left(1/ \sqrt{ 3} \right) = - {\pi \over 6} {1 \over 2} {\rm
\ell n} \ 3 - L(\pi /6) + L(\pi /2) - L(\pi /3)\ . $$
For this particular case, we furthermore have by duality $\lbrack$32$\rbrack$
$$ L(\pi /3) = {3 \over 2}L(\pi /6) + {\pi \over 12} {\rm \ell n} \ 2\ ,\ \ \
\ L(\pi /2) = {\pi \over 2} {\rm \ell n} \ 2\ , $$
so that
$$ {\rm Ti}_2 \left(1/ \sqrt{ 3} \right) = - {\pi \over 12} {\rm \ell n} \ 3 +
\left({\pi \over 2} - {\pi \over 12} \right) {\rm \ell n} \ 2 - {5 \over 2}
L(\pi /6)\ . \eqno (H.25) $$
Finally, collecting together the expressions (H.15a), (H.19), (H.24) and
(H.25),
we arrive at
$$ \lambda  = -h^{\prime}( 0) = 2 \sqrt{ 3} \left({\pi \over 6} {\rm \ell n} \
2 - L \left({\pi \over 6} \right) \right)\ , $$
Q.E.D.

\vfill\eject

\centerline{{\bf REFERENCES}}

\vskip 17pt

\item{$\lbrack$1$\rbrack$}M.E. Fisher, {\it Rev. Mod. Phys.} {\bf 46} (1974)
597.

\item{$\lbrack$2$\rbrack$}See, e.g., J. Zinn-Justin, \lq\lq Quantum Field
Theory and Critical
Phenomena\rq\rq , Clarendon Press, Oxford (1989).

\item{$\lbrack$3$\rbrack$}M.E. Fisher, \lq\lq General Scaling Theory for
Critical Points\rq\rq , in Proc.
Nobel Symposium $ 24^{ {\rm th}} $ (1973), on \lq\lq Collective Properties of
Physical Systems\rq\rq ;

\item{\nobreak\ \nobreak\ }G.A. Baker, in \lq\lq Phase Transitions and
Critical Phenomena\rq\rq , Vol.9, C.
Domb and J.L. Lebowitz eds. (Academic Press, London, 1984) p.234.

\item{$\lbrack$4$\rbrack$}D. Stauffer, M. Ferer and M. Wortis, {\it Phys. Rev.
Lett.} {\bf 29}
(1972) 345.

\item{$\lbrack$5$\rbrack$}M. Ferer, M.A. Moore and M. Wortis, {\it Phys. Rev.}
{\bf B8} (1973)
5205.

\item{$\lbrack$6$\rbrack$}V. Privman, P.C. Hohenberg and A. Aharony, in
\lq\lq Phase Transitions
and Critical Phenomena\rq\rq , Vol.14, C. Domb and J. Lebowitz eds. (Academic
Press, London, 1991) p.1.

\item{$\lbrack$7$\rbrack$}P.G. de Gennes, {\it Phys. Lett.} {\bf A38} (1972)
339, \lq\lq Scaling Concepts
in Polymer Physics\rq\rq\ (Cornell University Press, Ithaca, New York, 1979).

\item{$\lbrack$8$\rbrack$}J. des Cloizeaux, {\it J. Physique} {\bf 36} (1975)
281.

\item{$\lbrack$9$\rbrack$}J. des Cloizeaux and G. Jannink, \lq\lq Polymers in
Solution, their
Modeling and Structure\rq\rq\ (Clarendon Press, Oxford, 1989).

\item{$\lbrack$10$\rbrack$}J.J. Prentis, {\it J. Chem. Phys.} {\bf 76} (1982)
1574.

\item{$\lbrack$11$\rbrack$}V. Privman and S. Redner, {\it J. Phys. A. Math.
Gen.} {\bf 18} (1985)
L781.

\item{$\lbrack$12$\rbrack$}V. Privman and J. Rudnick, {\it J. Phys. A. Math.
Gen.} {\bf 18} (1985)
L789.

\item{$\lbrack$13$\rbrack$}I.G. Enting and A.J. Guttmann, {\it J. Phys. A.
Math. Gen.} {\bf 18}
(1985) 1007; {\it ibid.} {\bf 25} (1992) 2791, and references therein.

\item{$\lbrack$14$\rbrack$}J.L. Cardy, {\it J. Phys. A. Math. Gen.} {\bf 21}
(1988) L797.

\item{$\lbrack$15$\rbrack$}J.L. Cardy and A.J. Guttmann, {\it J. Phys. A.
Math. Gen.} {\bf 26}
(1993) 2485.

\item{$\lbrack$16$\rbrack$}S.F. Edwards, {\it Proc. Phys. Soc. Lond.} {\bf 85}
(1965) 613.

\item{$\lbrack$17$\rbrack$}J. des Cloizeaux, {\it J. Physique} {\bf 42} (1981)
635.

\item{$\lbrack$18$\rbrack$}B. Duplantier, {\it J. Physique} {\bf 47} (1986)
569.

\item{$\lbrack$19$\rbrack$}M. Benhamou and G. Mahoux, {\it J. Phys. Lett.}
{\bf 46} (1985) L689.

\item{$\lbrack$20$\rbrack$}A. Gervois and H. Navelet, {\it J. Math. Phys.}
{\bf 27} (1986) 682.

\item{$\lbrack$21$\rbrack$}A. Gervois and H. Navelet, {\it J. Math. Phys.}
{\bf 27} (1986) 688.

\item{$\lbrack$22$\rbrack$}C. Bervillier, {\it Phys. Rev.} {\bf B14} (1976)
4964.

\item{$\lbrack$23$\rbrack$}M.C. Chang and A. Houghton, {\it Phys. Rev.} {\bf
B21} (1980) 1881.

\item{$\lbrack$24$\rbrack$}Y. Okabe and K. Ideura, {\it Prog. Theor. Phys.}
{\bf 66} (1981) 1959.

\item{$\lbrack$25$\rbrack$}J.F. Nicoll and P.C. Albright, {\it Phys. Rev.}
{\bf B31} (1985) 4576.

\item{$\lbrack$26$\rbrack$}C. Bagnuls, C. Bervillier, D.I. Meiron and B.G.
Nickel, {\it Phys.
Rev.} {\bf B35} (1987) 3585.

\item{$\lbrack$27$\rbrack$}B. Duplantier, {\it J. Chem. Phys.} {\bf 86} (1987)
4233.

\item{$\lbrack$28$\rbrack$}J. des Cloizeaux, {\it J. Physique} {\bf 43} (1982)
1743\nobreak\ ; M. Muthukumar
and B.G. Nickel, {\it J. Chem. Phys.} {\bf 80} (1984) 5839.

\item{$\lbrack$29$\rbrack$}B.H. Zimm, W.H. Stockmayer and M. Fixman, {\it J.
Chem. Phys.} {\bf 21}
(1953) 1716.

\item{$\lbrack$30$\rbrack$}F. David, B. Duplantier and E. Guitter, {\it Nucl.
Phys. B $\lbrack$FS$\rbrack$}
{\bf 394} (1993) 555.

\item{$\lbrack$31$\rbrack$}C. Itzykson and J.-B. Zuber, \lq\lq Quantum Field
Theory\rq\rq , (McGraw Hill,
New York, 1980).

\item{$\lbrack$32$\rbrack$}I.S. Gradshteyn and I.M. Ryzhik, \lq\lq Table of
Integrals, Series, and
Products\rq\rq , (Academic, New York, 1981).

\item{$\lbrack$33$\rbrack$}B. Duplantier, {\it J. Physique} {\bf 47} (1986)
1865.

\item{$\lbrack$34$\rbrack$}B. Duplantier, in \lq\lq Fundamental Problems in
Statistical Mechanics
VII\rq\rq , H. van Beijeren, Editor, (Elsevier, Amsterdam, 1990).

\item{$\lbrack$35$\rbrack$}A.D. Sokal, \lq\lq Static Scaling Behavior of
High-Molecular-Weight
Polymers in Dilute Solution: A Reexamination\rq\rq\ NYU preprint
NYU-TH-93/05/01.

\item{$\lbrack$36$\rbrack$}E.R. Hansen, \lq\lq A Table of Series and
Products\rq\rq , (Prentice-Hall,
Inc., Englewood Cliffs, N.J., 1975).

\item{$\lbrack$37$\rbrack$}B. Duplantier, {\it Nucl. Phys.} {\bf B275}
$\lbrack$FS17$\rbrack$ (1986) 319.

\item{$\lbrack$38$\rbrack$}B. Duplantier and P. Pfeuty, {\it J. Phys.} {\bf
A15} (1982) L127.

\item{$\lbrack$39$\rbrack$}C. Itzykson, J.M. Luck, {\it J. Phys.} {\bf A19}
(1986) 211.

\item{$\lbrack$40$\rbrack$}J.C. Le Guillou and J. Zinn-Justin, {\it J.
Physique Lett.}
(Paris) {\bf 46} (1985) L137; {\it J. Physique} (Paris) {\bf 48} (1987) 19;
{\bf 50} (1989)
1365.

\item{$\lbrack$41$\rbrack$}C. Domb and F.T. Hioe, {\it J. Chem. Phys.} {\bf
51} (1969) 223.

\item{$\lbrack$42$\rbrack$}C. Bagnuls and C. Bervillier, {\it Phys. Rev.}
{\bf B32} (1985) 7209.

\item{$\lbrack$43$\rbrack$}C. Bervillier and C. Godr\`eche, {\it Phys. Rev.}
{\bf B21} (1980) 5427.

\item{$\lbrack$44$\rbrack$}M.N. Barber and B.W. Ninham, \lq\lq Random and
Restricted Walks\rq\rq ,
(Gordon and Breach, New York, 1970).

\item{$\lbrack$45$\rbrack$}\lq\lq Higher Transcendental Functions\rq\rq ,
Vol.1, Bateman Manuscript
Project (McGraw-Hill, New York, 1953).

\item{$\lbrack$46$\rbrack$}\lq\lq Tables of Integral Transforms\rq\rq , Vol.1,
Bateman Manuscript Project
(McGraw-Hill, New York, 1954).

\item{$\lbrack$47$\rbrack$}L. Lewin, \lq\lq Dilogarithms and Associated
Functions\rq\rq , (McDonald,
London, 1958).

\vfill\eject

\centerline{{\bf FIGURE CAPTIONS}}

\vskip 17pt

\item{Fig.1.\nobreak\ :}Rooted Feynman diagrams contributing to $
\superone{{\cal Z}}(S) $
to order $ b^2, $ the diagrams (0), (1), (2a), (2b) corresponding respectively
to $ \superone{{\cal Z}}^{(0)} $
(3.3a), $ \superone{{\cal Z}}^{(1)} $ (3.3b), and $ \superone{{\cal
Z}}^{(2)}_{(a)} $ (3.3c), $ \superone{{\cal Z}}^{(2)}_{(b)} $
(3.3d).

\item{Fig.2.\nobreak\ :}Polymer Feynman diagrams corresponding to the first
terms of
the expansion of $ {\cal Z} \left(S,\vec k \right), $ namely $ {\cal Z}^{(0)}
$ (4.6), $ {\cal Z}^{(1)} $ (4.7), and $ {\cal Z}^{(2)}_{(a,b,c)} $ (4.9),
(4.10) and (4.11). The size $ s_i $ of the $ i^{ {\rm th}} $ segment along the
thick polymer
line is $ s_i\equiv x_iS, $ $ 0\leq x_i\leq 1. $

\item{Fig.3.\nobreak\ :}Cut ring diagram, with flows of momenta, corresponding
to the
Fourier representation (B.2).

\listrefs
\draftend
\end